\documentclass[twocolumn]{aastex701}

\usepackage{xcolor}
\usepackage{amsmath}
\usepackage{affils}
\usepackage{enumitem}
\usepackage{booktabs}

\newcommand{\HI}{\ion{H}{1}}

\newcommand{\simpropto}{\mathrel{\vcenter{
  \offinterlineskip\halign{\hfil$##$\cr
    \propto\cr\noalign{\kern2pt}\sim\cr\noalign{\kern-2pt}}}}}

\newcommand{\kms}{{\rm km\ s^{-1}}}

\newcommand{\logmstar}[1][]{$\log_{10}({\rm M_\star/M_\odot})$#1}

\newcommand{\sagabg}{SAGAbg-A}
\newcommand{\sagalocal}{SAGAbg-SMF}
\newcommand{\aat}{AAT/2dF}
\newcommand{\mmt}{MMT/Hectospec}

\newcommand{\bolshoip}{\textit{Bolshoi-Planck}}

\defcitealias{kadofong2024a}{Paper I}
\defcitealias{kadofong2024b}{Paper II}

\newcommand{\ntot}[0]{3097}
\newcommand{\nfield}[0]{1562}
\newcommand{\nsat}[0]{1093}
\newcommand{\niso}[0]{908}
\newcommand{\nbga}[0]{28281}
\newcommand{\zmax}[0]{0.05}

\usepackage[english]{babel}
\addto\extrasenglish{%
}
\addto\extrasenglish{%
}

\newcommand{\rrr}[1]{#1}
\newcommand{\mathrrr}[1]{#1}

\defcitealias{paperone}{Paper I}
\defcitealias{papertwo}{Paper II}

\graphicspath{{./}{PaperIII_figures/}}

\begin{document}
\shortauthors{Kado-Fong et al.}

%\title{SAGAbg III: The Stellar Mass Function, Stellar-to-Halo Mass Relation, and Quenched Fraction of Field Dwarfs}
%\title{SAGAbg III: Mass and Star Formation Demographics in a Census of Low-Redshift Dwarfs}

%\title{SAGAbg III: Inventorying the Mass and Star Formation Demographics of Low-Redshift Dwarfs}
\title{SAGAbg III: Environmental Stellar Mass Functions, Self-Quenching, and the Stellar-to-Halo Mass Relation in the Dwarf Galaxy Regime}

\author[0000-0002-0332-177X]{Erin Kado-Fong}
\affiliation{Physics Department, Yale Center for Astronomy \& Astrophysics, PO Box 208120, New Haven, CT 06520, USA}
\email[show]{erin.kado-fong@yale.edu}

\author[0000-0002-1200-0820]{Yao-Yuan Mao}
\affiliation{Department of Physics and Astronomy, University of Utah, Salt Lake City, UT 84112, USA}
\email{yymao@astro.utah.edu}

\author[0000-0002-8320-2198]{Yasmeen Asali}
\affiliation{Department of Astronomy, Yale University, New Haven, CT 06520, USA}
\email{yasmeen.asali@yale.edu}

\author[0000-0002-7007-9725]{Marla~Geha}
\affiliation{Department of Astronomy, Yale University, New Haven, CT 06520, USA}
\affiliation{Physics Department, Yale Center for Astronomy \& Astrophysics, PO Box 208120, New Haven, CT 06520, USA}
\email{marla.geha@yale.edu}

\author[0000-0003-2229-011X]{Risa H. Wechsler}
\affiliation{Kavli Institute for Particle Astrophysics and Cosmology and Department of Physics, Stanford University, Stanford, CA 94305, USA}
\affiliation{SLAC National Accelerator Laboratory, Menlo Park, CA 94025, USA}
\email{rwechsler@stanford.edu}

\author[0000-0002-4739-046X]{Mithi A. C. de los Reyes}
\affiliation{Department of Physics and Astronomy, Amherst College, 25 East Drive, Amherst, MA 01002}
\email{mdelosreyes@amherst.edu}

\author[0000-0001-8913-626X]{Yunchong Wang}
\affiliation{Kavli Institute for Particle Astrophysics and Cosmology and Department of Physics, Stanford University, Stanford, CA 94305, USA}
\email{ycwang19@stanford.edu}

\author[0000-0002-1182-3825]{Ethan O. Nadler}
\affiliation{Department of Astronomy \& Astrophysics, University of California, San Diego, La Jolla, CA 92093, USA}
\email{enadler@ucsd.edu}

\author[0000-0002-3204-1742]{Nitya Kallivayalil}
\affiliation{Department of Astronomy, University of Virginia, 530 McCormick Road, Charlottesville, VA 22904, USA}
\email{njk3r@virginia.edu}

\author[0000-0002-9599-310X]{Erik J. Tollerud}
\affiliation{Space Telescope Science Institute, 3700 San Martin Drive, Baltimore, MD 21218, USA}
\email{etollerud@stsci.edu}

\author[0000-0001-6065-7483]{Benjamin Weiner}
\affiliation{Department of Astronomy and Steward Observatory, University of Arizona, Tucson, AZ 85721, USA}
\email{bjw@as.arizona.edu}

\correspondingauthor{Erin Kado-Fong} 
%\email{erin.kado-fong@yale.edu}
  
\date{\today}

\begin{abstract}
%Understanding the physical processes that govern low-mass galaxy assembly is key not only to building a complete picture of the dwarf galaxy population, but to answering fundamental questions in extragalactic astrophysics such as the 
%Assembling a complete physical picture of how low-mass galaxies evolve in isolation is key to answering large open questions in galaxy evolution, including how baryonic feedback interacts with the dark matter structure of low-mass haloes, and how massive galaxies influence the evolution of its dwarf satellites. 
Recent efforts have extended our view of the number and properties of satellite galaxies beyond the Local Group firmly down to $\rm M_\star\sim 10^6 M_\odot$. A similarly complete view of the field dwarf population has lagged behind.
Using the background galaxies sample from the Satellites Around Galactic Analogs (SAGA) Survey at $z<0.05$, we take inventory of the dwarf population down to $\rm M_\star \sim 5\times10^6 M_\odot$ using three metrics: the stellar mass function (SMF) as function of environment, the stellar-to-halo mass relation (SHMR) of dwarf galaxies inferred via abundance matching, and the quenched fraction of highly isolated dwarfs.
We find that the low-mass SMF shape shows minimal environmental dependence, with the field dwarf SMF described by a low-mass power-law index of $\alpha_1=-1.44\pm0.09$ down to $\rm M_\star \sim 5\times10^6 M_\odot$, and that the quenched fraction of isolated dwarfs drops monotonically to $f_{q} \sim 10^{-3}$ at $\rm M_\star \sim \rm 10^{8.5} M_\odot$.
Though slightly steeper than estimates from \HI{} kinematic measures, our inferred SHMR agrees with literature measurements of satellite systems, consistent with minimal environmental dependence of the SHMR in the probed mass range.
Finally, although most contemporary cosmological simulations against which we compare accurately predict the \sagalocal{} SHMR, we find that big-box cosmological simulations largely over-predict isolated galaxy quenched fractions via a turnaround in $f_q(\rm M_\star)$ at $\rm 10^8\lesssim M_\star/M_\odot\lesssim 10^9$, underscoring the complexities in disentangling the drivers of galaxy formation and the need for systematic multidimensional observations of the dwarf population across environments.

%Our inferred field SHMR bridges previous abundance matching and empirical modeling results at its high mass end, and mass modeling of individual dwarfs at its low mass end. 
%This SHMR is also in statistical agreement with the majority of contemporary cosmological simulations, indicating a multiplicity in viable pathways to achieving observed galaxy formation efficiencies down to low halo masses.

\end{abstract}

\section{Introduction}
The galaxy stellar mass function (SMF), the differential number density of galaxies as a function of their stellar mass, is a key observable by which we inventory the content of the Universe. Though there are substantial uncertainties in the estimation of stellar masses at both the individual galaxy and population level, observed SMFs have shown remarkable consistency across the literature for massive low-redshift galaxies \citep{peng2010, baldry2012, weigel2016, wright2017, driver2022}.

Measurements of the SMF at low stellar masses, however, have been elusive due to the difficulty in obtaining large samples of so-called ``dwarf'' galaxies, typically defined as galaxies with stellar masses $\rm M_\star \lesssim 10^9 M_\odot$. While these galaxies are the most common type of galaxy at all redshifts, they remain at the observational frontier of extragalactic astronomy, both because they are intrinsically faint systems, and because obtaining reliable redshifts from photometry alone can be challenging for systems outside the resolved star regime. 
As such, constructing statistical samples of dwarf galaxies over large volumes remains a resource-intensive endeavor.

In the nearby ($z\sim 0$) and low-redshift ($z\lesssim 0.2$) Universe, searches for dwarf galaxies associated with more massive galaxies have led to increasingly precise stellar mass function measurements of satellite dwarfs \citep{chiboucas2013, bennet2019, crnojevic2019, carlsten2022, dolivadolinsky2023, sagaiii}.
%Similarly, many of the closest and most well-studied dwarf galaxies lie within the environmental influence of the Milky Way and/or Local Group \citep{simon2019, drlicawagner2020}. 
Measurements of the SMF for field dwarfs outside the influence of a massive host have remained observationally challenging. This is due in part to the difficulty of selecting low-mass, low-redshift objects from imaging alone \citep{sagaii, wu2022, darraghford2023, danieli2024}.
Nevertheless, it i  s of substantial importance to understand the SMF across extragalactic environments due to the key role that the SMF plays in constraining galaxy formation models and probing the stellar-to-halo mass relation \citep{behroozi2013, read2017, moster2018, behroozi2019, nadler2020, manwadkar2022, sagav}.

The relationship between the baryonic and dark matter content of low-mass galaxies has been a key, if elusive, ingredient in our understanding of both cosmology and extragalactic astrophysics. Dwarf galaxies have long been identified as key testing grounds of $\Lambda$CDM \citep{lyndenbell1976, kunkel1976, maschenko2008, boylankolchin2011, boylankolchin2012, mcgaugh2012, tollerud2014, bullock2017, sales2022};
%; notably, well-known challenges to $\Lambda$CDM such as the ``Core-Cusp'' problem \citep[e.g.,][]{maschenko2008}, the ``Too Big to Fail`` problem \citep[e.g.,][]{boylankolchin2011, boylankolchin2012}, the ``Plane of Satellites Problem'' \citep[e.g.,][]{lyndenbell1976, kunkel1976, tollerud2014} and the ``Diversity'' problem \citep[e.g.,][]{mcgaugh2012} have manifested in the dwarf mass regime \citep[for a review, see][]{bullock2017, sales2022}. 
many accepted or proposed solutions to these challenges invoke baryonic feedback prescriptions to redistribute dark matter or modulate the galaxy--halo connection \citep{navarro1996, dicintio2014, gnedin2002, governato2012, madau2014, mashchenko2008, onorbe2015, read2016, buck2019, munshi2021, mostow2024, muni2025}. The connection between dark matter and baryons in low-mass galaxies is thus key not only to understanding the structure and evolution of dwarf galaxies, but to our understanding of the nature of dark matter at large.

Like the stellar mass function, measuring the stellar-to-halo mass relation (SHMR) of the dwarf galaxy population outside the influence of a more massive host is a long-standing observational challenge. This problem is, again, largely technical in nature: nearby dwarf galaxies for which we can measure dark matter halo masses via kinematics are typically associated with the Milky Way or Local Group \citep[see, e.g.][]{vandenbosch2001, stark2009,oh2015,read2017}. Individual measurements of halo masses via kinematic means also incur large uncertainties, and are necessarily directly measurable only out to the radius at which reliable kinematic properties can be securely measured.
% moreover, individual halo mass measurements often have large uncertainties, so this technique is not really (currently) viable for statistically constraining P(M* | M_halo); rather, it measures P(M_halo | M*) for a limited sample of objects; more on this below.
Recent progress has been made in measuring halo profiles of dwarf galaxies from weak lensing; however, this relies on stacked photometric samples rather than individual spectroscopic measurements
%and a recent weak lensing measurement of dwarfs relies on a photometric sample 
\citep{thornton2024}. 

In this work, we leverage the background spectra taken as part of the Satellites Around Galactic Analogs (SAGA) Survey to measure the stellar mass function of field dwarf galaxies down to $\rm M_\star \sim 5\times10^6 M_\odot$. 
The recently completed SAGA Survey executed a census of the satellite systems of 101 nearby Milky Way-like analogs with the goal of placing the MW satellite system into cosmological context \citep{sagai, sagaii, sagaiii, sagaiv, sagav}.  To identify the $\sim 400$ satellites of the SAGA hosts identified in the main survey, more than 40000 total spectra were taken of candidate dwarf satellites; due to the photometric selection of these candidates, the objects that comprise the SAGA Background Sample (SAGAbg) are preferentially low-mass, low-redshift galaxies. 

We presented the SAGAbg in \citetalias{paperone} and \citetalias{papertwo} of this series, where we leveraged the high completeness of the SAGAbg sample for low-redshift dwarfs to estimate the mass-loading factor of dwarf galaxies and the late-time evolution of the Star-Forming Sequence, respectively. Here, we further leverage our characterization of the selection function of the SAGAbg galaxies to measure the stellar mass function of dwarf galaxies out to $z\sim 0.05$. 

Our approach to constructing SMF as a function of environment is presented across two sections:  \autoref{s:data}, which delineates the \sagalocal{} sample within the SAGA Survey, and  \autoref{s:smf}, which describes the observational corrections we derive to measure the underlying SMF. We compute spectroscopic targeting corrections that account for the effective targeting volume of the SAGA Survey for each galaxy in the sample \autoref{s:smf:completeness}, and establish that photometric targeting incompleteness is highly subdominant in the construction of our observation-corrected SMF \autoref{s:smf:photcomplete}. 

We fit the form of the SMF and use the measured SMF to derive two further results in \autoref{s:model}: the stellar-to-halo mass relation of field dwarf galaxies down to $\rm M_\star \sim 5\times10^6 M_\odot$ via abundance matching in \autoref{s:model:am}, and  the mass-dependence of quenching in isolated dwarf galaxies down to $\rm M_\star \sim 4\times 10^8 M_\odot$ in \autoref{s:model:fq}. To guide the reader through this methodology, we present this outline in schematic form in \autoref{s:appendix:schematic}.

These measurements push SMF and SHMR estimates of field dwarfs into the classical dwarf regime ($\rm 10^5<M_\star<10^7M_\odot$, \citealt{bullock2017}), both extending our understanding of this extragalactic mass inventory in central galaxies and providing environmental context to previous measurements of the SMF and SHMR in satellite dwarf systems. 

We adopt a flat $\Lambda$CDM cosmology with $\Omega_m=0.3$ ($\Omega_\Lambda=0.7$), and
$H_0=70\ \kms{}\ {\rm Mpc^{-1}}$. We use a \cite{kroupa2001} initial mass function (IMF) unless
otherwise specified, and convert literature results that use other IMFs to a 
\cite{kroupa2001} IMF
as stated in the text.

\begin{figure*}[ht!]
    \centering
    \includegraphics[width=\linewidth]{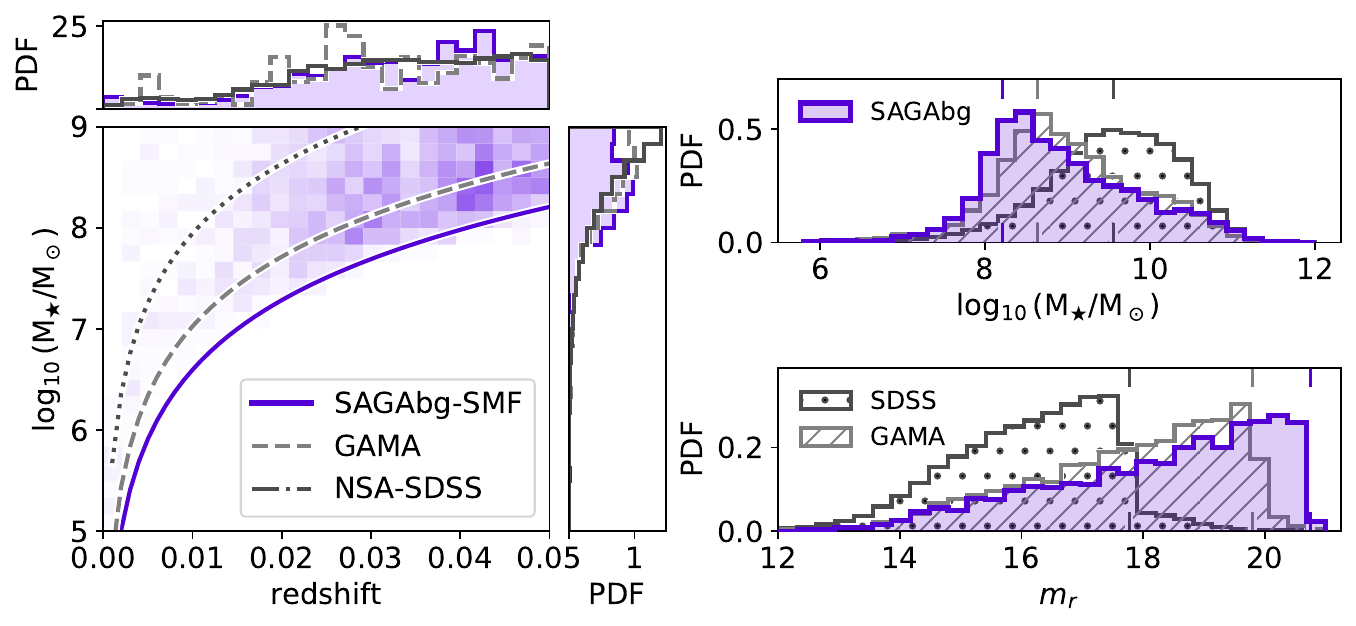}
    \caption{
        The \sagalocal{} sample compared to previous wide-field surveys over the joint distribution in redshift and stellar mass of dwarfs (\rrr{purple 2D histogram,} left), as well as the full stellar mass range (right top) and $r$-band magnitude (right bottom). In each panel, the purple curves correspond to \sagalocal{}, grey dashed and hatched curves to GAMA, and dark grey dotted curves to the NASA Sloan Atlas analysis of SDSS spectra. 
        In the main panel at left, we show the mean stellar mass at the survey limiting magnitude as estimated from the mean mass-to-light relation of the \sagalocal{} galaxies. In both panels at right, we show this approximate stellar mass completeness at $z=0.05$ (top) and $r$-band magnitude limit (bottom) as colored ticks for each survey. %As expected of a survey that goes $\sim 1$ [$\sim 3$] mag deeper in $r$-band limiting magnitude than GAMA [SDSS] (bottom right) with a photometric selection function heavily weighted towards low surface-brightness objects, the SAGAbg sample is distributed towards lower stellar mass at fixed redshift than previous widefield efforts.
    }
    \label{f:selection}
\end{figure*}

\section{The SAGA background sample}\label{s:data}
The SAGAbg sample, as presented in \citetalias{paperone}, is operationally defined as any spectrum within the SAGA spectroscopic catalog that is \textit{not} associated with a SAGA host. 
In the present work, 
we introduce the \sagalocal{} sample: the set of all SAGAbg galaxies with a secure redshift at $0<z<\zmax$ that are not a SAGA satellite, but lie within the targeting-complete footprint of a SAGA host (a projected distance of $<300$ kpc at the host redshift, or $\sim 0.5$ deg).

\subsection{\sagalocal{}}
\sagalocal{}, a set of \ntot{} galaxies at $5\times 10^6 < \rm M_\star < 2\times 10^{11} M_\odot$, contains many of the same low-redshift galaxies as the \sagabg{} sample introduced in \citetalias{paperone} of this series, but does not impose the same spectrum source requirements as \sagabg{}. 
\citetalias{paperone} and \citetalias{papertwo} of this series focused on SAGAbg-A, the subset of all SAGAbg galaxies for which redshifts were obtained either from archival wide-field surveys 
or from dedicated SAGA fiber spectroscopy with \aat{} or \mmt{}. In \autoref{t:samplenumbers} we outline the sizes and differences of the various SAGAbg samples.

We set an upper redshift limit of $z=0.05$ for the \sagalocal{} sample for three reasons: environment classification accuracy, SMF redshift evolution, and low-mass SAGAbg completeness. 
First, we use the HyperLEDA database \citep{hyperleda} to assign environmental 
classifications based on physical proximity to any $M_{K_s}<-23$ galaxy. For an extended source magnitude limit of $m_{K_s}=13.5$, the 2MASS survey is complete 
down to $M_{K_s}=-23$ out to $z=0.046$, modulo $K$-corrections. We thus cannot accurately diagnose the environment of the \sagalocal{} sample beyond $z\sim0.05$.

Second, the stellar mass function is known both theoretically and observationally to evolve with redshift \citep[see, e.g.,][]{muzzin2013}.
In \citetalias{kadofong2024b} of this series, we defined the $z\sim0$ approximation as the redshift range over which negligible galaxy evolution has occurred at the observational precision of that sample. We then demonstrated that for the \sagabg{} sample, the maximum redshift at which the $z\sim0$ approximation holds for the Star-Forming Sequence (SFS) is $z=0.05$. We therefore adopt $z=\zmax{}$ as the maximum redshift for the \sagalocal{} sample. 

Finally, the SAGAbg sample mass sensitivity is strongly redshift-dependent, with the limiting detectable mass rising to $\rm M_\star \sim 10^8 M_\odot$ at $z=0.05$. Because the focus of this work is on the dwarf galaxy population, we therefore restrict our analysis to $z<0.05$.

\subsection{Stellar Mass Estimates}\label{s:smf:smass}

We use the photometric stellar mass-to-light ratio and color prescriptions of \cite{delosreyes2024}, which are updated calibrations tuned for unbiased
performance in stellar mass recovery of simulated galaxies at $\rm M_\star < 10^9 M_\odot$:
\begin{equation}
    % mass = 1.433*gminusr + 0.00153*Mg**2 -0.335*Mg + 2.072
    {M}_\star = 1.433(g-r) + 0.00153 M_g^2 - 0.335 M_g + 2.072,
\end{equation}
where $M_g$ is the restframe $g$-band absolute magnitude. The updated stellar mass prescriptions induce a moderate systematic offset in our stellar mass estimates
with respect to the color--mass relation used in \cite{sagaiii}, with a median difference of $\rm \langle \log_{10}(M_\star/M_\odot)_{SAGA,DR3} - \log_{10}(M_\star/M_\odot)_{dlR}\rangle_{50} = -0.13$ over our sample.

%These updated prescriptions also report a stellar mass uncertainty of $\sim\! 0.1$ dex for an individual galaxy; we incorporate this per-galaxy stellar mass uncertainty into our estimate of the stellar mass function via bootstrap resampling.

\begin{figure*}
    \centering
    \includegraphics[width=\linewidth]{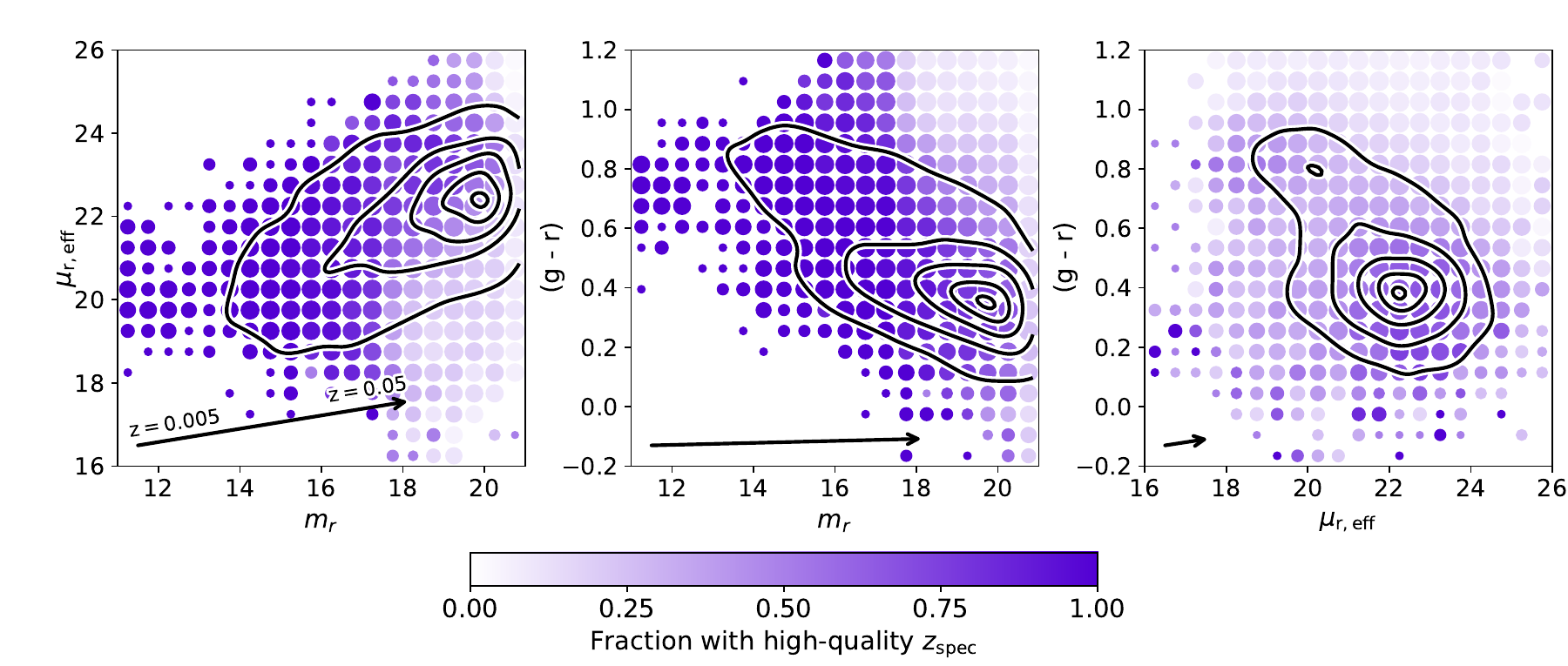}
    \caption{
        The fraction of galaxies that have received a high quality redshift in the parent SAGA photometric catalog as a function of $r$-band magnitude, $r$-band surface brightness, and $(g-r)$ color. The size of each point corresponds to the number of galaxies that contribute to each bin in photometric parameter space. 
        The black contours enclose between 10\% and 95\% of the observed photometric properties of the \sagalocal{} sample, with each contour evenly spaced in enclosed fraction. 
        The black arrow in the corner of each panel shows the median difference in each space of galaxies in our sample between $z=0.005$ and $z=0.05$ due to observational effects alone.
    }
    \label{f:completeness}
\end{figure*}

\subsection{Environmental classifications}\label{s:data:environment}
In order to measure the SMF as a function of environment, we must first establish environmental criteria for the sample.
%In order to measure the SMF and SHMR of field dwarfs from the \sagalocal{} sample, the field galaxies in \sagalocal{} must first be identified. 
To do so, we flag potential massive galaxies in the vicinity of the \sagalocal{} galaxies
by querying the HyperLEDA database for galaxies at heliocentric 
velocities $v<15000\ \kms{}$ ($z\lesssim 0.05$) and absolute $Ks$-band magnitudes 
of $M_{Ks}<-23$ ($\log_{10} M_\star/M_\odot\gtrsim 10.4$). Again, we emphasize that galaxies classified as satellites of the SAGA hosts themselves (i.e., the SAGA satellites presented in \citealt{sagaiii}) are not included in this sample.

We define three environmental classifications for galaxies in the \sagalocal{} sample at $\rm M_\star < 10^{10} M_\odot$, where the upper limit of the stellar mass range is set by the approximate minimum host mass defined by the \cite{geha2012} criterion:
\begin{description}
    \item[Field] galaxies are those at $\rm M_\star < 10^{10} M_\odot$ at $c|z - z_{\rm massive}| > V_{\rm 200, massive}$ and $R_{\rm proj}(z_{\rm massive}) > 1$ Mpc of any $M_{K_s} < -23$ galaxy.    
    \item[Satellite] galaxies are those at $\rm M_\star < 10^{10} M_\odot$ that do not meet the criteria of a field galaxy. Because we define potential hosts as galaxies with $M_{K_s}<-23$ (\logmstar[$\sim10$]), we do not assign field or satellite status to galaxies above this stellar mass threshold.
    \item[Isolated] A more isolated subset of the field galaxies, following \cite{geha2012}. These galaxies must be at $c|z - z_{\rm massive}| > 1000\ \kms{}$ and $R_{\rm proj}(z_{\rm massive}) > 1.5$ Mpc of a $M_{K_s} < -23$ galaxy.
\end{description}

\rrr{Here, $c$ is the speed of light, $z$ is the redshift of each \sagalocal{} galaxy, and $z_{\rm massive}$ is the redshift of the potential host.} This classification scheme requires an estimate of \rrr{halo velocity} $V_{200}$ for the potential hosts in the sample. To approximate $V_{200}$, 
we adopt a $Ks$-band 
mass-to-light ratio of ${M}_\star / L_{K_s} = 1 {M}_\odot/L_\odot$ and compute stellar masses from 2MASS $Ks$-band absolute magnitudes. We then use the
stellar-to-halo mass relation of \cite{behroozi2019} to estimate \rrr{halo masses,} $M_{200}$,   from
$M_\star$. Finally, we relate $M_{200}$ to $V_{200}$ for an NFW halo \citep{nfw} by:
\begin{equation}\label{e:v200}
    \begin{split}
        {M}_{200} &= \frac{V_{200}^3}{10 H_0 G},\\
    \end{split}
\end{equation}
which follows from taking $\rho_{\rm crit}(t) = \left(8\pi G H(t)\right)/3$, and relating the circular velocity of a NFW halo at $R_{200}$ to ${M}_{200}$ and $R_{200}$ via
$V_{200}^2 = G {M}_{200}/R_{200}$.

We refer to measurements over all galaxies in the \sagalocal{} sample as ``\sagalocal{} (all)'' -- however, it should be noted that the SAGA survey fields were selected to avoid known background clusters. As such, our satellite and full samples are skewed away from the richest group and cluster environments.

\subsection{Comparison to Existing Wide-field Surveys}\label{s:data:sample}
We place the \sagalocal{} sample in the context of other wide-field spectroscopic surveys in \autoref{f:selection} by comparing the \sagalocal{} galaxies to the redshift, apparent $r$-band magnitude, and stellar mass distributions of the Galaxy Mass and Assembly \citep[GAMA DR4,][]{driver2022} and the SDSS-derived
NASA Sloan Atlas \citep[\textsf{nsa\textunderscore{}v0\textunderscore{}1\textunderscore{}2},][]{nsa}. 
To minimize systematic offsets, we recompute stellar masses for both the NSA and GAMA catalogs using the prescription of \cite{delosreyes2024}. In both catalogs, we adopt Petrosian magnitudes measured from SDSS imaging\footnote{We draw catalog photometry for \autoref{f:selection} from the \textsf{nsa\_v0\_1\_2} and \textsf{InputCatAv07} for SDSS and GAMA, respectively.} to compute stellar masses.

%The SAGA selection function was furthermore tuned to construct a complete sample of nearby dwarfs, as reflected by the overall increase in successfully redshifted galaxies at lower stellar masses, especially at the lowest redshifts ($z<0.01$) where the SAGA photometric selection is highly complete for low-mass galaxies \citep{sagaiii}.
As expected from its fainter $r$-band limits, the \sagalocal{} sample reaches lower in stellar mass than GAMA and SDSS at fixed redshift. %In the bottom right of \autoref{f:selection} we show that the SAGA Survey reaches $\sim\!1$ magnitude deeper in the $r$-band than the GAMA Survey ($m_{r,\rm lim}=19.8$) and approximately $\sim\! 3$ magnitudes deeper than the NSA ($m_{r,\rm lim}=17.77$). 
To make a rough comparison of the stellar mass completeness limit across surveys, we estimate the stellar mass that corresponds to the limiting $r$-band magnitude of each survey ($m_{r,\rm lim}=\{20.75,19.8,17.77\}$ for SAGA, GAMA, and SDSS, respectively) as a function of redshift based off the mean stellar mass-to-light ratio of the \sagalocal{} galaxies.  The results of this estimate are shown in \autoref{f:selection} as a function of redshift in the main left panel, and at $z=0.05$ in the top right panel. 

Although the SAGA Survey covers a smaller on-sky area ($\Omega_{\rm SAGA}\sim 90\ \rm deg^2$) than GAMA ($\Omega_{\rm GAMA} \sim 180\rm \ deg^2$) and SDSS ($\Omega_{\rm SDSS}\sim 8500\rm\ deg^2$), the fainter targeting limit and dwarf-focused photometric targeting of the SAGA Survey produce a low-mass, $z<0.05$ galaxy sample that is comparable in number to GAMA at \logmstar[$\lesssim 8$] ($N_{\rm SAGAbg-SMF}/N_{\rm GAMA,z<0.05}\approx 0.8$, where $N_{\rm SAGAbg-SMF}=473$).

This rough estimate allows us to compare the spectroscopic targeting limit between surveys as a function of stellar mass, but does not fully describe the implications of the SAGA selection function (which is not a purely magnitude-limited survey like GAMA and SDSS).
Because the SAGA selection function is of tantamount importance to our measurement of the stellar mass function, we will defer our analysis of the \sagalocal{} photometric selection to a standalone discussion in \autoref{s:smf:completeness}.

\section{Measuring the Stellar Mass Function}\label{s:smf}
With our \sagalocal{} sample in hand, we can now consider the measurement of the SMF as a function of environment. 
For readability, we separate our discussion of SMF measurement methodology into several sections: first, we construct corrections for spectroscopic incompleteness in 
\autoref{s:smf:completeness}. These corrections account for the multivariate spectroscopic selection function imposed by the SAGA photometric targeting scheme. This targeting scheme means in practice that the SAGA Survey targeting deviates substantially from a
magnitude-limited sample for which classical $1/V_{\max}$ corrections --- that is, corrections based on the maximum volume $V_{\rm max}$ over which some galaxy can be observed --- are valid. 

Next, we estimate corrections for photometric surface brightness incompleteness in 
\autoref{s:smf:photcomplete}. These corrections account for the existence of galaxies 
within the survey footprint at our mass and redshift range of interest that were not 
included in the original photometric targeting catalogs. As we will demonstrate, the photometric completeness correction is small compared to the spectroscopic completeness correction (and does not enter into our estimate of the isolated dwarf quenched fraction), but can play a significant role in setting the slope of the SMF at the lowest masses we consider in this work.
In \autoref{s:smf:sky}, we derive corrections for the 
size of the SAGA Survey on-sky footprint.

Finally, in \autoref{s:smf:final}, we bring the considerations developed throughout the
section together to construct the framework with which we measure the stellar mass 
function of the \sagalocal{} sample. To guide the reader, we include a schematic figure of this methodology in \autoref{s:appendix:schematic} that summarizes the selection and correction functions constructed to derive the SMFs presented in this work.

\begin{deluxetable}{rccl}
\tablecaption{SAGAbg Sample Descriptions}
\tablewidth{0pt}
\tablehead{
\colhead{Sample} & & \colhead{$\rm N_{gal}$} & \colhead{Description}
}
\startdata
\sagabg{} && \nbga{} & \citetalias{kadofong2024a} \\
\sagalocal{} & All & \ntot{} & \autoref{s:data:sample} \\
 &Field & \nfield{} &  \autoref{s:data:environment} \\
 &Satellite & \nsat{} &  \autoref{s:data:environment}\\
 &Isolated & \niso{} &  \autoref{s:data:environment}\\
\enddata
\tablecomments{
    A summary of the samples discussed in this work. Both \sagabg{ } and \sagalocal{} are 
    drawn from the SAGAbg sample, but \sagalocal{} is not strictly a subset of \sagabg{}. 
}
\end{deluxetable}\label{t:samplenumbers}

\subsection{Spectroscopic Completeness Corrections}\label{s:smf:completeness}
The SAGA Survey targeting strategy was multidimensional, with 
tiered targeting priorities based on host proximity, surface brightness, $r$-band magnitude, and $g-r$ color. 
The targeting selection also changed substantially over the course of the survey as the photometric properties of the SAGA satellites became better understood \citep{sagai, sagaii, sagaiii}.

As such, we modify the traditional $1/V_{\rm max}$ method to calculate incompleteness corrections by computing the effective comoving volume, $V_{\rm eff}$, over which each individual galaxy in the sample would have been 
included in the sample from the probability that it would have been included as a redshifted SAGA background galaxy, ${\rm Pr[obs}|\vec\theta_i(z)]$, given its expected photometric properties as a function of redshift $\theta_i(z)$. 

The SAGA host volumes are explicitly excluded from the SAGAbg sample, and so we include an additional term ${\rm Pr[SAGAbg}|z]$ such that ${\rm Pr[SAGAbg}|z]=0$ within the redshift range corresponding to the SAGA satellite volume. This yields an effective volume defined as:

\begin{equation}\label{e:veff}
    V_{{\rm eff},i} = \int_0^{z_{\rm max}} \frac{dV}{dz}{\rm Pr[SAGAbg}|z]{\rm Pr[obs}|\vec\theta_i(z)] dz,
\end{equation}
where $dV/dz$ is the derivative of comoving volume with respect to redshift and $z_{\rm max}=0.05$. $\vec \theta_i(z)$ refers to the predicted photometric properties of the $i$\textsuperscript{th} galaxy in the sample as a function of redshift: 
\begin{equation}
    \vec \theta_i(z)=\left[ \mu_{r}^p(z,\mu_{r,i}), m_{r}^p(z, m_{r,i}),(g-r)^p(z, (g-r)_i)\right]
\end{equation} 
where the $p$ superscript indicates the predicted values of the $r$-band surface brightness, $r$-band 
magnitude, and $(g-r)$ color as a function of redshift. Here, the redshift dependence of the photometric properties reflects observational considerations and not physical redshift evolution.
In the magnitude-limited case, our definition of $V_{\rm eff}$ becomes equivalent to the classic $V_{\rm max}$, modulo the exclusion of the SAGA host volume.

We empirically determine the spectroscopic success probability over photometric parameter space,
${\rm Pr[obs}|\vec\theta]$, by measuring the fraction of galaxies in the base photometric catalog that were redshifted (SAGA DR3 redshift quality flag of \textsf{ZQUALITY}$\geq3$; see \citealt{sagaiii}) in bins of surface brightness (bins defined as $16<\mu_r<26$ mag arcsec$^{-2}$, $d\mu_r=0.5$), magnitude ($10<m_r<21$, $dm_r=0.55$), and $(g-r)$ color ($-0.2<g-r<2$, $d(g-r)=0.11$).  The success fraction is shown in \autoref{f:completeness} in each projection of the three-dimensional space considered. In each panel, the size of the marker reflects the relative number of sources with the given photometric properties in the SAGA base catalog. The black contours show the distribution of the \sagalocal{} sample as observed, while the black arrow in the bottom left indicates the average direction that a galaxy moves as a function of redshift due to observational effects.

As expected, the SAGA Survey completeness shows a substantial color dependence; redder galaxies are less likely to be included in the sample. This is unsurprising given that: (1) the photometric targeting strategy of the SAGA Survey prioritizes bluer galaxies that are more likely to be low-redshift satellites of SAGA hosts \citep{sagaiii}, and (2) red galaxies are more likely to be quiescent galaxies for which reliable redshifts must be based on absorption features, which have a lower success rate at fixed object magnitude. 
%There is also a subdominant but statistically significant dependence on surface brightness, which is again expected due to the survey targeting strategy.  

%A more apparently surprising result is that, 
Below the SDSS spectroscopic targeting limit, the SAGA Survey is more incomplete for higher surface brightness objects. These high surface brightness galaxies are ostensibly easier to spectroscopically confirm, however, the SAGA Survey was substantially less likely to target them because the probability for these objects to be satellites of a SAGA host is low. The low completeness of high surface-brightness objects can then be understood as a photometric selection choice, rather than an observational constraint. %As illustrated by these two points, we can then say that the \sagalocal{} sample deviates significantly from a magnitude-limited sample, necessitating the use of the $V_{\rm eff}$ strategy described in \autoref{e:veff}.

With an empirical mapping of ${\rm Pr[obs}|\vec\theta]$ in hand, we can now consider the observational redshift dependence of the photometric properties of the galaxies in our sample. Our estimate of $\vec\theta_i(z)$ includes three observational effects: the $1/r^2$ drop-off in spectral flux, the redshift dependence of observed angular size, and the $g$ and $r-$band $k$-corrections for each galaxy. We estimate $k$-corrections for each galaxy as a function of redshift using the photometric relations of \cite{chilingarian2010}.  Between $z=0.005$ and $z=0.05$,
the change in $r$-band magnitude dominates ($\langle |\Delta m_r|| \rangle=6.7$) over the change in $r$-band surface brightness ($\langle |\Delta \mu_r|\rangle = 1.1$) and $g-r$ color ($\langle \Delta |g-r|\rangle = 0.02$). 

The probability that a galaxy will be classified as a SAGA satellite, ${\rm Pr[SAGAbg}|z]$, is straightforward to compute: for each galaxy, we exclude redshifts 
within 275 $\kms{}$ of the SAGA host in whose footprint the galaxy was observed.

\subsection{Exploring the need for photometric completeness corrections}\label{s:smf:photcomplete}

Unlike \citetalias{paperone} and \citetalias{papertwo} of this series, which dealt with the relative selection function of the SAGA background galaxies as defined with respect to the lowest redshift sub-sample of the SAGA background galaxies, the work at present concerns the absolute completeness of the \sagalocal{} selection function. 

As such, we consider whether a correction is neeeded for galaxies that were not identified as sources in the DESI Imaging catalog, which was used as SAGA's original photometric targeting catalog. To do so, we construct a simple model based off of the observed mass-size and mass-to-light ratio distributions of the \sagalocal{} galaxies to predict the distribution of apparent magnitude and surface brightness for the low-mass end of \sagalocal{} sample and estimate the occurrence fraction of galaxies brighter than our spectroscopic targeting limit of $m_r<20.75$ that may be missing from our \rrr{photometric targeting catalog}.

\rrr{Here we sketch out the form of the model and incompleteness correction results; a} full explanation of \rrr{this} incompleteness exercise can be found in \autoref{s:appendix:sbdetection}. \rrr{To estimate the relative number density of galaxies brighter than our spectroscopic targeting limit that are below our photometric catalog detection limits, we would like to know the distribution in apparent magnitude and surface brightness of the underlying galaxy population. And more specifically, at a 5$\sigma$ limiting depth of $m_r=23.9$ \citep{dey2019}, the photometric catalogs are substantially deeper than our spectroscopic limit in apparent magnitude; as such, we focus on the expected surface brightness incompleteness.}

\rrr{We forward model the joint distribution over stellar mass, surface brightness, and redshift of the population using the size-mass relation of Asali et al., 2025 (submitted) and an assumed mass-to-light ratio of ${\rm M_\star}/L_r=0.9 {\rm M}_\odot/L_\odot$ (the median value of the observed \sagalocal{} galaxies). Through Monte Carlo resampling, we can then estimate the probability that galaxies at $m_r<20.75$ of a given stellar mass and redshift will be fainter than our nominal surface brightness limit of $\mu_{r, lim}=24.9$ mag arcsec$^{-2}$, as approximated from the DESI Legacy Imaging survey depth and seeing.}

\rrr{This model for photometric incompleteness explicitly assumes a lognormal form to the mass-size relation; as such, a hidden population of low surface brightness galaxies (LSBGs) could cause an underestimate of our photometric incompleteness. Such a possibility has been explored in depth by \cite{sagaiii}, wherein it was concluded via a comparison to the Systematically Measuring Ultra-diffuse Galaxies \citep[SMUDGes,][]{zaritsky2023} catalog as well as visual inspection of the SAGA fields that no substantial population of LSBGs brighter than the spectroscopic targeting limit was missed in the SAGA photometric catalogs.}

We find that the effects of photometric incompleteness are small down to our stellar mass limit of $\rm M_\star = 5\times10^6 M_\odot$. \rrr{Indeed, when integrated over redshift, the photometric incompleteness correction factor is no more than 1.4 at ${\rm M_\star>5\times10^6 M_\odot}$, and no more than 1.1 at ${\rm M_\star>5\times10^8 M_\odot}$. This} completeness correction for any given mass bin is smaller than the counting uncertainty incurred in that bin. However, because there is a mass dependence to the correction, we choose to include it in our SMF analysis despite the fact that the spectroscopic completeness correction dominates at all masses.

\subsection{Survey footprint}\label{s:smf:sky}
Because the main science goal of the SAGA Survey was to complete
a census of the satellite systems around nearby MW-like hosts, the
survey has a very high completeness of targets at $10<r_{\rm proj}<300$ kpc
in projected distance from a SAGA host. 
This high spatial completeness makes 
estimating the total area of the SAGA Survey straightforward:
the total area is simply the sum of the on-sky area between $10<r_{\rm proj}<300$ kpc from each host for a total of $\Omega_{\rm SAGA}=86.67$ deg$^2$
over 102 hosts (101 SAGA Survey DR3 hosts and 1 additional complete host).
This produces a final
sample of \ntot{} galaxies in our \sagalocal{} sample, 
as tabulated in \autoref{t:samplenumbers}.

\subsection{Constructing the SMF}\label{s:smf:final}

%With the corrections discussed above in hand, we can now compute the completeness-corrected stellar mass function of the \sagalocal{} galaxies.

We define the stellar mass function $\phi(\rm M_\star)$ to be the differential number 
density of galaxies with respect to \logmstar[]. In practice, we bin our galaxy sample by stellar mass such that a bin contains no fewer than 10 objects and is no narrower than 
$\Delta m \equiv \Delta(\log_{10}(\rm M_\star/M_\odot))=0.1$. In each bin, the differential number density of the $i^{\rm th}$ bin, $\phi(\rm M_{\star,i})$, is:

\begin{equation}
    \phi({\rm M}_{\star,i}) = f_{\rm phot}({\rm M}_{\star,i})\frac{\Omega_{\rm sky}}{\Omega_{\rm SAGA}}\sum_j^N \frac{1}{V_{{\rm eff},j} \Delta m_i}, 
\end{equation}
where $\Omega_{\rm sky} = 4\pi$ sr, $j=0,1,...,N$ indexes the $N$ galaxies within the $i^{\rm th}$ stellar mass bin and $\Omega_{\rm SAGA}$ is the on-sky footprint of the SAGA Survey as discussed in 
\autoref{s:smf:sky}.

We consider two sources of uncertainty in our differential number densities: counting uncertainties due to the finite size of our galaxy sample, and weighting uncertainties due to our empirical method of determining the \sagalocal{} selection function.

To estimate counting uncertainties on $\phi(\rm M_\star)$ when the raw number of galaxies observed in the sample is small or zero, we model the posterior probability of the underlying counts $\lambda$ as a Gamma distribution with shape parameter $k+0.5$ for $k$ observed counts. This formulation is proportional to a Poisson distribution centered on $k$ observed counts with a Jeffreys' prior where $P(\lambda)\propto\lambda^{-1/2}$, allowing us to defined reasonable confidence intervals when the number of observed galaxies is low.

Because our photometric catalog is finite in size, this empirical method to estimate the spectroscopic completeness function is itself uncertain; we quantify this uncertainty as the 68\% confidence interval on the  proportion of galaxies with a high-quality redshift assuming a Jeffreys prior on a binominal counting population. We propagate this uncertainty in the individual galaxy weights to our measurement of the SMF as:
\begin{equation}
    \begin{split}
    \sigma_{upper, i}^2 &= \sqrt{\sigma_{{\rm weights},i}^2 + \left[W_i\left(\frac{f(0.84,C_i + 0.5)}{C_i} - 1\right)\right]^2},\\
    \sigma_{lower, i}^2 &= \sqrt{\sigma_{{\rm weights},i}^2 + \left[W_i\left(1 - \frac{f(0.16,C_i + 0.5)}{C_i} \right)\right]^2},
    \end{split}
\end{equation}
where for each stellar mass bin $i$, $\sigma_{{\rm weights},i}$ is the uncertainty of the weights themselves, $C_i$ is the observed galaxy counts in the $i^{\rm th}$ bin, $W_i$ is the sum of the weights associated with the galaxies in the $i^{\rm th}$ bin, and $f(\alpha,k)$ is the percent point function of the Gamma distribution with $k=C_i+0.5$.
%by adding the upper ($84^{\rm th}$ percentile $-$ 50\textsuperscript{th} percentile) and lower ($50^{\rm th}$ percentile $-$ 16\textsuperscript{th} percentile) uncertainties on the sum of the weights ($\sigma_{\rm weights,i}$) to the counting uncertainties from the raw SMF counts ($\sigma_{\rm counting,i}$) in each stellar mass bin $i$ considered. 

It should be noted that this is an ad hoc method to preserve asymmetric uncertainties; in our case, we find that uncertainty budget is strongly dominated by the SMF counts themselves over the uncertainties from the weights and so $\sigma_{SMF} \approx \sigma_{\rm counting}$. This statement can be seen visually in \autoref{f:completeness}: the \sagalocal{} sample (black contours) occupies regions of photometric parameter space that are well-sampled by the photometric and spectroscopic catalogs. %This holds true even as the \sagalocal{} galaxies move through observational parameter space (black arrows). 
Thus, the uncertainty in our estimate of galaxy differential number densities is dominated by the counting statistics on the galaxies themselves, rather than our uncertainty on the volume over which each galaxy would be observed by the SAGA Survey.

\section{Inferring SMF parameters, the SHMR, and Quenched Fractions}\label{s:model}
Having established our measurement of the SMF from the SAGA Survey, we now proceed with three steps. First, we fit standard functional forms to facilitate comparison with existing literature. Second, we present these results in a format that will make our measurements more accessible for future studies. Finally, we apply abundance matching techniques to estimate the relationship between stellar and halo mass across our sample.
%With our measurement of the SMF over the SAGA Survey in hand, we will now fit standard functional forms to compare to literature results, make our measurement of the SMF more easily interpretable against future literature results, and perform abundance 
%matching to estimate the relationship between stellar and halo mass over our sample.

\subsection{Fitting the Stellar Mass Function}\label{s:model:fit}
%For the environment-dependent subsamples, we find that a single Schechter function is a %sufficient functional descriptor:
%\begin{equation}\label{e:singleschechter}
%    \phi(\rm M_\star) \equiv \frac{dN}{d\log_{10}M_\star} = \ln10\left[  \phi_{0,1}\left%( \frac{\rm M_\star}{\rm M_{ch}} \right)^{1+\alpha_1}\right],    
%\end{equation}
%where $\phi_{0,1}$ describes the normalization of the SMF $\phi(\rm M_\star)$, $\rm M_%{ch}$ its characteristic mass and $\alpha_1$ its mass dependence. The subscript %indicates that these parameters are associated with the first (and, in this instance, %only) component of the Schechter function.

We fit the \sagalocal{} stellar mass functions with a double Schechter function \citep[following, e.g.][]{weigel2016, driver2022}, defined as:

\begin{equation}
    \begin{split}
    \phi(M_\star) \equiv \frac{dN}{d\log_{10}M_\star} &= \ln10 \left[ \phi_{0,1}\left( \frac{M_\star}{M_{ch}}\right)^{1+\alpha_1} \right. \\
    &+ \left. \phi_{0,2}\left( \frac{M_\star}{M_{ch}}\right)^{1+\alpha_2} \right]e^{\frac{-M_\star}{M_{ch}}},
    \end{split}
\end{equation}
where $\phi_{0,1}$ describes the normalization of the SMF $\phi(\rm M_\star)$, $\rm M_{ch}$ its characteristic mass and $\alpha_1$ its mass dependence. The subscript indicates whether these parameters are associated with the first or second component of the Schechter function.

\rrr{In our fiducial fitting procedure, we fit a double Schechter function directly to the binned, observed SMF assuming a Gaussian likelihood.} We estimate SMF fit parameters using the \textsf{emcee} \citep{foremanmackey2013} implementation of 
Markov Chain Monte Carlo sampling using 32 walkers.
Because we do not sample the SMF well at stellar masses above the expected value of 
$\rm M_{ch}$, we 
adopt a Gaussian prior centered on $\log_{10}(\rm M_{ch}/M_\odot)=10.8$ and 
$\sigma=2$ based on literature fits to the SMF at higher masses \citep{peng2010, baldry2012}. We assume a log-uniform prior for $\phi_{0}$ bounded by $10^{-4}<\phi_0<10^{-1}$ Mpc$^{-1}$, and a uniform prior for $\alpha$ bounded by $-3<\alpha<0$.
%\footnote{We assume prior bounds of ${\rm Pr}[(\log_{10}(\phi_0)]\sim U[-2,2]$ and ${\rm Pr}[\alpha]\sim U[-3,3]$. We sample $\phi_0$ logarithmically as the density of galaxies must be non-negative.}. 

To prevent walkers from settling at local extrema, we reinitialize our sampler after 3000 steps using a normally distributed initial walker position whose standard deviation is set by the difference between the 60\textsuperscript{th} and 40\textsuperscript{th} percentiles for each parameter in the preceding chain. We repeat until convergence is achieved, as indicated by a maximum Gelman--Rubin statistic of no larger than $R=1.2$ across the fit parameters.

%def predict_fn ( m, phi, alpha ):
%            return np.log10(fixed_component(10.**m) + functions.logschechter(10.**m, 10.**phi*1e-3, fixed_coeffs[1], alpha ))
% nd = np.log(10.)*phi_ast *(m/M_ast)**(alpha+1.)*np.e**(-m/M_ast)

\subsection{Abundance matching}\label{s:model:am}
To interpret the implications of our SMF results in the context of the galaxy--halo connection of dwarf galaxies, we make an estimate of the stellar-to-halo mass relation of our field galaxy sample via simple abundance matching of halo and galaxy number densities. 

Following classical abundance matching \citep[e.g.][]{conroy2006}, we map halo masses to stellar masses by finding the halo and stellar mass pair ($M_{\rm peak}$,$M_\star$) such that
\begin{equation}
    \rm \int_{M_{\rm peak}}^\infty \phi_{h}(M)dM = \int_{M_\star}^\infty \phi_\star(M)dM,
\end{equation} 
where $\phi_h({\rm M},\epsilon)$ and $\phi_\star({\rm M})$ are the peak halo mass and stellar mass functions, respectively. The peak halo mass is the maximum halo mass reached by each halo and subhalo over cosmic time. In our fiducial method, we assume that there is no scatter in the SHMR (that is, exact top-down galaxy-halo matching). We have additionally verified that allowing for scatter in the SHMR does not change the inferred SHMR at the 1$\sigma$ level, but that the \sagalocal{} sample does not provide meaningful constraints on this scatter, which is an expected result given the size of the sample \citep{allen2019, monzon2024}. Thus, though we note that the log-normal constant scatter model allows for SHMR scatter that is consistent with previous results \citep{behroozi2019}, we choose to adopt the fiducial model in the remainder of this work.  

We derive our halo mass function from the \textsf{BolshoiP} simulation, which has a box size of 250 Mpc/$h$ on a side with a mass resolution of $1.55\times 10^8 {M}_\odot/h$.\footnote{The cosmology adopted in \textsf{BolshoiP} is consistent with our adopted cosmology ($h=0.7$, $\Omega_m=0.3$).} Because this value is not small with respect to the lowest halo masses probed by the sample, we choose to fit a functional form to the simulated HMF rather than using the HMF directly.
We also choose to use a functional form in our abundance matching approach because the simple matching scheme we adopt here does not account for environmentally driven mass loss that has been shown to be necessary for subhalo abundance matching \citep[see, e.g.][]{jiang2021, danieli2023, monzon2024}.

The form of the halo velocity (and mass) function adopted by \cite{klypin2011} for the \textsf{bolshoip} simulations behaves approximately as a power-law at the expected halo masses of the \sagalocal{} sample, which are much lower than the characteristic halo velocity (mass) at the knee of the halo mass function. 
We thus adopt fit a power-law form to the HMF over $10.5<\rm M_h/M_\odot< 11.5$, the mass regime where the \textsf{BolshoiP} simulations are well-resolved ($\rm M_h\gtrsim 10^9 M_\odot$). Below this minimum halo mass, we assume that the HMF is well-approximated by an extrapolation of the best-fit power-law down to $\log_{10}(\rm M_h/M_\odot)\sim 9$.

In the main text, we perform abundance matching for the full \sagalocal{} sample -- in \autoref{s:appendix:field_shmr}, we carry out an additional exercise wherein we apply a modified abundance matching technique to the \sagalocal{} field galaxies alone.

%We multiply the halo number density by $f(\epsilon)$, the fraction of the sample that has a given environmental classification. 

%We convert the circular velocity function of distinct (i.e. central) halos determined by \cite{klypin2011} from the \textsf{BolshoiP} \citep{bolshoip} dark-matter only simulation to a halo mass function using their best-fit conversions presented in their Equations 8 \& 16. 
\begin{figure*}[ht!]
    \centering
    \includegraphics[width=\linewidth]{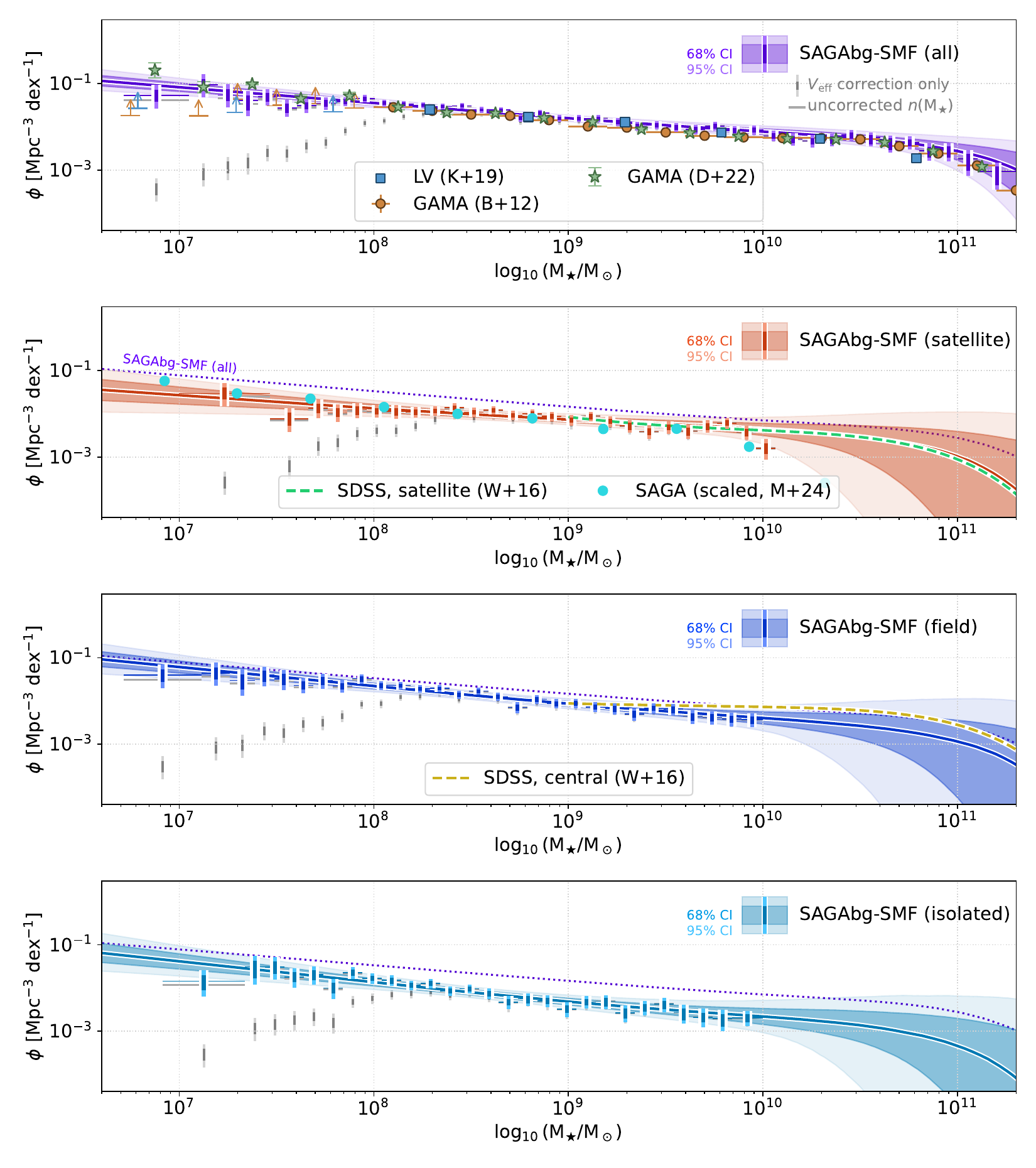}
    \caption{
    Completeness-corrected galaxy number densities (colored bars) and best-fit stellar mass function (shaded curves) for the SAGAbg sample as a function of environment. The measurements are compared, where possible, with results from the literature. Completeness un-corrected number densities $n(\rm M_\star)$ are shown as grey lines in each panel; number densities obtained when only the photometric incompleteness term is applied are shown as grey crosses.  
    We fit the SAGAbg SMF with a double Schechter function;
    the shaded regions show the 68\% and 95\% confidence intervals of the fit, while the error bars show the 68\% and 95\% confidence intervals for the binned counts. Each panel shows the best-fit SMF of the full \sagalocal{} sample as the dotted purple curve as a visual reference for both slope and overall normalization.
    }
    \label{f:smf}
\end{figure*}

\setlength{\tabcolsep}{3pt}
\begin{deluxetable*}{cccccccc}
\tablecaption{Double Schechter fit parameters for \sagalocal{} and recent literature results}
\tablewidth{0pt}
\tablehead{
    \colhead{Sample}&
    %\colhead{$\rm \log_{10}(\frac{M_\star^{min}}{M_\odot})$\tablenotemark{a}}&
    %\colhead{$\rm \log_{10}(\frac{M_\star^{max}}{M_\odot})$\tablenotemark{a}}&
    \colhead{$\rm M_{\star}^{min}$\tablenotemark{a}}&
    \colhead{$\rm M_{\star}^{max}$\tablenotemark{a}}&    
    \colhead{$10^{3}\phi_{0,1}$} & 
    \colhead{$\alpha_1$} & 
    \colhead{$10^{3}\phi_{0,2}$} & 
    \colhead{$\alpha_2$} &     
    \colhead{$\log_{10}\left(\frac{\rm M_{ch}}{\rm M_\odot}\right)$} \\
    \colhead{} &   
    \colhead{[dex]} &
    \colhead{[dex]} & 
    \colhead{[Mpc$^{-3}$]} &
    \colhead{} &
    \colhead{[Mpc$^{-3}$]} &
    \colhead{} &    
    \colhead{} 
}
\startdata
\midrule
\sagalocal{}: all & 6.7 & 11.3 & $0.97^{+0.62}_{-0.38}$ & $-1.37^{+0.05}_{-0.06}$ & $2.30^{+1.86}_{-1.04}$ & $-0.41^{+0.29}_{-0.39}$ & $11.21^{+0.55}_{-0.45}$ \\
\sagalocal{}: satellite & 10 & 7 & $0.64^{+0.70}_{-0.36}$ & $-1.32^{+0.11}_{-0.13}$ & $3.13^{+3.39}_{-1.74}$ & $-0.52^{+0.38}_{-0.36}$ & $10.81^{+0.83}_{-0.96}$ \\
\sagalocal{}: field & 6.7 & 10 & $0.49^{+0.59}_{-0.27}$ & $-1.44^{+0.09}_{-0.09}$ & $2.42^{+3.81}_{-1.52}$ & $-0.41^{+0.29}_{-0.39}$ & $10.99^{+0.69}_{-0.93}$ \\
\sagalocal{}: isolated & 6.9 & 10 & $0.28^{+0.33}_{-0.14}$ & $-1.48^{+0.11}_{-0.13}$ & $1.45^{+2.14}_{-0.93}$ & $-0.42^{+0.30}_{-0.37}$ & $10.68^{+0.77}_{-0.79}$ \\
\midrule
%\textit{Wide-field spectroscopic surveys: global SMFs} & & & & & & \\
\citealt{driver2022} (GAMA, all) & 6.75 & 11.5 &  $0.63^{+0.10}_{-0.09}$ & $-1.53^{+0.03}_{-0.03}$ & $3.65^{+0.14}_{-0.13}$ & $-0.47^{+0.07}_{-0.07}$ & $10.75^{+0.02}_{-0.02}$\\
\citealt{baldry2012} (GAMA, all) & 8 & 11.6 & $0.8^{+0.23}_{-0.23}$& $-1.47^{+0.05}_{-0.05}$ & $4.0^{+0.34}_{-0.34}$ & $-0.4^{+0.18}_{-0.18}$ & $10.66 ^{+0.05}_{-0.05}$\\
\citealt{wright2017} (GAMA, all) & 7.5 & 11.7 & $0.6^{+0.1}_{-0.1}$& $-1.5^{+0.15}_{-0.15}$ & $2.9^{+0.4}_{-0.4}$ & $-0.62^{+0.15}_{-0.15}$ & $10.8 ^{+0.2}_{-0.2}$\\
\citealt{peng2010} (SDSS) & 8.8 & 11.8 & $0.66^{+0.09}_{-0.09}$ & $-1.6 ^{+0.12}_{-0.12}$ & $4.0^{+0.12}_{-0.12}$ & $0.52^{+0.04}_{-0.04}$ & $10.67^{+0.01}_{-0.01}$\\
%\midrule
%\textit{Environment-specific SMFs} & & & & & & \\
\citealt{dolivadolinsky2023} \rrr{(M31 sats)}\tablenotemark{b} & 4.1 & 8.7 &-- & $-1.5^{+0.1}_{-0.1}$\tablenotemark{c} & -- & -- & -- \\
\citealt{nadler2020} (MW sats)\tablenotemark{b,d} & 2.5 & 7.4 &-- & [-1.46, -1.39]\tablenotemark{c} & -- & -- & -- \\
\citealt{bennet2019} \rrr{(M101 sats)}\tablenotemark{b} & $5.2$ & $10.3$ & &-- & $-1.1^{+0.1}_{-0.1}$\tablenotemark{c} & -- & -- \\
\citealt{crnojevic2019} \rrr{(Cen A sats)}\tablenotemark{b} & $5.1$ & $10.3$ & &-- & $-1.1^{+0.2}_{-0.2}$\tablenotemark{c} & -- & -- \\
\citealt{chiboucas2013} \rrr{(M81 sats)}\tablenotemark{b} & $\sim 5$ & $\sim 10.5$ & &-- & $-1.27^{+0.04}_{-0.04}$\tablenotemark{c} & -- & -- \\
\citealt{sbaffoni2025} (GAMA groups) & 8.3 & 12 & -- & $-1.5^{+0.13}_{-0.13}$ & -- & $-0.81 ^{+0.04}_{-0.04}$ & $10.87 ^{+0.01}_{-0.01}$ \\
\citealt{weigel2016} (SDSS sats) & 9 & 11.5 & $0.04^{+0.08}_{-0.03}$ & $-1.83^{+0.23}_{-0.23}$ & $1.2_{-0.9}^{+3.7}$ & $-0.84^{+0.08}_{-0.08}$ & $10.71^{+0.02}_{-0.02}$\\
\citealt{morgan2025} (Virgo cluster) & 7 & 11.8 &-- & $-1.35^{+0.02}_{-0.02}$ & -- & -- &$11.33^{+0.17}_{-0.11}$ \\
\citealt{cuillandre2025} (Perseus cluster) & 7 & 11.5& -- & $-1.20^{+0.02}_{-0.02}$ & -- & -- & $11.2^{+0.1}_{-0.1}$ \\
%\midrule
%\textit{Field, centrals, \& non-groups} & & & & & & \\
\citealt{sbaffoni2025} (GAMA ungrouped) & 8.3 & 12 & -- & $-1.57^{+0.03}_{-0.03}$ & -- & $-0.22 ^{+0.06}_{-0.06}$ & $10.41 ^{+0.01}_{-0.01}$ \\
\citealt{weigel2016} (SDSS centrals) & 9 & 12 & $0.9_{-0.20}^{+0.25}$ & $-1.16 ^{+0.05}_{-0.05}$ & $1.6^{+0.9}_{-0.6}$& $-0.51 ^{+0.09}_{-0.09}$ & $10.80 ^{+0.01}_{-0.01}$\\
\enddata
\tablecomments{Best-fit parameters for the stellar mass functions presented in \autoref{f:smf}, as well as some recent literature measurements that include a low-mass component to the SMF. For our results, the marginalized posterior of each parameter is summarized by the 50\textsuperscript{th} percentile in each table entry (super- and subscript values showing the 68\% credible interval). We report the uncertainties in the same way as the reference authors. We additionally provide a Machine-Readable Table with our best-fit estimates of $\phi(\rm M_\star)$ in \autoref{t:smfmrt} to properly account for covariance between parameters in our posterior reporting.
\tablenotetext{a}{The approximate limiting stellar masses for each measured SMF are given as $\log_{10}[\rm M_\star/M_\odot]$. This value should only be taken as an approximation for the literature measurements, and the reader should consult the reference to understand the full mass completeness of each SMF.}
\tablenotetext{b}{\rrr{Satellite luminosity function} results are given as a function of $L_V$; we estimate the expected limiting stellar mass and low-mass SMF by assuming a constant mass-to-light ratio of ${\rm M}_\star/L_V=1 {\rm M_\odot}/L_\odot$. }
\tablenotetext{\rrr{c}}{Following the reporting of \cite{nadler2020}, the brackets give the 95\% credible interval for the low-mass slope.}
}
\end{deluxetable*}\label{t:smf}

\subsection{Measuring galaxy quenched fractions}\label{s:model:fq}
From our measurement of the SMF of isolated low-mass galaxies in \autoref{f:smf}, we can also measure the volume-corrected quenched fraction of these isolated systems. We follow \cite{sagaiv} in our definition of quenched; that is, we require that quenched galaxies have an H$\alpha$ equivalent width of $\rm EW(H\alpha)<2\ \AA$, and a NUV specific star formation rate of $\rm SFR_{NUV}/M_\star < 10^{-11}\ yr^{-1}$.  Conversely, we classify a galaxy as star-forming if either star formation indicator surpasses its threshold for quenched classification. 
We restrict our analysis to \logmstar[$\geq8.5$]; below this stellar mass, the uncertainty on measured H$\alpha$ equivalent widths for at least half of the considered spectra are greater than $2\rm \AA$. 

Unlike the SAGA satellite sample and \sagabg{} sample analyzed in \citetalias{paperone} and \citetalias{papertwo} of this series, not all of the \sagalocal{} galaxies have an associated spectrum, as non-satellite galaxies with secure archival redshifts were not necessarily re-observed. 
As such, for this analysis we consider only the intersection between the \sagabg{} and \sagalocal{} samples, reducing the number of isolated galaxies considered from 936 to 803. 
To bolster this number, we additionally include the $z<0.05$ isolated galaxies from the equatorial GAMA survey in the survey regions that do not overlap with the SAGA footprint, adding 2347 galaxies to the sample. 

The effective volume $V_{\rm eff}$ over which each GAMA galaxy is observable is computed in the same way as the \sagalocal{} SMFs described in \autoref{s:smf}, where the GAMA stellar masses and photometric properties are calculated as in \autoref{s:data:sample}, and the spectroscopic completeness function is calculated as the ratio between the differential surface density of spectrosopic targets with respect to $r$-band magnitude in GAMA and that of the deeper SAGAbg sample down to the SAGAbg magnitude limit. In practice, this approach yields a sigmoidal completeness function with a boundary near the GAMA magnitude limit (best-fit logistic function parameters: $m_{r,0}=19.8$, $k=1.5$ where the completeness can be described as ${\rm Pr[GAMA}|m_r] = 1-(1 + \exp\{-k(m_r-m_{r,0})\})^{-1}$. 

With these completeness functions in hand, we can directly calculate the volume-corrected quenched fractions of the \sagalocal{} and GAMA samples as a function of stellar mass. We find no evidence for a statistically significant shift between the surface brightness distribution of the star-forming and quenched dwarfs in the \sagalocal{} at fixed mass across environments, and so do not apply a photometric correction to either survey. Each galaxy in the sample is weighted by its corresponding effective volume in our calculation of the quenched fraction.  We will return to this metric and its implications in \autoref{s:discussion:fquench}.

\iffalse{}
We model the quenched fraction for isolated field dwarfs implied by the \sagalocal{} sample via logistic regression, where the probability of observing a quenched galaxy ($\rm Pr[q]$) conditioned on stellar mass is given by:
\begin{equation}\label{e:fquench}
    {\rm Pr}[q|{\rm \log_{10}(\frac{M_\star}{M_\odot})}] = \frac{A_1}{1 + \exp\left(-\frac{\log_{10}\left(\frac{M_\star}{M_\odot}\right) - \mu)}{s}\right)} + A_0,
\end{equation}
where $q\in \{0,1\}$ is a binary indicator where $q=1$ indicates quenched and $q=0$ indicates star-forming, $A_1$ gives the quenched fraction at the high-mass limit, $A_0$ gives the quenched fraction at the low-mass limit, and $\mu$ and $s$ describe the shape of the logistic curve. 

We adopt uniform priors over the physically allowable range for each parameter ($A_0+A_1<1$, $0<s<100$, $6<\mu<11$), and
infer the best-fit values for these parameters using the same Markov Chain Monte Carlo procedure described in \autoref{s:smf:fit}. We note that our choice of functional form for the quenched fraction cannot capture non-monotonic behavior such as a turnaround in quenched fraction at low stellar masses, but that such a turnaround is not implied by the \sagalocal{} sample when binned by stellar mass.
\fi{}

\begin{figure*}[ht!]
    \centering
    \includegraphics[width=0.9\linewidth]{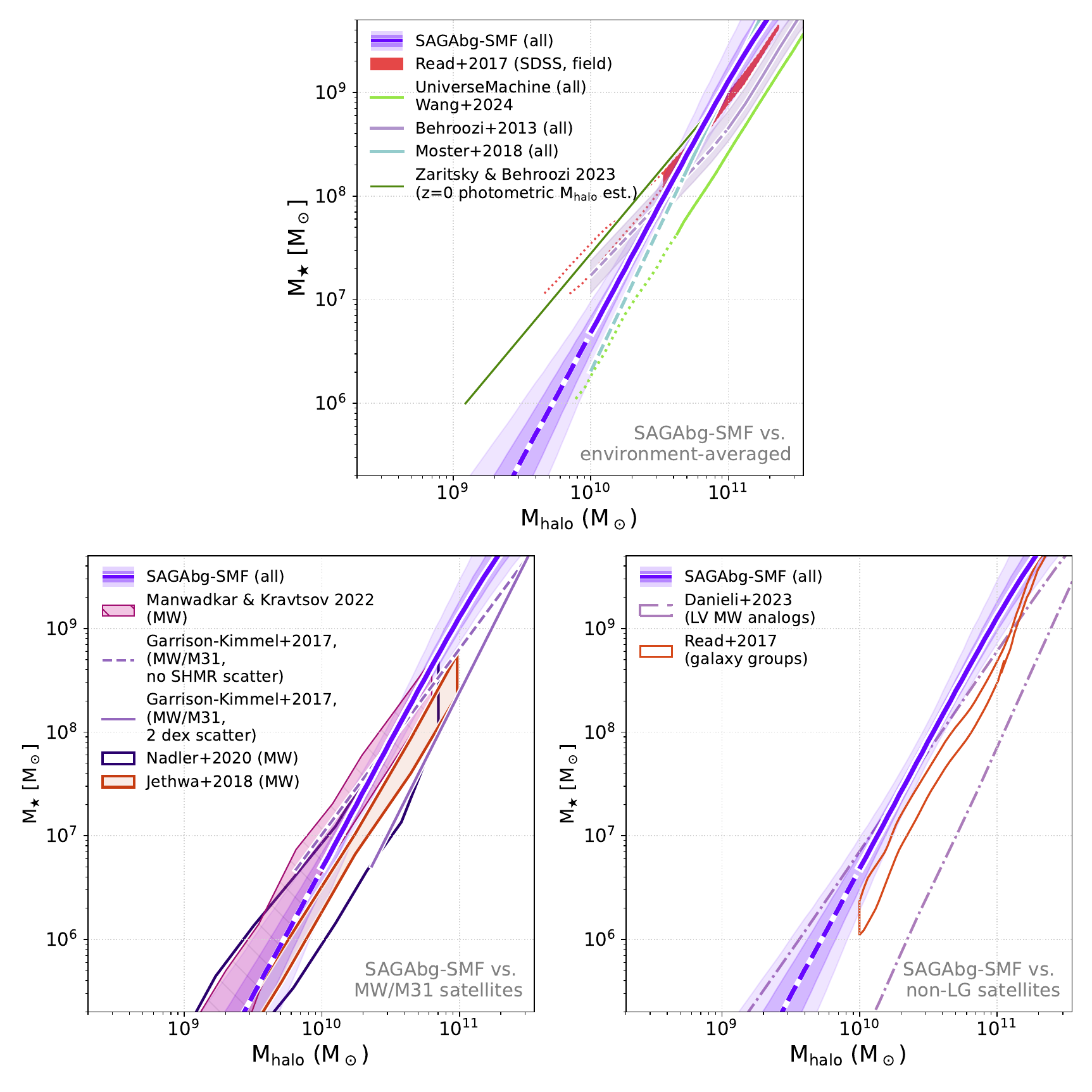}
    \caption{
    The \sagalocal{} stellar-to-halo mass relation (SHMR, blue) we infer from abundance matching of the \sagalocal{}field sample compared to the SHMR obtained from various abundance matching (AM) and empirical modeling (EM) studies in the literature. Clockwise from top, the panels show a comparison between the \sagalocal{} SHMR and relations inferred from environmentally-averaged samples, satellites from beyond the Local Group, and satellites of the Milky Way \& M31. Shaded regions show the 68\% and 95\% confidence intervals of our SHMR posterior. \textit{Top:} a comparison to empirical and semi-analytic modeling of higher mass populations. \textit{Left:} our field (blue) and satellite (red) SHMRs compared to studies of the MW \& M31 satellite systems. \textit{Right:} the same as at left, but for satellite environments beyond the MW/M31 systems. Unless noted otherwise, halo mass are given are peak halo masses ($\rm M_{peak}$).
    }
    \label{f:shmr_comparison_discussion}
\end{figure*}

\section{Results}
Here we present the stellar mass function of \sagalocal{} galaxies as a function of environment down to $\rm M_\star \sim 5\times10^6 M_\odot$ (\autoref{s:results:smf}); this measurement further enables a measurement of the stellar-to-halo mass relation of field dwarfs (\autoref{s:results:shmr}). 

\subsection{The stellar mass function}\label{s:results:smf}
We show the stellar mass function of the full, field, satellite, and isolated \sagalocal{} samples of \autoref{f:smf}, sequentially from the top panel. In each panel, our SMF measurements are shown by vertical bars (where the dark- and light-shaded bars show the 68\% and 95\% confidence intervals, respectively), whereas the dark- and light-shaded regions show the 68\% and 95\% confidence intervals, respectively, of a double Schechter model fit to the data. 

We also show the estimated SMF without volume corrections as grey errorbars in each panel (that is, the SMF one would arrive at by assuming $V_{\rm eff} = V(z=0.05)$ for all galaxies where $V(z)$ denotes comoving volume out to redshift $z$). In this mass range, the photometric completeness correction is small ($f_{\rm phot} \sim 2$) relative to the volume correction. However, we note that the two address different effects: the former accounts for galaxies missing from the parent photometric sample, while the latter adjusts the volume over which number densities are computed to reflect the actual detection limit for each galaxy. A further discussion of the mass-dependence of \sagalocal{} effective volume and the raw galaxy counts for each subsample are given in \autoref{s:appendix:veff}.
%The \sagalocal{} sample extends down to stellar masses of $\rm M_\star \sim 10^6 M_\odot$, but we find that the photometric completeness term dominates our estimate of the number density at these masses. Although we propagate our uncertainty within the incompleteness model to our reported uncertainty on the SMF, we choose to show the estimated differential number densities in this regime as unfilled rectangles to reflect the systematic uncertainty incurred by the model correction. We do note that in this regime, our predicted number densities are in statistical agreement with previous measurements from the literature.

In the top panel, we compare the SMF of the full \sagalocal{} sample to previous measurements of the SMF from other spectroscopic surveys. This represents an important consistency check if we are to then consider SMFs based on environmental classifications.
As shown by the red points, our measured stellar mass function is consistent with lower limits estimated from previous GAMA \citep{baldry2012} and Local Volume \citep{karachentsev2019} estimates, as well as the most recent GAMA survey measurement \citep{driver2022}. 

There is evidence for a local cosmic under-density at comoving distances less than $\sim 100$ Mpc \citep[$z\sim 0.02$,][]{bohringer2020}, which encompasses a substantial fraction of our total redshift range. It has also been recently suggested by \cite{xu2025} using an analysis of photometric objects near DESI Year One spectroscopic targets that the stellar mass function may be more steeply rising that is implied by surveys whose low-mass SMF is primarily constrained by nearby galaxies. To estimate whether local large-scale structure in the vicinity of the Local Group and the presence of the SAGA hosts may bias our measurements, we  compute the stellar mass function of the full SAGAbg sample at $0.03<z<0.1$ in \autoref{s:appendix:highzsmf} using the same methodology presented above. At this higher redshift range, our limiting stellar mass rises to $\rm M_\star \sim 10^8 M_\odot$; in the overlapping stellar mass range, we see good agreement between the \sagalocal{} SMF and that of the SAGAbg galaxies at $0.03<z<0.1$, indicating that the local under-density does not significantly affect our measured SMF. 

To place our measured SMFs in literature context, in \autoref{t:smf} we give the best-fit parameters for the \sagalocal{} fit to a double Schechter function along with recent results from spectroscopic surveys (middle section) and recent measurements of the low-mass SMF as a function of environment (lower section, ordered from highest to lowests density environments).

We caution that a direct comparison of parameter inferences should be tempered by two sources of systematic uncertainty: the mass range of each study, and possible mass-dependent stellar mass estimation biases. First, for the observed range of SMF shapes seen in the literature, the low-mass slope dominates the overall shape of the SMF only at $\rm M_\star \lesssim 10^8\ M_\odot$. Second, as noted by \cite{wright2017}, a mass-dependent bias in stellar mass estimations can introduce substantial systematic uncertainty in the slope of the SMF -- it has furthermore been shown recently that such mass-dependent biases may be exacerbated in the dwarf mass regime \citep{delosreyes2024}.

Nevertheless, it is instructional to compare our results to those in the literature. Two main points emerge: first, we find broad consistency between our results and those from the literature to the level of existing study-to-study variation. The faint-end slope that we infer ($\alpha_1$) \rrr{at $\rm M_\star> 5\times 10^6 M_\odot$} from our \rrr{double Schechter fit} is fairly consistent with previous measurements from GAMA \citep{baldry2012} and SDSS \citep{peng2010} (zero lies within the 73\% and 84\% credible intervals, respectively, of the difference between our posterior for $\alpha_1$ and their reported values, assuming Gaussian uncertainties for literature values). We do find a significantly shallower low-mass slope than \cite{driver2022}, though we note that our directly measured SMFs are consistent, as shown in \autoref{f:smf}. 

Second, there is no clear evidence in the literature for a correlation between environment and \rrr{the} faint-end SMF slope \rrr{in the mass range explored by the \sagalocal{} sample}. With our environmentally-stratified \sagalocal{} sample, we now consider the question of environmental dependence for a set of homogeneously selected and measured SMFs.

In the second panel of \autoref{f:smf}, we show the stellar mass function of satellite galaxies. Again, the SAGA satellites themselves are explicitly excluded from the sample --- the satellites considered here are associated with galaxies other than the SAGA hosts. We overplot the SDSS-based results of \cite{weigel2016} as the aqua curve, and the scaled results of the SAGA satellite mass function \citep{sagaiii} as aqua points. 
Scaling the SAGA satellite mass function is required because the number densities of the SAGA satellite mass function refer specifically to the volume within 300 kpc of a MW-like host (as opposed to the stellar mass function, which is the number density of galaxies that are satellites of some host). 

We find generally good agreement in the three mass functions in the region of stellar mass overlap. At $\rm M_\star \gtrsim10^{10}M_\odot$, the divergence between the \cite{sagaiii} and \cite{weigel2016} results is expected from the choice of SAGA hosts (that is, a ``satellite'' of a MW-analog that exceeds the stellar mass of the MW-like host is ill-defined). One may note from \autoref{t:smf} that the measured low-mass slope of our SMF differs significantly from that of \cite{weigel2016} as a function of environment, but we emphasize that given the stark difference in mass range covered between the two samples, one should compare results as a function of directly predicted differential number density rather than best-fit values of $\alpha_1$.

In the third panel of \autoref{f:smf} we show the stellar mass function of only low-mass field galaxies, as defined in \autoref{s:data:environment}. By definition, here, our stellar mass function constraints truncate at $\rm M_\star=10^{10}M_\odot$. We again overplot the measurements of \cite{weigel2016}, now for central galaxies, in red. We find a somewhat lower field galaxy SMF at the high-mass end of our sample (that is, the \citealt{weigel2016} SMF is greater than the 95\% CI of our estimated number densities for $\rm M_\star \gtrsim 10^{9.2} M_\odot$), but we note that this may be due to the 
difference in our definition of ``field'' for galaxies that approach our massive host stellar mass limit.

Finally, in the lower panel of \autoref{f:smf}, we show the SMF of only isolated low-mass galaxies. As a reminder, these are a subset of field galaxies that are at least 1.5 Mpc in projected distance and 1000 $\kms{}$ in velocity difference from their nearest massive neighbor; these distance criteria follow those presented in \cite{geha2012} as the threshold separation at which environmental processing from massive hosts is negligible. We find no statistically significant difference in the shape of the isolated SMF as compared to the field SMF, consistent with a minimal impact of environmental processing on stellar mass assembly in the field sample.

\begin{figure}[t!]
    \centering
    \includegraphics[width=\linewidth]{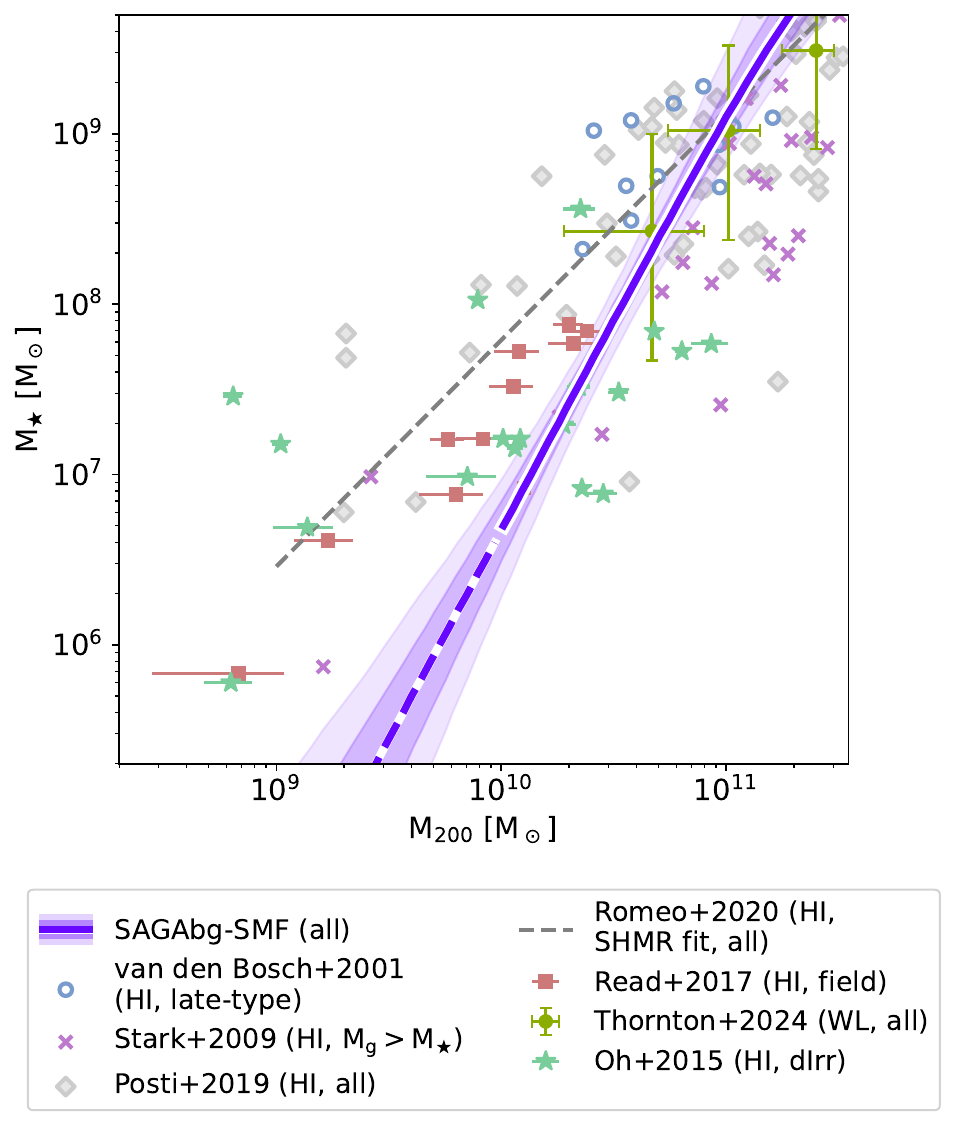}
    \caption{
    Our dwarf stellar-to-halo mass relation (SHMR, blue) from abundance matching of the \sagalocal{} sample compared
    to direct observational results obtained via \HI{} rotation curves and weak lensing. We show the relationship between stellar mass and peak halo mass; the offset between $\rm M_{peak}$ and $\rm M_{halo}($z=0$)$ should be small for the bulk of the sample considered.
    Shaded regions show the 68\% and 95\% confidence intervals of our SHMR posterior.
    Our results are generally consistent with previous findings in the literature up to study-to-study variability in the literature at high stellar masses, and are systematically lower than previous kinematic measurements at low stellar masses. The dashed line shows the extrapolation of our abundance matching results below our limiting stellar mass of $\rm M_\star = 5\times 10^6 M_\odot$.
    }
    \label{f:shmr_comparison}
\end{figure}

\begin{deluxetable*}{cccccc}
\tablecaption{\sagalocal{} Abundance matching-based estimate of the dwarf SHMR}
\tablewidth{0pt}
\tablehead{
 \colhead{\logmstar[]} & 
 \colhead{$\log_{10}(\langle \rm M_{halo}\rangle_{2.5})$} & 
 \colhead{$\log_{10}(\langle \rm M_{halo}\rangle_{16})$} & 
 \colhead{$\log_{10}(\langle \rm M_{halo}\rangle_{50})$} & 
 \colhead{$\log_{10}(\langle \rm M_{halo}\rangle_{84})$} & 
 \colhead{$\log_{10}(\langle \rm M_{halo}\rangle_{97.5})$} \\
}
\startdata
\textbf{6.75} & 9.93 & 9.99 & \textbf{10.05} & 10.11 & 10.16 \\
\textbf{7.25} & 10.16 & 10.20 & \textbf{10.24} & 10.29 & 10.33 \\
\textbf{7.75} & 10.36 & 10.40 & \textbf{10.44} & 10.47 & 10.51 \\
\textbf{8.25} & 10.55 & 10.59 & \textbf{10.63} & 10.67 & 10.72 \\
\textbf{8.75} & 10.74 & 10.79 & \textbf{10.84} & 10.89 & 10.95 \\
\textbf{9.25} & 10.93 & 10.99 & \textbf{11.06} & 11.12 & 11.19 \\
\textbf{9.75} & 11.13 & 11.22 & \textbf{11.31} & 11.38 & 11.47 \\
\enddata
\tablecomments{
The stellar-to-halo mass relation inferred from the \sagalocal{} sample via abundance matching. Bold columns show our best-fit SHMR as traced by the median of the predictive posterior of the SMF assuming a double Schechter form; other columns show uncertainty in the SHMR as quantified by the credible intervals of the posterior predictive distribution (not an estimate of SHMR scatter), as shown by the shaded regions in \autoref{f:shmr_comparison_discussion}, \autoref{f:shmr_comparison}, and \autoref{f:shmr_comparison_sims}. Each column gives the corresponding percentile of the posterior $P$ as $\log_{10}(\langle \rm M_{halo}\rangle_{P})$. The best-fit power-law to this relation is given by \autoref{e:shmr}.
}
\end{deluxetable*}\label{t:shmr}

\subsection{The stellar-to-halo mass relation}\label{s:results:shmr}
As described in \autoref{s:model:am}, we estimate the stellar-to-halo mass relation implied by our measured stellar mass functions using abundance matching. 

In \autoref{f:shmr_comparison_discussion}, we compare the results of our SHMR to stellar-to-halo mass relations estimated via semi-analytic and empirical modeling at left, and to subhalo abundance matching of nearby satellite systems at right. These two categories do not represent completely disjoint sets, but the visual separation is created to allow the reader to distinguish broadly between works that infer the SHMR using semi-analytic or empirical modeling across environments (top), subhalo and forward modeling of the MW/M31 systems (left), and satellite systems beyond the MW/M31 pair (right). 

%We also show the results of abundance matching for our \sagalocal{}  satellite sample at right, though we again caution that we are not directly modeling satellite environmental evolution \citep[as is common in practice in satellite LF modeling, see][]{jiang2021}. Many of these studies report the SHMR in terms of peak halo mass rather than $z=0$ halo mass; these quantities should be approximately equivalent for field dwarfs that have not experienced significant mass loss via processes such as tidal stripping \citep[see, e.g.][]{engler2021, christensen2024}. 
%Here, we find weak evidence of lower integrated galaxy formation efficiencies in satellite galaxies, as will be discussed in \autoref{s:discussion:environment}.

In \autoref{f:shmr_comparison}, we compare our inferred SHMR via abundance-matching to the state of the observational literature from \HI{} kinematic \citep{vandenbosch2001, stark2009, oh2015, read2017, posti2019, romeo2020} and weak lensing \citep{thornton2024} measurements. The most notable feature in our comparison is that our inferred SHMR lies systematically below mass modeling estimates at $ 5\times10^6 \lesssim  \rm M_\star \lesssim 5\times 10^7 M_\odot$ --- that is, our SHMR implies lower integrated galaxy formation efficiencies ($\rm M_\star/M_{halo}$) than kinematic estimates.  We will explore this divergence in greater detail in \autoref{s:discussion:kinematics}.
At higher masses,
the \sagalocal{} SHMR is in overall good agreement with weak lensing and high-mass \HI{} kinematic measurements at the level of existing study-to-study scatter in the literature.

We report a non-parametric description of our posterior distribution over the stellar-to-halo mass relation in \autoref{t:shmr}. The best-fit power-law\footnote{We do not report uncertainties on the power-law fit, as the non-parametric SHMR posterior is better suited for uncertainty analysis.} that corresponds to this relation is:
\begin{equation}\label{e:shmr}
    \rm \log_{10}(M_{halo}) = 0.40\log_{10}(M_\star)+7.36,
\end{equation}
for a stellar mass range of $5\times10^6\lesssim \rm M_\star <10^{10}$.

{}

\section{Discussion}
With our measurements of the galaxy stellar mass function, dwarf quenched fraction, and stellar-to-halo mass relation in hand, we now consider the implications of our measurements on the contemporary picture for dwarf galaxy star formation and assembly. 

%on the quenched fraction of isolated dwarfs (\autoref{s:results:fquench}), the ability of cosmological simulations to reproduce the observed SHMR down to \logmstar[$\sim 6$] (\autoref{s:discussion:sims}), and the evidence for or against an environmental dependence in the $\rm M_\star -M_{peak}$ relation of dwarfs (\autoref{s:discussion:environment}).
\subsection{Environmental variation in the low-mass SMF}\label{s:discussion:smfslope}
As summarized visually in \autoref{f:smf}, we find that the low-mass slope ($\alpha_1$ in \autoref{t:smf}) of the galaxy SMF is relatively insensitive to environment at the precision of our SMF measurements; we find that $\alpha_1$ is statistically consistent between the full, satellite, and field measurements of the \sagalocal{} SMF. This finding is in agreement with a recent analysis of the GAMA survey at higher masses (\logmstar[$\gtrsim8$]) who find the slope of the low-mass SMF to be insensitive to whether or not galaxies reside in groups \citep{sbaffoni2025}.

This minimal environmental dependence of the low-mass slope of the SMF is a result that has been implied by previous cross-sample analyses of satellite systems \citep[see, e.g.,][who compare the SAGA satellite SMF to the average GAMA SMF slope]{sagaiii} and clusters \citep[e.g.,][who also compare to the GAMA SMF]{morgan2025}, as well as a comparison between our results and the recent literature measurements of environmentally partitioned SMFs in \autoref{t:smf}, but is here shown with a homogeneous approach in both source data and analysis. 

%As we noted in \autoref{s:data:environment}, the SAGA survey fields were selected to avoid known clusters in their backgrounds -- the similarity in the low-mass shape of the SMF between the \sagalocal{} satellite sample and literature cluster SMFs, then, further indicates that the low-mass slope of the galaxy SMF is relatively insensitive to environment. 

Although the magnitude of the change between any given environmental bin is not statistically significant, we do see weak evidence for a steepening low-mass slope as a function of distance from a massive host: when drawing from the posterior distribution of $\alpha_1$ for each environmental classification considered, the satellite SMF ($\alpha_1=-1.32\pm0.12$) is the steepest in only $\sim\! 5\%$ of draws. Because star formation activity in dwarfs correlates strongly with environment, we now consider whether a difference in slope between the satellite and isolated dwarf samples can be explained by a correlation between star formation and low-mass SMF slope.

Using the same star formation rate indicators as described in \autoref{s:model:fq}, we classify all available \sagalocal{} galaxies as either star-forming or quenched, and measure the resulting SF-dependent SMFs with the same methodology as \autoref{s:smf}. The full details of this exercise can be found in \autoref{s:appendix:highzsmf}; we find that the low-mass slope of the star-forming SMF, as well as the star-forming satellite SMF, are consistent with $\alpha_1$ measured for the field and isolated \sagalocal{} samples. The quenched galaxies, meanwhile, are consistent with (and formally slightly shallower than) $\alpha_1$ measured for the satellite \sagalocal{} SMF. 

We therefore suggest that the slight change in slope between the isolated and satellite \sagalocal{} SMFs is attributable to the difference in star formation properties of the galaxies in these subsamples. 
This finding is consistent with work in literature with higher mass galaxies, where it was found that has found that the SMF at $\log_{10}(\rm M_\star/M_\odot)\gtrsim 9$ steepens for galaxies with higher specific star formation rate increases \citep{weigel2016} or bluer optical colors \citep{baldry2012}. 
We note that recent work on the Virgo cluster has found the opposite trend \citep{morgan2025}; though a unified analysis would be necessary to demonstrate this result, such a reversal could reflect a difference in the mass-dependence of the dominant quenching mechanisms experienced in clusters versus lower-density environments.

%This is also distinct behavior from the high-mass end of the SMF, where previous measurements from SDSS suggest a more steeply declining satellite SMF \citep{weigel2016} and, consistent with our findings, increasingly convergent central and satellite SMFs with decreasing stellar mass down to SDSS mass limits of \logmstar[$\sim 9$]. Indeed, as shown in the second and third panels of \autoref{f:smf}, our satellite and field SMFs are consistent with those of \cite{weigel2016} over the overlap in stellar mass between the two samples.

\subsection{Comparison to mass modeling of individual dwarfs}\label{s:discussion:kinematics}
Our abundance matching-estimate of the dwarf SHMR extends down to the lowest stellar masses for which stellar-to-halo mass ratios have been measured via \HI{} kinematic modeling. As discussed above, this sensitivity has revealed an apparent tension, wherein our abundance matching-derived SHMR implies a steeper SHMR than previous mass-modeling results.

%The discrepancy between our SHMR and kinematic measurements at low masses has previously been reported in the literature. 
Possibilities for such an apparent discrepancy have been previously reported in the literature. For example, \cite{rey2025} proposed that the inherent prerequisite of gas-richness in HI kinematic modeling biases the low-mass end of HI mass-modeling samples towards unusually late-forming systems with anomalously high stellar-to-halo mass ratios, while \cite{brook2015} suggested that the over-representation of LG dwarfs in \HI{} studies could bias these samples towards galaxies with more tidal processing.
Similarly, \cite{jethwa2018} suggested that sampling individual dwarf halo masses at fixed stellar mass biases the apparent SHMR to low-mass halos due to the steeply declining HMF. 

Another possibility, of course, is that the assumptions we make to perform abundance matching begin to break down at the low-mass end of our sample. This potential breakdown could include factors such as a rising, mass-dependent scatter in the dwarf SHMR (as opposed to our test of a constant lognormal scatter) or other violations to the assumption that stellar mass is rank-ordered with halo mass. Previous work predicts, for example, that an extrapolation of our best-fit SHMR down to lower mass ($\rm M_{halo} \lesssim 3\times10^9\ M_\star$) would likely be invalid due a steep decline in the fraction of halos that actually host a galaxy, breaking the rank-ordered assumption \citep{sawala2015}. It could also be the case that we under- or over-estimate the slope of the stellar or halo mass functions at the very lowest masses in our sample, where both the relative mass resolution of the simulation from which we estimate the HMF and our observational signal-to-noise in the SMF are lowest. 

Though we cannot comment more definitively here on the origin of a possible bias,
the hypothesis proposed in previous works that a discrepancy between simulated dwarf SHMRs and individual \HI{} mass modeling results is partially observational (or selection-related) in nature is supported by the discrepancy between individual field dwarf modeling and our statistical measurement of the field dwarf SHMR.

\subsection{Environmental effects on the dwarf SHMR}\label{s:discussion:environment}

Our abundance matching-based estimate 
of the dwarf SHMR extends more than an order of magnitude lower in stellar mass than previous statistical measurements of dwarfs \citep{read2017, sagav} based on shallower SDSS data, allowing the sample to act as a bridge between previous modeling efforts to infer the dwarf SHMR and observational measures of dwarf halo masses.
%, meaning that our \sagalocal{} SHMR can effectively 
%bridge the gap between abundance matching measurements of the SHMR at higher masses, and Nearby Universe measurements of satellite systems, as shown in \autoref{f:shmr_comparison_discussion}.

The literature on abundance matching in the low-mass regime has previously been dominated by higher density environments (satellites, groups, and clusters), where the efficiency of low-mass galaxy identification is bolstered by the known distance of the more massive neighbors.
Recent simulations have suggested that the $z=0$ stellar mass to peak halo mass relations of dwarfs varies with environment due to differences in halo assembly history \citep{christensen2024}; as these effects manifest in the relationship between the stellar mass and peak halo mass, the predicted effect is separate from differences in the expected environmental dependence of the ${\rm M_\star}(z=0)/{\rm M_{halo}}(z=0)$ due to preferential mass loss from the dark matter halo from processes such as tidal and ram-pressure stripping \citep[see, e.g.][]{engler2021}. 
To test this prediction, we compare our abundance matched SHMR of dwarfs to the 
abundance matching-driven SHMRs inferred for satellites of the Milky Way \citep{jethwa2018, nadler2020}, SDSS groups \citep{read2017}, and nearby MW-like galaxies \citep{danieli2023}. 

We do not find strong evidence for a systematic difference between the SHMR we infer from abundance matching the \sagalocal{} sample and that of satellite systems from the literature. This result is consistent with minimal environmental evolution of the low-mass SHMR, though we caution that a direct comparison against literature results may incur substantial systematics due both to the galaxy-halo modeling scheme adopted and to variation in the nature of the SHMR being fit (e.g., MW-like systems in \citealt{danieli2023}, galaxy groups in \citealt{read2017}, and the MW itself in \citealt{manwadkar2022}). To test whether this result could be attributable to the presence of satellite dwarfs in the \sagalocal{} sample, we perform an additional abundance matching exercise to estimate the SHMR of field dwarfs alone and find no statistically significant difference with the SHMR inferred for the full sample (\autoref{s:appendix:field_shmr}). 

A minimal environmental dependence of the dwarf SHMR is potentially in tension with previous predictions for satellite galaxies to show higher $\rm M_\star/M_{peak}$ values compared to their field counterparts (\citealt{christensen2024} using the MARVELOUS simulation suite), as the magnitude of the predicted difference in SHMR at a host separation of 50 kpc (satellite-like) and 1 Mpc (field-like) would nominally be statistically significant given our SHMR uncertainties. 
That being said, other observational modeling --- for example, \citealt{garrisonkimmel2017} modeling of the Local Group and Local Field mass functions) --- do imply that Local Field dwarfs may be characterized by a steeper SHMR than Local Group satellites. 

Due to the variation in SHMR inferred for dwarf systems across the literature, we thus suggest that precisely ruling in or out an environmental dependence of the SHMR will require joint modeling of satellite and field systems where the same methodology for environment selection and SHMR inference is used for the observed and simulated galaxy samples. This should be possible in the near future with the large dwarf galaxy samples that probe the full range of environments avaialable from DESI \citep{darraghford2023}.

%There is weak evidence for field dwarfs to be shifted to higher $\rm M_\star/M_{\rm peak}$ with respect to the SHMR of dwarf satellites seen in the literature, and from abundance matching of the present sample. However, at the precision of the statistical and systematic uncertainties, we find that our results are consistent with minimal environmental dependence in the dwarf galaxy SHMR.  This result is separate from environmental effects that modulate halo masses after infall such as tidal and ram-pressure stripping by the host halo \citep{zolotov2012}, and from group pre-processing prior to infall to the $z=0$ host \citep{wetzel2015}. This result  previous results that found that satellite galaxies have elevated integrated formation efficiencies compared to their field counterparts (\citealt{christensen2024} using the MARVELOUS simulation suite, and \citealt{garrisonkimmel2017} using Local Group and Local Field mass functions) ---  
%at the precision of the statistical and systematic uncertainties, however, we do not claim strong evidence for an environmental effect on the stellar-to-halo mass relation of dwarf galaxies. 
\subsection{Comparison to Cosmological Simulations}
\begin{figure}[ht!]
    \centering
    \includegraphics[width=\linewidth]{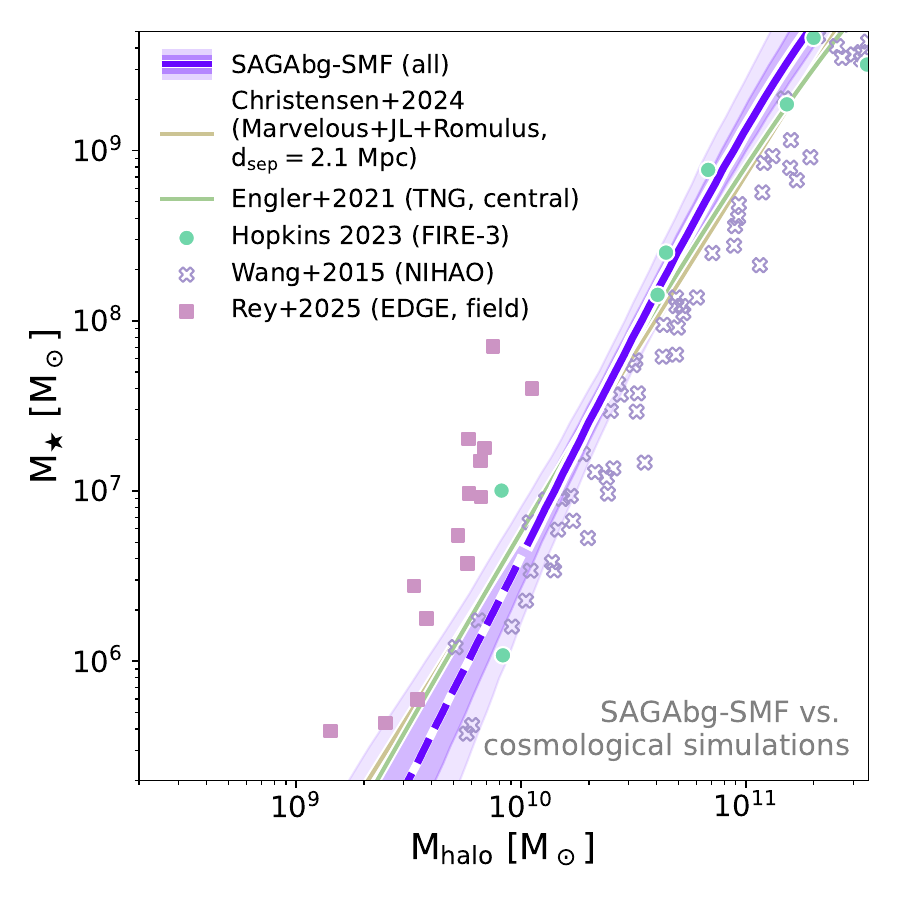}
    \caption{
     Stellar-to-halo mass relation (SHMR, blue) from abundance matching of the \sagalocal{} sample compared to the SHMR obtained from various cosmological simulations in the literature. We show halo mass as peak halo virial mass for the \sagalocal{} sample; \cite{christensen2024} and FIRE-3  citep{hopkins2023} also quote SHMRs as a function of peak halo mass, while NIHAO \citep{wang2015} and EDGE \citep{rey2025} quote $z=0$ $\rm M_{200}$ halo masses. TNG100 \citep{engler2021} instead gives the total dynamical mass associated with the halo. For our purposes (that is, for dwarf galaxies that are being compared to simulated galaxies largely in the field), these measures should be approximately comparable.
    }
    \label{f:shmr_comparison_sims}
\end{figure}

As shown in  \autoref{f:shmr_comparison_sims}, we find
that most contemporary cosmological simulations successfully reproduce our dwarf field SHMR down to our limiting mass and precision. 
Of the numerical works considered, only the NIHAO \citep{wang2015} and EDGE \citep{rey2025} simulations deviate from the 95\% confidence interval of our measured SHMR at any mass. The NIHAO SHMR, furthermore, shows a similar scaling to the \sagalocal{} SHMR ($\rm M_{\rm halo}\propto M_\star^{0.43}$ in the NIHAO suite); the log-additive offset from our results may reflect systematics in stellar and/or halo mass calibrations or definitions, rather than differences in galaxy formation physics.

Given the general success of simulations in predicting our observed SHMR, it is worth asking how these simulations capture the physics that modulate $M_\star/M_{\rm peak}$, and what we can learn from the dwarf SHMR. For dwarfs that are sufficiently massive to have evaded formation suppression from reionization or the cosmic UV background \citep{fitts2017}, star formation and stellar feedback are thought to be the main modulators of galaxy formation efficiency \citep{bullock2017}.
The nature of simulated star formation-driven outflows --- and, in particular, the multiphase nature of the outflows --- that serve to regulate star formation can depend sensitively on numerical resolution. Resolution-based challenges in simulating stellar feedback manifest largely in two regimes: resolving supernova remnant evolution when the outflow is launched \citep{dale2015, hopkins2018, smith2018, chaikin2022}, and --- for codes that adopt fixed-mass rather than fixed-spatial resolution --- when the low-density outflow reaches large galactocentric radii \citep{kim2020, bennett2024}. 

This is especially true in light of the fact that, to achieve the large volumes necessary to assemble a cosmological sample of simulated galaxies, most of the simulations referenced in \autoref{f:shmr_comparison_sims} invoke some subgrid model or effective equation of state to model stellar feedback and regulate star formation.
Thus, while the low-mass SHMR is a key observable that quantifies overall galaxy formation efficiency, there appears to be a multiplicity of avenues through which a realistic SHMR can be generated down to $\rm M_\star\gtrsim10^6M_\odot$. As such, understanding the mechanisms of the regulation requires a multi-pronged approach with other tracers of feedback, as we will discuss in the following section.

\subsection{The role of feedback in setting the low-mass SHMR}\label{s:discussion:sims}
%Having established that our \sagalocal{} SHMR is in good agreement with previous 
%abundance matching/empirical modeling estimates at the high-mass end of the 
%\sagalocal{} stellar mass range and individual mass modeling of nearby dwarfs,
%we can now compare our observational estimate of the SHMR for field dwarfs to contemporary predictions in the literature.
%It is widely accepted that stellar feedback processes such as supernovae-driven outflows and winds dominate the integrated galaxy formation efficiency of the low-mass galaxy population --- particularly at $\rm M_{halo}\gtrsim 10^{10}M_\odot$, above which star formation suppression by reionization and the subsequent cosmic UV background are not expected to play major roles \citep{fitts2017} ---  
Though it is widely accepted that stellar feedback processes such as supernovae-driven outflows and winds dominate the integrated galaxy formation efficiency of the low-mass galaxy population, the mechanisms behind this regulation are still a highly active area in the literature \citep[for recent reviews, see][]{naab2017,thompson2024}. 
Quantifying the relative roles of ejective feedback via strongly mass-loaded winds (characterized by high mass-loading factors $\eta_m \equiv \rm\dot M_{out}/ SFR$) and preventative feedback via strongly energy-loaded winds (characterized by high energy-loading factors $\eta_E\equiv  \dot E_{\rm out}/\rm SFR$) is in particular crucial to understanding the physical mechanisms that modulate formation efficiency in the dwarf regime.

In \citetalias{paperone} of this Series, we demonstrated that the low-mass galaxies in the SAGAbg-A sample at $z<0.2$ are well-described by a mass-loading factor of $\eta_m=0.92$. Indeed,
there is growing observational support for the idea that bright ($10^8 \lesssim \rm M_\star \lesssim 10^9\ M_\odot$) dwarfs are characterized by relatively low (order unity) mass-loading factors relative to the high $\eta_m$ schemes that many of the above simulations employ \citep[see, e.g.,][]{heckman2015,chisholm2017,mcquinn2019,marasco2023,wang2024,watts2024,zheng2024}.

If we posit that the \sagalocal{} sample is described by weakly mass-loaded winds, we can consider further implications for how the galaxy--halo connection is regulated in this mass regime. A complete characterization of the mass- and energy-loading factors  of the \sagalocal{} galaxies would require a more comprehensive joint analysis with numerical experiments. Here, however, we note that when considered in conjunction with the mass-loading factor measured in \citetalias{paperone}, the declining galaxy formation efficiency of the \sagalocal{} SHMR is most naturally described by a model in which preventative star formation feedback, cosmic ray pressure and/or high energy-loading factors efficiently regulates galaxy formation at low mass. 

This picture aligns well with recent results on preventative and ejective feedback, including work by \cite{carr2023}, who use an analytic gas-regulator model to argue that the SHMR is dominated by an energy-loading factor that rises steeply with declining halo mass and only weakly dependent on the mass-loading factor; by \cite{martinalvarez2025}, who showed that the inclusion of non-thermal pressure support sources such as cosmic rays can support both bursty star formation and moderately ($\eta_m\sim1$) mass-loaded outflows; and by \cite{bennett2024}, who used the \textsc{Arkenstone} wind-launching model to show that regulation via preventative feedback from high-specific energy winds (i.e. star formation efficiently prevents gas from accreting) is able to reproduce the stellar-to-halo mass relation seen in simulations with highly ejective feedback schemes.

\begin{figure*}[ht!]
   % \centering
    \includegraphics[width=\linewidth]{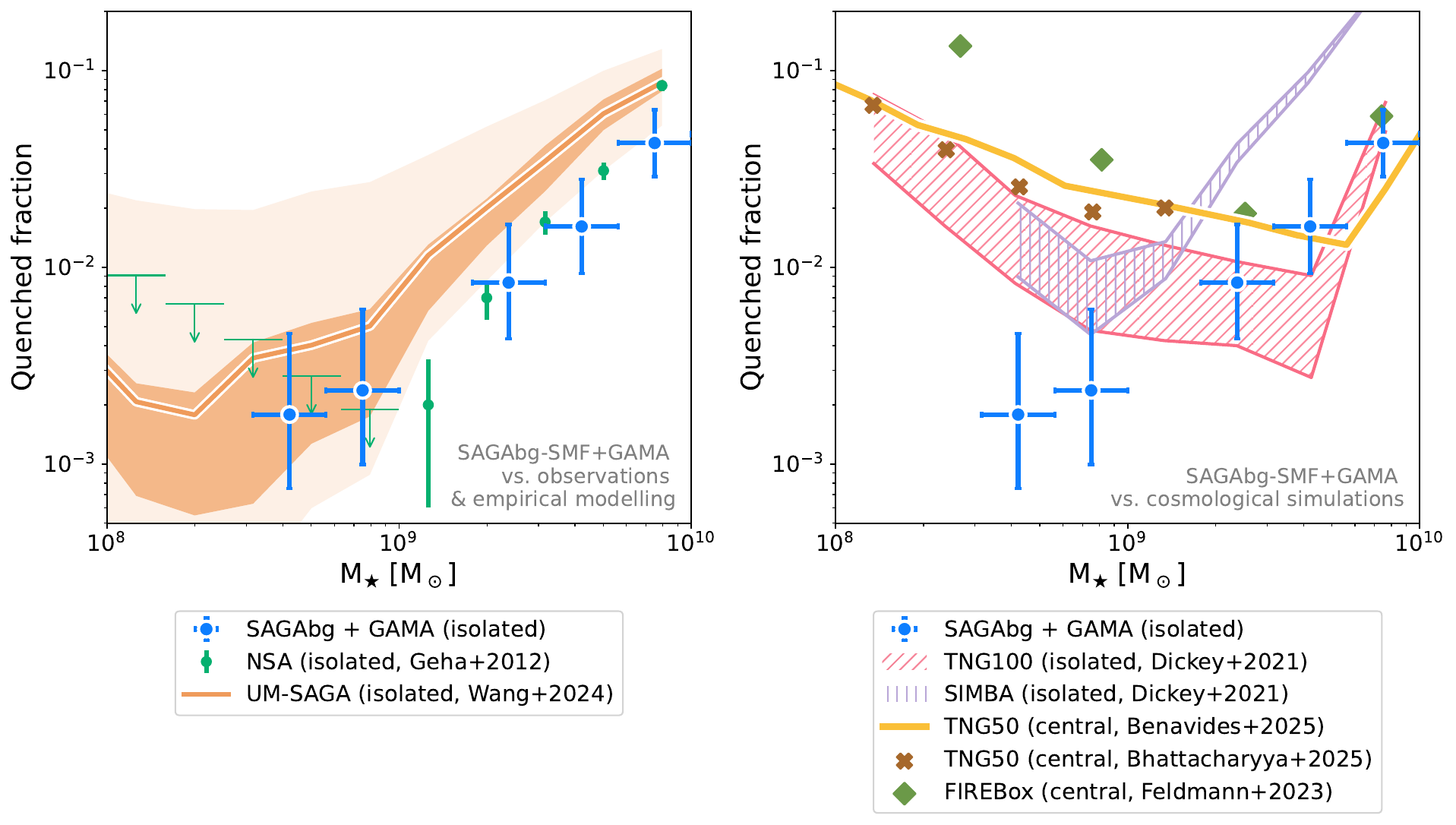}
    \caption{
    Quenched fraction for the combined \sagalocal{} and GAMA samples 
    as a function of stellar mass (blue errorbars). \textit{Left:} Comparison to previous observational measurements from the NSA \citep{geha2012} in green, where upper limits (unfilled triangles) show uncertainties from binomial counting statistics alone, and the semi-empirical model median (68\%/ 95\% CI) predictions from UM-SAGA\citep{sagav} in orange. 
    \textit{Right:} Comparison to recent predictions from cosmological simulations. We show predictions for isolated and central galaxies (as noted in the legend) from Illustris-TNG100 \citep[red hatches,][]{dickey2021}, SIMBA \citep[purple vertical hatches,][]{dickey2021}, TNG50 \citep[gold line and brown crosses,][respectively]{benavides2025, bhattacharyya2025}, FIREBox \citep[green diamonds,][]{feldmann2023}. The cosmological simulations shown here consistently predict a turnaround in isolated galaxy quenching efficiency at $\rm 10^8 \lesssim  M_\star/M_\odot \lesssim 10^{10}$, however several higher resolution studies predict that the quenched fraction does not rise until much lower masses (see text).
    }
    \label{f:fquench}
\end{figure*}

\subsection{The quenched fraction of isolated dwarfs}\label{s:discussion:fquench}
The presence of nearby, massive galaxies is understood to play a predominant role in the quenching of star formation in low-mass galaxies; observationally, self-quenching mechanisms have been demonstrated to be highly inefficient at $\rm M_\star\lesssim 10^9 M_\odot$ \citep{geha2012}. Numerical studies have since suggested that both internal feedback processes \citep{dickey2021}, large-scale effects from the cosmic web \citep{benavides2025}, and backsplash populations from massive hosts \citep{bhattacharyya2025} have the potential to contribute substantively to the assembly of an isolated, quenched dwarf population.
Indeed, a growing number of isolated, quenched dwarfs have since been discovered below this mass threshold, but it has remained challenging to pin down quenching mechanisms for individual systems without a broader statistical picture of the quenched population \citep{makarov2012, karachentsev2015, polzin2021, kadofong2024a, carleton2024, li2024}. 

%The quenched fraction implied by the \sagalocal{} and GAMA samples is shown in  \autoref{f:fquench}, along with previous measurements from SDSS \citep{geha2012}, UniverseMachine \citep{behroozi2019} semi-empirical modeling of the \cite{geha2012} (SDSS isolated dwarfs) and \cite{sagaiii} (SAGA satellite) samples. As shown by \cite{dickey2021}, our measured quenched fractions are substantially lower than predictions from the Illustris-TNG100, EAGLE, and SIMBA cosmological simulations.

\iffalse{}
The best-fit logistic function is given by the parameters:
\begin{equation}
    \begin{split}
    A_0&=5_{-3}^{+7}\times10^{-4}\\
    A_1&=0.2_{-0.1}^{+0.3}\\
    \mu&=10.2_{-0.3}^{+0.4}\\
    s&=0.28_{-0.07}^{+0.06}\\         
    \end{split}
\end{equation}
as defined in \autoref{e:fquench}. We also report the uncertainty of our inferred quenched fraction as quantified from credible intervals of the posterior predictive distribution  in \autoref{t:fquench}.
\fi{} 

The quenched fraction implied by the \sagalocal{} and GAMA samples is shown in  \autoref{f:fquench}, along with previous measurements from SDSS \citep{geha2012}. Our measurement of the quenched fraction demonstrates the existence of a small population of quenched and isolated dwarfs down to \logmstar[$=8.5$], contextualizing serendipitous discoveries and allowing for a comparison to theoretical predictions of $f_q(\rm M_\star)$, the mass-dependent quenched fraction of isolated systems. We find a total of 13 quenched, isolated galaxies at \logmstar[$<10$] between the two samples; two of the \sagalocal{} galaxies are additionally at \logmstar[$<9$] (all of the GAMA quenched and isolated galaxies have stellar masses of $10^9<\rm M_\star/M_\odot<10^{10}$). 

In addition to previous observations of quenched, isolated systems, we also compare to semi-empirical modeling results at left, and recent results from cosmological simulations at right. 
In particular, at left we show the UniverseMachine semi-empirical modeling of quenched, isolated systems by \cite{sagav}, UM-SAGA, which uses both the SAGA satellite data and \cite{geha2012} isolated dwarf sample to update the \textsf{UniverseMachine} environmental quenching model.
We find that the UM-SAGA model is broadly consistent with observational results: the 68\% confidence interval of the UniverseMachine prediction overlaps with the 68\% credible interval of our quenched fraction over the majority of the domain (and our measurement is contained within the 95\% CI of the UniverseMachine prediction at all stellar masses considered).

Conversely, mock observations of isolated Illustris-TNG100 and SIMBA \citep{dave2019} made by \cite{dickey2021} and star-formation splits of central galaxies in Illustris-TNG50 \citep{benavides2025, bhattacharyya2025} and FIREBox \citep{feldmann2023} predict a substantively different trend for $f_q(\rm M_\star)$. At higher masses, FIREBox, TNG50 and TNG100 successfully reproduces the quenched fraction of isolated galaxies, but predict a turnaround in $f_q(\rm M_\star)$ at a factor of a few about $\rm M_\star= 10^9 M_\odot$. SIMBA, conversely, predicts a systematically higher quenched fraction for more massive galaxies but intersects with our measurement just below $\rm M_\star=10^9 M_\odot$ before hinting at a turnaround in $f_q(\rm M_\star)$. The other simulation considered in \cite{dickey2021}, EAGLE \citep{crain2015, schaye2015, schaller2015, mcalpine2016} and SIMBA, predicts a similar turnaround behavior of $f_q(\rm M_\star)$, but maintains an overall higher quenched fraction at all masses ($f_q>0.1$, above the upper y-limit of \autoref{f:fquench}); an analysis of the NewHorizon simulations by \cite{rhee2024} also predict higher quenched fractions (albeit for dwarfs in the field, not necessarily isolated to the degree that we consider), with quenched fraction increasing for decreasing stellar mass at $\rm M_\star \lesssim 10^{10} M_\odot$.

\rrr{One caveat to our comparison is potential for systematics originating from the simulation works' definition of galaxy quenching.  \cite{dickey2021} created mock spectra from their simulated galaxies to closely emulate the observational selection of \cite{geha2012}; \cite{benavides2025} defined quenched galaxies as any system with an instantaneous specific SFR $<10^{-11}$ yr$^{-1}$; \cite{bhattacharyya2025} defined quenched galaxies as any system at least 1 dex below the Star-Forming Sequence at its $z=0$ stellar mass; and \cite{feldmann2023} defined quenched as a galaxy with a 100 Myr-averaged specific star formation rate of $<10^{-10.5}$ yr$^{-1}$ (they also consider a lower threshold of $<10^{-11}$ yr$^{-1}$, which results in a lower $f_q$ than their fiducial values but is still elevated relative to our observations).}

\rrr{Despite variations in how each work defines the threshold for quenched galaxy classification, we do not see strong evidence that this variation could account for the discrepancy between the simulated and observed quenched fractions. We in particular highlight that the results of \cite{dickey2021}, who explicitly created mock-observed spectra for a near apples-to-apples comparison with observatiodsns, produces the same trend as seen in the works that define quenching based on intrinsic SFR. We also note that \cite{benavides2025} and \cite{bhattacharyya2025}, who both use the TNG50 simulations, produce consistent $f_q(\rm M_\star)$ relations despite choosing different quenched classifiers.}

This consistency indicates that many contemporary cosmological simulations systematically predict a turnaround in self-quenching efficiency at around $\rm 10^8<M_\star /M_\odot <10^{10}$, resulting in apparent over-quenching of isolated dwarfs at $\rm M_\star \sim 4\times 10^8 M_\odot$ compared both to our sample and previous upper limits placed by \cite{geha2012}. 
The apparent over-quenching \rrr{in simulations} is of particular interest because the cosmological simulations considered in \autoref{f:shmr_comparison_sims} are able to broadly and consistently reproduce the dwarf SHMR, which should also trace galaxy formation feedback physics. 
One potential mechanism to explain both the simulated SHMR and $f_q(\rm M_\star)$ predictions is that over-vigorous stellar feedback (as suggested in \citealt{dickey2021}) is coupled with elevated star formation rates in low-mass progenitors at higher redshifts (as suggested observationally by \citetalias{papertwo} of this Series), allowing for a build-up of a simulated quiescent dwarf population while still producing a $M_\star(z=0) /M_{\rm peak}$ relation in line with observations.

This being said, however, it is also important to recall the potential role of numerical effects, particularly when the total galaxy mass becomes close to the mass resolution of the simulation\footnote{Of the simulations considered by \citealt{dickey2021}, the TNG100 baryonic mass resolution is $\rm M_{bary}\sim 2\times10^6 M_\odot$, and the SIMBA resolution is an order of magnitude higher at $\rm M_{bary}\sim 2\times10^7M_\odot$. TNG50 and FIREBox around two orders of magnitude higher resolution in bayronic mass resolution compared to TNG100 \citep{pillepich2019, feldmann2023}}. 
The potential for a tight coupling between resolution and emergent galaxy properties can be straightforwardly understood for the case of stellar feedback, where both the energetics of individual supernovae and their effect on the multiphase interstellar medium are largely implemented via subgrid models \citep{burger2025}.

As such, both variations in the presumed physical models for stellar feedback and the coupled numeric effects of their implementation can substantially impact star formation structure and behavior in simulated galaxy. These factors are expected to be particularly strong at dwarf masses, both because the overall number of resolution elements associated with the galaxy is smaller, and because stellar feedback is expected to dominate galaxy formation efficiency \citep{rodriguezcardoso2025}.
Indeed, \cite{dickey2021} find that increasing simulation resolution can substantially impact their predicted quenched fractions, though the general trend of quenched fraction as a function of mass remains largely intact \citep[][their Figure 7]{dickey2021}. 

%Although it is applied to a void environment that many not be typical for our isolated galaxy sample, 
%RW -- void and iolated seem the same for our purposes here...
High resolution simulations of dwarfs also point towards lower quenched fractions in isolated systems at these masses: the EDGE simulations \citep{kim2024, rey2025} and MARVEL-ous Dwarfs + DC JUSTICE LEAGUE + ROMULUS25 suite \citep{christensen2024} predict ubiquitously star-forming populations down to $\rm M_\star \sim 10^7 M_\odot$, and FIRE \citep{fitts2017} zoom-in simulations of $\rm M_{halo}\sim 10^{10} M_\odot$ galaxies find two out of 14 galaxies to be self-quenched in the field at $\rm 10^5\lesssim M_\star/M_\odot\lesssim 10^7$. 
Similarly, the DARKLIGHT semi-empirical modeling by \cite{kim2024} --- which is constrained by both observations and, at $\rm M_\star \lesssim 10^7\ M_\odot$, by the high-resolution EDGE cosmological simulations \citep{pontzen2021, rey2025}) --- predicts a quenched fraction of $f_q=0$ for their model galaxies at $\rm M_\star \gtrsim 5\times 10^6\ M_\odot$. Assuming binomial counting uncertainties with a Jeffreys' prior, we arrive at a 68\% upper limit of $f_q\lesssim 0.01$ between $\rm 10^8 \lesssim M_\star/M_\odot \lesssim 10^9$, which is statistically consistent with our observations.
This apparent relation between simulation resolution and isolated dwarf quenched fraction implies that the stellar mass turnaround in $f_q(\rm M_\star)$ may be impacted by both numerical resolution and the included physics \citep{kim2024}.

That cosmological simulations can successfully reproduce the \sagalocal{} dwarf SHMR while over-predicting the quenched fraction of isolated dwarfs underscores the complexity of building stellar feedback models to modulate dwarf evolution. This complexity leads to the need for multidimensional observations --- such as the mass-loading factor of \citetalias{paperone}, the Star-Forming Sequence time evolution of \citetalias{papertwo}, and the SHMR and quenched fractions presented here --- to identify and distinguish between physically reasonable  recipes for extragalactic astrophysical processes. 

%There are only seven quenched and isolated galaxies in the combined GAMA \& \sagalocal{} samples below the SDSS empirical self-quenching limit of \logmstar[$=9$] \citep{geha2012}.
%This small number is unsurprising, given the low expected quenched fraction for isolated dwarf galaxies. 

\section{Conclusions}
In this work, we leverage the background spectra associated with the Satellites Around Galactic Analogs (SAGA) survey to compute three key diagnostics of the low-mass galaxy population:
\begin{itemize}
    \item The galaxy stellar mass function (SMF) down to $\rm M_\star \sim 5\times10^6 M_\odot$ separated by extragalactic environment (\autoref{s:results:smf}; \autoref{f:smf} and best-fit model parameters given in \autoref{t:smf}).
    \item The stellar-to-halo mass relation of field dwarfs down to $\rm M_\star \sim 5\times10^6 M_\odot$ (\autoref{s:results:shmr}, \autoref{f:shmr_comparison} and tabulated in \autoref{t:shmr}).
    \item The quenched fraction of isolated field dwarfs with measurements down to $\rm M_\star = 4\times 10^8 M_\odot$  (\autoref{s:discussion:fquench}; \autoref{f:fquench}).
\end{itemize}

Our measurements with \sagalocal{} provide new statistical leverage to connect $\rm M_\star/M_{halo}$ measured for individual LV dwarfs with that of previous wide-field surveys, compare the SHMR of dwarf satellites with that of field dwarfs, and contextualize serendipitous discoveries of quenched field dwarfs below the empirical self-quenching mass scale of \logmstar[$\sim 9$]. In particular:
\begin{itemize}
    %\item The field dwarf SHMR bridges the gap between previous statistical measurements of the field SHMR for bright dwarfs, and with measurements of $\rm M_\star/M_{\rm halo}$ for individual dwarfs at lower masses.
    \item We observe minimal environmental dependence of the dwarf galaxy SMF, with some evidence that the steeper (shallower) SMF observed in the isolated dwarfs (satellites) \rrr{at $\rm 10^7 \lesssim M_\star/M_\odot \lesssim 10^{10}$} can be explained by a correlation between SMF shape and star formation activity. (\autoref{s:discussion:smfslope}). 
    \item Our inferred SHMR may be steeper than implied by mass modeling of individual galaxies at \logmstar[$\sim 7$]; such a discrepancy may imply a systematic bias in kinematic measurements of dwarf halo masses, or a divergence between the low-mass SHMR and abundance matching assumptions (\autoref{s:results:shmr}, \autoref{s:discussion:kinematics}). 
    \item We do not find strong evidence for higher integrated galaxy formation efficiency ($\rm M_\star/M_{\rm peak}$) in satellite dwarfs as compared to field dwarfs, as has been recently predicted from cosmological simulations (\autoref{s:discussion:environment}).
    \item In combination with the near-unity mass-loading factor measured in \citetalias{paperone} of this series, the \sagalocal{} SHMR is naturally explained by high energy-loading factors and largely preventative star formation feedback.
    Despite this, the field dwarf SHMR is consistent with many cosmological simulations (at the 68\% CI level) that employ a largely ejective feedback scheme (\autoref{s:discussion:sims}).
    %\item In conjunction with our mass-loading measurement from \citetalias{paperone} of this series, the SHMR of field dwarfs is naturally explained by a picture in which high specific energy winds efficiently suppress galaxy formation efficiency via largely preventative feedback.
    \item The \sagalocal{} includes two quenched, isolated galaxies down to $\rm M_\star \sim 4\times 10^8 M_\odot$, implying that the isolated quenched fraction continues to drop below $\rm M_\star \sim 10^9 M_\odot$. This is contrast to contemporary big-box cosmological simulations, which largely predict a turnaround in $f_q(\rm M_\star)$ at these masses -- even when they accurately reproduce the dwarf SHMR (\autoref{s:discussion:fquench}).
\end{itemize}

The field dwarf population presents an opportune testbed in which to identify how and where competing methods to regulate galaxy and halo assembly succeed or fail.
As summarized above, physically reasonable numerical recipes for galaxy formation can simultaneously reproduce the stellar-to-halo mass relation of field dwarfs, and over-predict the quenched fraction of dwarfs in isolation. This mismatch emphasizes the complexity of evaluating prescriptions for baryonic feedback and cycling at galactic scales --- especially given that we lack \textit{ab initio} theories for many of the key astrophysical processes involved --- and the need for multidimensional observations to differentiate between competing theories for dwarf galaxy formation regulation.

In the three papers of this series, we have used the SAGAbg sample to build this kind of observational dataset: in \citetalias{paperone}, we showed that the SAGAbg dwarfs are characterized by moderately mass-loaded winds, while in \citetalias{papertwo}, we demonstrated that the Star-Forming Sequence of dwarfs has evolved significantly at $z<0.2$. Finally, in this work, we measured the integrated galaxy formation efficiency and quenched fraction of dwarfs outside the influence of a massive halo.

Each of these works aims to illuminate a different aspect of the dwarf formation problem to drive towards a complete picture of low-mass assembly across environment and cosmic time. As discussed in \autoref{s:discussion:sims} and \autoref{s:discussion:fquench}, the complexity of the problem calls for a holistic view of the dwarf population to advance our understanding of the processes that govern the assembly of low-mass galaxies and their halos.
Further elucidating the physical mechanisms of dwarf regulation will require not just a broader, but a multi-pronged view of the low-mass Universe. Fortunately, one need not wait overly long; with new and upcoming wide-field views of properties including but not limited to star formation \citep{kulkarni2021, danieli2024}, dwarf halo structure \citep{luo2024},  and interstellar and circumgalactic medium enrichment \citep{mishra2024}, and the significant expansion in the number of spectroscopically identified galaxies from large surveys like DESI \citep[e.g.][]{darraghford2023}, our view of the low-mass Universe is set to not only exponentially expand in number, but greatly broaden in parameter space over the coming years.

\begin{acknowledgements}

The ideas underpinning this work took shape during and benefited greatly from discussion at the Dwarf Galaxies, Star Clusters, and Streams Workshop hosted by the Kavli Institute for Cosmological Physics at the University of Chicago. The authors would like to thank Drummond Fielding, Pratik Gandhi, Stacy Kim, Andrey Kravtsov, Sergio Martin-Alvarez, Viraj Pandya, and Andrew Pontzen for insightful discussions that contributed substantially to the development of this work. \rrr{The authors would additionally like to thank the anonymous referee for their helpful feedback and suggestions that substantially improved the quality of this manuscript.}

M.G.\ and Y.A.\ were supported in part by a grant to M.G~from the Howard Hughes Medical Institute (HHMI) through the HHMI Professors Program.

% SAGA Standard short acknowledgments
This research used data from the SAGA Survey (Satellites Around Galactic Analogs; sagasurvey.org). The SAGA Survey is a galaxy redshift survey with spectroscopic data obtained by the SAGA Survey team with the Anglo-Australian Telescope, MMT Observatory, Palomar Observatory, W. M. Keck Observatory, and the South African Astronomical Observatory (SAAO). The SAGA Survey also made use of many public data sets, including: imaging data from the Sloan Digital Sky Survey (SDSS), the Dark Energy Survey (DES), the GALEX Survey, and the Dark Energy Spectroscopic Instrument (DESI) Legacy Imaging Surveys, which includes the Dark Energy Camera Legacy Survey (DECaLS), the Beijing-Arizona Sky Survey (BASS), and the Mayall z-band Legacy Survey (MzLS); 
redshift catalogs from SDSS, DESI, the Galaxy And Mass Assembly (GAMA) Survey, the Prism Multi-object Survey (PRIMUS), the VIMOS Public Extragalactic Redshift Survey (VIPERS), the WiggleZ Dark Energy Survey (WiggleZ), the 2dF Galaxy Redshift Survey (2dFGRS), the HectoMAP Redshift Survey, the HETDEX Source Catalog, the 6dF Galaxy Survey (6dFGS), the Hectospec Cluster Survey (HeCS), the Australian Dark Energy Survey (OzDES), the 2-degree Field Lensing Survey (2dFLenS), 
and the Las Campanas Redshift Survey (LCRS); HI data from the Arecibo Legacy Fast ALFA Survey (ALFALFA), the FAST all sky HI Survey (FASHI), and HI Parkes All-Sky Survey (HIPASS); and compiled data from the NASA-Sloan Atlas (NSA), the Siena Galaxy Atlas (SGA), the HyperLeda database, and the Extragalactic Distance Database (EDD). The SAGA Survey was supported in part by NSF collaborative grants AST-1517148 and AST-1517422 and Heising–Simons Foundation grant 2019-1402. SAGA Survey’s full acknowledgments can be found at \url{https://sagasurvey.org/ack}.
%%%%%%%%%%%%%%%%%%%%%%%%%%%%%%%%%%%%%%%%%

% BolshoiP acknowledgements
The CosmoSim database used in this paper is a service by the Leibniz-Institute for Astrophysics Potsdam (AIP). The MultiDark database was developed in cooperation with the Spanish MultiDark Consolider Project CSD2009-00064.

The authors gratefully acknowledge the Gauss Centre for Supercomputing e.V. (www.gauss-centre.eu) and the Partnership for Advanced Supercomputing in Europe (PRACE, www.prace-ri.eu) for funding the MultiDark simulation project by providing computing time on the GCS Supercomputer SuperMUC at Leibniz Supercomputing Centre (LRZ, www.lrz.de). The Bolshoi simulations have been performed within the Bolshoi project of the University of California High-Performance AstroComputing Center (UC-HiPACC) and were run at the NASA Ames Research Center.
%%%%%%%%%%%%%%%%%%%%%%%%%%%%%%%%%%%%%%%%%

% GAMA acknowledgements
GAMA is a joint European-Australasian project based around a spectroscopic campaign using the Anglo-Australian Telescope. The GAMA input catalogue is based on data taken from the Sloan Digital Sky Survey and the UKIRT Infrared Deep Sky Survey. Complementary imaging of the GAMA regions is being obtained by a number of independent survey programmes including GALEX MIS, VST KiDS, VISTA VIKING, WISE, Herschel-ATLAS, GMRT and ASKAP providing UV to radio coverage. GAMA is funded by the STFC (UK), the ARC (Australia), the AAO, and the participating institutions. The GAMA website is https://www.gama-survey.org/ .

%%%%%%%%%%%%%%%%%%%%%%%%%%%%%%%%%%%%%%%%%

% BolshoiP acknowledgements
The CosmoSim database used in this paper is a service by the Leibniz-Institute for Astrophysics Potsdam (AIP). The MultiDark database was developed in cooperation with the Spanish MultiDark Consolider Project CSD2009-00064.

The authors gratefully acknowledge the Gauss Centre for Supercomputing e.V. (www.gauss-centre.eu) and the Partnership for Advanced Supercomputing in Europe (PRACE, www.prace-ri.eu) for funding the MultiDark simulation project by providing computing time on the GCS Supercomputer SuperMUC at Leibniz Supercomputing Centre (LRZ, www.lrz.de). The Bolshoi simulations have been performed within the Bolshoi project of the University of California High-Performance AstroComputing Center (UC-HiPACC) and were run at the NASA Ames Research Center.
\end{acknowledgements}

\software{Astropy \citep{astropy:2013, astropy:2018, astropy2022}, matplotlib \citep{Hunter:2007}, SciPy \citep{scipy2020}, the IPython package \citep{PER-GRA:2007}, NumPy \citep{van2011numpy}, 
pandas \citep{pandas2022},
Astroquery \citep{astroquery}, extinction \citep{barbary2021}}

\bibliography{sagabgiii.bib, software, quenched_field_dwarfs, massloading, dwarf_LCDM, surveys}{}

\begin{thebibliography}{}
\expandafter\ifx\csname natexlab\endcsname\relax\def\natexlab#1{#1}\fi
\providecommand{\url}[1]{\href{#1}{#1}}
\providecommand{\dodoi}[1]{doi:~\href{http://doi.org/#1}{\nolinkurl{#1}}}
\providecommand{\doeprint}[1]{\href{http://ascl.net/#1}{\nolinkurl{http://ascl.net/#1}}}
\providecommand{\doarXiv}[1]{\href{https://arxiv.org/abs/#1}{\nolinkurl{https://arxiv.org/abs/#1}}}

\bibitem[{{Allen} {et~al.}(2019){Allen}, {Behroozi}, \& {Ma}}]{allen2019}
{Allen}, M., {Behroozi}, P., \& {Ma}, C.-P. 2019, \mnras, 488, 4916, \dodoi{10.1093/mnras/stz2067}

\bibitem[{{Astropy Collaboration} {et~al.}(2013){Astropy Collaboration}, {Robitaille}, {Tollerud}, {Greenfield}, {Droettboom}, {Bray}, {Aldcroft}, {Davis}, {Ginsburg}, {Price-Whelan}, {Kerzendorf}, {Conley}, {Crighton}, {Barbary}, {Muna}, {Ferguson}, {Grollier}, {Parikh}, {Nair}, {Unther}, {Deil}, {Woillez}, {Conseil}, {Kramer}, {Turner}, {Singer}, {Fox}, {Weaver}, {Zabalza}, {Edwards}, {Azalee Bostroem}, {Burke}, {Casey}, {Crawford}, {Dencheva}, {Ely}, {Jenness}, {Labrie}, {Lim}, {Pierfederici}, {Pontzen}, {Ptak}, {Refsdal}, {Servillat}, \& {Streicher}}]{astropy:2013}
{Astropy Collaboration}, {Robitaille}, T.~P., {Tollerud}, E.~J., {et~al.} 2013, \aap, 558, A33, \dodoi{10.1051/0004-6361/201322068}

\bibitem[{{Astropy Collaboration} {et~al.}(2022){Astropy Collaboration}, {Price-Whelan}, {Lim}, {Earl}, {Starkman}, {Bradley}, {Shupe}, {Patil}, {Corrales}, {Brasseur}, {N{\"o}the}, {Donath}, {Tollerud}, {Morris}, {Ginsburg}, {Vaher}, {Weaver}, {Tocknell}, {Jamieson}, {van Kerkwijk}, {Robitaille}, {Merry}, {Bachetti}, {G{\"u}nther}, {Aldcroft}, {Alvarado-Montes}, {Archibald}, {B{\'o}di}, {Bapat}, {Barentsen}, {Baz{\'a}n}, {Biswas}, {Boquien}, {Burke}, {Cara}, {Cara}, {Conroy}, {Conseil}, {Craig}, {Cross}, {Cruz}, {D'Eugenio}, {Dencheva}, {Devillepoix}, {Dietrich}, {Eigenbrot}, {Erben}, {Ferreira}, {Foreman-Mackey}, {Fox}, {Freij}, {Garg}, {Geda}, {Glattly}, {Gondhalekar}, {Gordon}, {Grant}, {Greenfield}, {Groener}, {Guest}, {Gurovich}, {Handberg}, {Hart}, {Hatfield-Dodds}, {Homeier}, {Hosseinzadeh}, {Jenness}, {Jones}, {Joseph}, {Kalmbach}, {Karamehmetoglu}, {Ka{\l}uszy{\'n}ski}, {Kelley}, {Kern}, {Kerzendorf}, {Koch}, {Kulumani}, {Lee}, {Ly}, {Ma}, {MacBride}, {Maljaars}, {Muna}, {Murphy}, {Norman},
  {O'Steen}, {Oman}, {Pacifici}, {Pascual}, {Pascual-Granado}, {Patil}, {Perren}, {Pickering}, {Rastogi}, {Roulston}, {Ryan}, {Rykoff}, {Sabater}, {Sakurikar}, {Salgado}, {Sanghi}, {Saunders}, {Savchenko}, {Schwardt}, {Seifert-Eckert}, {Shih}, {Jain}, {Shukla}, {Sick}, {Simpson}, {Singanamalla}, {Singer}, {Singhal}, {Sinha}, {Sip{\H{o}}cz}, {Spitler}, {Stansby}, {Streicher}, {{\v{S}}umak}, {Swinbank}, {Taranu}, {Tewary}, {Tremblay}, {de Val-Borro}, {Van Kooten}, {Vasovi{\'c}}, {Verma}, {de Miranda Cardoso}, {Williams}, {Wilson}, {Winkel}, {Wood-Vasey}, {Xue}, {Yoachim}, {Zhang}, {Zonca}, \& {Astropy Project Contributors}}]{astropy2022}
{Astropy Collaboration}, {Price-Whelan}, A.~M., {Lim}, P.~L., {et~al.} 2022, \apj, 935, 167, \dodoi{10.3847/1538-4357/ac7c74}

\bibitem[{{Baldry} {et~al.}(2012){Baldry}, {Driver}, {Loveday}, {Taylor}, {Kelvin}, {Liske}, {Norberg}, {Robotham}, {Brough}, {Hopkins}, {Bamford}, {Peacock}, {Bland-Hawthorn}, {Conselice}, {Croom}, {Jones}, {Parkinson}, {Popescu}, {Prescott}, {Sharp}, \& {Tuffs}}]{baldry2012}
{Baldry}, I.~K., {Driver}, S.~P., {Loveday}, J., {et~al.} 2012, \mnras, 421, 621, \dodoi{10.1111/j.1365-2966.2012.20340.x}

\bibitem[{{Barbary}(2021)}]{barbary2021}
{Barbary}, K. 2021, {extinction: Dust extinction laws}, Astrophysics Source Code Library, record ascl:2102.026.
\newblock \doeprint{2102.026}

\bibitem[{{Behroozi} {et~al.}(2019){Behroozi}, {Wechsler}, {Hearin}, \& {Conroy}}]{behroozi2019}
{Behroozi}, P., {Wechsler}, R.~H., {Hearin}, A.~P., \& {Conroy}, C. 2019, \mnras, 488, 3143, \dodoi{10.1093/mnras/stz1182}

\bibitem[{{Behroozi} {et~al.}(2013){Behroozi}, {Wechsler}, \& {Conroy}}]{behroozi2013}
{Behroozi}, P.~S., {Wechsler}, R.~H., \& {Conroy}, C. 2013, \apj, 770, 57, \dodoi{10.1088/0004-637X/770/1/57}

\bibitem[{{Benavides} {et~al.}(2025){Benavides}, {Navarro}, {Sales}, {P{\'e}rez}, \& {Bidaran}}]{benavides2025}
{Benavides}, J.~A., {Navarro}, J.~F., {Sales}, L.~V., {P{\'e}rez}, I., \& {Bidaran}, B. 2025, \apj, 985, 86, \dodoi{10.3847/1538-4357/adced0}

\bibitem[{{Bennet} {et~al.}(2019){Bennet}, {Sand}, {Crnojevi{\'c}}, {Spekkens}, {Karunakaran}, {Zaritsky}, \& {Mutlu-Pakdil}}]{bennet2019}
{Bennet}, P., {Sand}, D.~J., {Crnojevi{\'c}}, D., {et~al.} 2019, \apj, 885, 153, \dodoi{10.3847/1538-4357/ab46ab}

\bibitem[{{Bennett} {et~al.}(2024){Bennett}, {Smith}, {Fielding}, {Bryan}, {Kim}, {Springel}, \& {Hernquist}}]{bennett2024}
{Bennett}, J.~S., {Smith}, M.~C., {Fielding}, D.~B., {et~al.} 2024, arXiv e-prints, arXiv:2410.12909, \dodoi{10.48550/arXiv.2410.12909}

\bibitem[{{Bhattacharyya} {et~al.}(2025){Bhattacharyya}, {Peter}, \& {Leauthaud}}]{bhattacharyya2025}
{Bhattacharyya}, J., {Peter}, A. H.~G., \& {Leauthaud}, A. 2025, The Open Journal of Astrophysics, 8, 43, \dodoi{10.33232/001c.137130}

\bibitem[{{Blanton} {et~al.}(2011){Blanton}, {Kazin}, {Muna}, {Weaver}, \& {Price-Whelan}}]{nsa}
{Blanton}, M.~R., {Kazin}, E., {Muna}, D., {Weaver}, B.~A., \& {Price-Whelan}, A. 2011, \aj, 142, 31, \dodoi{10.1088/0004-6256/142/1/31}

\bibitem[{{B{\"o}hringer} {et~al.}(2020){B{\"o}hringer}, {Chon}, \& {Collins}}]{bohringer2020}
{B{\"o}hringer}, H., {Chon}, G., \& {Collins}, C.~A. 2020, \aap, 633, A19, \dodoi{10.1051/0004-6361/201936400}

\bibitem[{{Boylan-Kolchin} {et~al.}(2011){Boylan-Kolchin}, {Bullock}, \& {Kaplinghat}}]{boylankolchin2011}
{Boylan-Kolchin}, M., {Bullock}, J.~S., \& {Kaplinghat}, M. 2011, \mnras, 415, L40, \dodoi{10.1111/j.1745-3933.2011.01074.x}

\bibitem[{{Boylan-Kolchin} {et~al.}(2012){Boylan-Kolchin}, {Bullock}, \& {Kaplinghat}}]{boylankolchin2012}
---. 2012, \mnras, 422, 1203, \dodoi{10.1111/j.1365-2966.2012.20695.x}

\bibitem[{{Brook} \& {Di Cintio}(2015)}]{brook2015}
{Brook}, C.~B., \& {Di Cintio}, A. 2015, \mnras, 450, 3920, \dodoi{10.1093/mnras/stv864}

\bibitem[{{Buck} {et~al.}(2019){Buck}, {Macci{\`o}}, {Dutton}, {Obreja}, \& {Frings}}]{buck2019}
{Buck}, T., {Macci{\`o}}, A.~V., {Dutton}, A.~A., {Obreja}, A., \& {Frings}, J. 2019, \mnras, 483, 1314, \dodoi{10.1093/mnras/sty2913}

\bibitem[{{Bullock} \& {Boylan-Kolchin}(2017)}]{bullock2017}
{Bullock}, J.~S., \& {Boylan-Kolchin}, M. 2017, \araa, 55, 343, \dodoi{10.1146/annurev-astro-091916-055313}

\bibitem[{{Burger} {et~al.}(2025){Burger}, {Springel}, {Ostriker}, {Kim}, {Jeffreson}, {Smith}, {Pakmor}, {Hassan}, {Fielding}, {Hernquist}, {Bryan}, {Somerville}, {Bennett}, \& {Weinberger}}]{burger2025}
{Burger}, J.~D., {Springel}, V., {Ostriker}, E.~C., {et~al.} 2025, arXiv e-prints, arXiv:2502.13244, \dodoi{10.48550/arXiv.2502.13244}

\bibitem[{{Carleton} {et~al.}(2024){Carleton}, {Ellsworth-Bowers}, {Windhorst}, {Cohen}, {Conselice}, {Diego}, {Zitrin}, {Archer}, {McIntyre}, {Kamieneski}, {Jansen}, {Summers}, {D'Silva}, {Koekemoer}, {Coe}, {Driver}, {Frye}, {Grogin}, {Marshall}, {Nonino}, {Pirzkal}, {Robotham}, {Ryan}, {Ortiz}, {Tompkins}, {Willmer}, {Yan}, \& {Holwerda}}]{carleton2024}
{Carleton}, T., {Ellsworth-Bowers}, T., {Windhorst}, R.~A., {et~al.} 2024, \apjl, 961, L37, \dodoi{10.3847/2041-8213/ad1b56}

\bibitem[{{Carlsten} {et~al.}(2022){Carlsten}, {Greene}, {Beaton}, {Danieli}, \& {Greco}}]{carlsten2022}
{Carlsten}, S.~G., {Greene}, J.~E., {Beaton}, R.~L., {Danieli}, S., \& {Greco}, J.~P. 2022, \apj, 933, 47, \dodoi{10.3847/1538-4357/ac6fd7}

\bibitem[{{Carr} {et~al.}(2023){Carr}, {Bryan}, {Fielding}, {Pandya}, \& {Somerville}}]{carr2023}
{Carr}, C., {Bryan}, G.~L., {Fielding}, D.~B., {Pandya}, V., \& {Somerville}, R.~S. 2023, \apj, 949, 21, \dodoi{10.3847/1538-4357/acc4c7}

\bibitem[{{Chaikin} {et~al.}(2022){Chaikin}, {Schaye}, {Schaller}, {Bah{\'e}}, {Nobels}, \& {Ploeckinger}}]{chaikin2022}
{Chaikin}, E., {Schaye}, J., {Schaller}, M., {et~al.} 2022, \mnras, 514, 249, \dodoi{10.1093/mnras/stac1132}

\bibitem[{{Chiboucas} {et~al.}(2013){Chiboucas}, {Jacobs}, {Tully}, \& {Karachentsev}}]{chiboucas2013}
{Chiboucas}, K., {Jacobs}, B.~A., {Tully}, R.~B., \& {Karachentsev}, I.~D. 2013, \aj, 146, 126, \dodoi{10.1088/0004-6256/146/5/126}

\bibitem[{{Chilingarian} {et~al.}(2010){Chilingarian}, {Melchior}, \& {Zolotukhin}}]{chilingarian2010}
{Chilingarian}, I.~V., {Melchior}, A.-L., \& {Zolotukhin}, I.~Y. 2010, \mnras, 405, 1409, \dodoi{10.1111/j.1365-2966.2010.16506.x}

\bibitem[{{Chisholm} {et~al.}(2017){Chisholm}, {Orlitov{\'a}}, {Schaerer}, {Verhamme}, {Worseck}, {Izotov}, {Thuan}, \& {Guseva}}]{chisholm2017}
{Chisholm}, J., {Orlitov{\'a}}, I., {Schaerer}, D., {et~al.} 2017, \aap, 605, A67, \dodoi{10.1051/0004-6361/201730610}

\bibitem[{{Christensen} {et~al.}(2024){Christensen}, {Brooks}, {Munshi}, {Riggs}, {Van Nest}, {Akins}, {Quinn}, \& {Chamberland}}]{christensen2024}
{Christensen}, C.~R., {Brooks}, A.~M., {Munshi}, F., {et~al.} 2024, \apj, 961, 236, \dodoi{10.3847/1538-4357/ad0c5a}

\bibitem[{{Conroy} {et~al.}(2006){Conroy}, {Wechsler}, \& {Kravtsov}}]{conroy2006}
{Conroy}, C., {Wechsler}, R.~H., \& {Kravtsov}, A.~V. 2006, \apj, 647, 201, \dodoi{10.1086/503602}

\bibitem[{{Crain} {et~al.}(2015){Crain}, {Schaye}, {Bower}, {Furlong}, {Schaller}, {Theuns}, {Dalla Vecchia}, {Frenk}, {McCarthy}, {Helly}, {Jenkins}, {Rosas-Guevara}, {White}, \& {Trayford}}]{crain2015}
{Crain}, R.~A., {Schaye}, J., {Bower}, R.~G., {et~al.} 2015, \mnras, 450, 1937, \dodoi{10.1093/mnras/stv725}

\bibitem[{{Crnojevi{\'c}} {et~al.}(2019){Crnojevi{\'c}}, {Sand}, {Bennet}, {Pasetto}, {Spekkens}, {Caldwell}, {Guhathakurta}, {McLeod}, {Seth}, {Simon}, {Strader}, \& {Toloba}}]{crnojevic2019}
{Crnojevi{\'c}}, D., {Sand}, D.~J., {Bennet}, P., {et~al.} 2019, \apj, 872, 80, \dodoi{10.3847/1538-4357/aafbe7}

\bibitem[{{Cuillandre} {et~al.}(2025){Cuillandre}, {Bolzonella}, {Boselli}, {Marleau}, {Mondelin}, {Sorce}, {Stone}, {Buitrago}, {Cantiello}, {George}, {Hatch}, {Quilley}, {Mannucci}, {Saifollahi}, {S{\'a}nchez-Janssen}, {Tarsitano}, {Tortora}, {Xu}, {Bouy}, {Gwyn}, {Kluge}, {Lan{\c{c}}on}, {Laureijs}, {Schirmer}, {Abdurro'uf}, {Awad}, {Baes}, {Bournaud}, {Carollo}, {Codis}, {Conselice}, {De Lapparent}, {Duc}, {Ferr{\'e}-Mateu}, {Gillard}, {Golden-Marx}, {Jablonka}, {Habas}, {Hunt}, {Mei}, {Miville-Desch{\^e}nes}, {Montes}, {Nersesian}, {Peletier}, {Poulain}, {Scaramella}, {Scialpi}, {Sola}, {Stephan}, {Ulivi}, {Urbano}, {Z{\"o}ller}, {Aghanim}, {Altieri}, {Amara}, {Andreon}, {Auricchio}, {Baldi}, {Balestra}, {Bardelli}, {Bender}, {Biviano}, {Bodendorf}, {Bonino}, {Branchini}, {Brescia}, {Brinchmann}, {Camera}, {Capobianco}, {Carbone}, {Carretero}, {Casas}, {Castander}, {Castellano}, {Castignani}, {Cavuoti}, {Cimatti}, {Congedo}, {Conversi}, {Copin}, {Courbin}, {Courtois}, {Cropper}, {Da Silva}, {Degaudenzi},
  {De Lucia}, {Di Giorgio}, {Dinis}, {Douspis}, {Dubath}, {Duncan}, {Dupac}, {Dusini}, {Farina}, {Farrens}, {Ferriol}, {Fotopoulou}, {Frailis}, {Franceschi}, {Galeotta}, {Gillis}, {Giocoli}, {G{\'o}mez-Alvarez}, {Grazian}, {Grupp}, {Guzzo}, {Haugan}, {Hoar}, {Hoekstra}, {Holmes}, {Hook}, {Hormuth}, {Hornstrup}, {Hudelot}, {Jahnke}, {Jhabvala}, {Keih{\"a}nen}, {Kermiche}, {Kiessling}, {Kilbinger}, {Kitching}, {Kohley}, {Kubik}, {Kuijken}, {K{\"u}mmel}, {Kunz}, {Kurki-Suonio}, {Lahav}, {Le Mignant}, {Ligori}, {Lilje}, {Lindholm}, {Lloro}, {Maino}, {Maiorano}, {Mansutti}, {Marggraf}, {Markovic}, {Martinet}, {Marulli}, {Massey}, {Maurogordato}, {McCracken}, {Medinaceli}, {Melchior}, {Mellier}, {Meneghetti}, {Merlin}, {Meylan}, {Mohr}, {Mora}, {Moresco}, {Moscardini}, {Nakajima}, {Nichol}, {Niemi}, {Padilla}, {Paltani}, {Pasian}, {Pedersen}, {Percival}, {Pettorino}, {Pires}, {Polenta}, {Poncet}, {Popa}, {Pozzetti}, {Raison}, {Renzi}, {Rhodes}, {Riccio}, {Romelli}, {Roncarelli}, {Saglia}, {Sapone}, {Schneider},
  {Schrabback}, {Secroun}, {Seidel}, {Serrano}, {Simon}, {Sirignano}, {Sirri}, {Skottfelt}, {Stanco}, {Tallada-Cresp{\'\i}}, {Taylor}, {Teplitz}, {Tereno}, {Toledo-Moreo}, {Tutusaus}, {Valentijn}, {Valenziano}, {Vassallo}, {Verdoes Kleijn}, {Wang}, {Weller}, {Zucca}, {Burigana}, \& {Scottez}}]{cuillandre2025}
{Cuillandre}, J.~C., {Bolzonella}, M., {Boselli}, A., {et~al.} 2025, \aap, 697, A11, \dodoi{10.1051/0004-6361/202450808}

\bibitem[{{Dale}(2015)}]{dale2015}
{Dale}, J.~E. 2015, \nar, 68, 1, \dodoi{10.1016/j.newar.2015.06.001}

\bibitem[{{Danieli} {et~al.}(2023){Danieli}, {Greene}, {Carlsten}, {Jiang}, {Beaton}, \& {Goulding}}]{danieli2023}
{Danieli}, S., {Greene}, J.~E., {Carlsten}, S., {et~al.} 2023, \apj, 956, 6, \dodoi{10.3847/1538-4357/acefbd}

\bibitem[{{Danieli} {et~al.}(2024){Danieli}, {Kado-Fong}, {Huang}, {Luo}, {Li}, {Kelvin}, {Leauthaud}, {Greene}, {Mintz}, {Lin}, {Li}, {Baldassare}, {Banerjee}, {Bhattacharyya}, {Blanco}, {Brooks}, {Cai}, {Chen}, {Cruz}, {Geda}, {Guan}, {Johnson}, {Kannawadi}, {Kim}, {Li}, {Lupton}, {Mace}, {Medina}, {Pan}, {Peter}, {Read}, {C{\'o}rdova Rosado}, {Seifert}, {Wasleske}, \& {Wick}}]{danieli2024}
{Danieli}, S., {Kado-Fong}, E., {Huang}, S., {et~al.} 2024, arXiv e-prints, arXiv:2410.01884, \dodoi{10.48550/arXiv.2410.01884}

\bibitem[{{Darragh-Ford} {et~al.}(2023){Darragh-Ford}, {Wu}, {Mao}, {Wechsler}, {Geha}, {Forero-Romero}, {Hahn}, {Kallivayalil}, {Moustakas}, {Nadler}, {Nowotka}, {Peek}, {Tollerud}, {Weiner}, {Aguilar}, {Ahlen}, {Brooks}, {Cooper}, {de la Macorra}, {Dey}, {Fanning}, {Font-Ribera}, {Gontcho A Gontcho}, {Honscheid}, {Kisner}, {Kremin}, {Landriau}, {Levi}, {Martini}, {Meisner}, {Miquel}, {Myers}, {Nie}, {Palanque-Delabrouille}, {Percival}, {Prada}, {Schlegel}, {Schubnell}, {Tarl{\'e}}, {Vargas-Maga{\~n}a}, {Zhou}, \& {Zou}}]{darraghford2023}
{Darragh-Ford}, E., {Wu}, J.~F., {Mao}, Y.-Y., {et~al.} 2023, \apj, 954, 149, \dodoi{10.3847/1538-4357/ace902}

\bibitem[{{Dav{\'e}} {et~al.}(2019){Dav{\'e}}, {Angl{\'e}s-Alc{\'a}zar}, {Narayanan}, {Li}, {Rafieferantsoa}, \& {Appleby}}]{dave2019}
{Dav{\'e}}, R., {Angl{\'e}s-Alc{\'a}zar}, D., {Narayanan}, D., {et~al.} 2019, \mnras, 486, 2827, \dodoi{10.1093/mnras/stz937}

\bibitem[{{de los Reyes} {et~al.}(2024){de los Reyes}, {Asali}, {Wechsler}, {Geha}, {Mao}, {Kado-Fong}, {Pucha}, {Grant}, {Gandhi}, {Manwadkar}, {Engelhardt}, {Munshi}, \& {Wang}}]{delosreyes2024}
{de los Reyes}, M. A.~C., {Asali}, Y., {Wechsler}, R., {et~al.} 2024, arXiv e-prints, arXiv:2409.03959, \dodoi{10.48550/arXiv.2409.03959}

\bibitem[{{Dey} {et~al.}(2019){Dey}, {Schlegel}, {Lang}, {Blum}, {Burleigh}, {Fan}, {Findlay}, {Finkbeiner}, {Herrera}, {Juneau}, {Landriau}, {Levi}, {McGreer}, {Meisner}, {Myers}, {Moustakas}, {Nugent}, {Patej}, {Schlafly}, {Walker}, {Valdes}, {Weaver}, {Y{\`e}che}, {Zou}, {Zhou}, {Abareshi}, {Abbott}, {Abolfathi}, {Aguilera}, {Alam}, {Allen}, {Alvarez}, {Annis}, {Ansarinejad}, {Aubert}, {Beechert}, {Bell}, {BenZvi}, {Beutler}, {Bielby}, {Bolton}, {Brice{\~n}o}, {Buckley-Geer}, {Butler}, {Calamida}, {Carlberg}, {Carter}, {Casas}, {Castander}, {Choi}, {Comparat}, {Cukanovaite}, {Delubac}, {DeVries}, {Dey}, {Dhungana}, {Dickinson}, {Ding}, {Donaldson}, {Duan}, {Duckworth}, {Eftekharzadeh}, {Eisenstein}, {Etourneau}, {Fagrelius}, {Farihi}, {Fitzpatrick}, {Font-Ribera}, {Fulmer}, {G{\"a}nsicke}, {Gaztanaga}, {George}, {Gerdes}, {Gontcho}, {Gorgoni}, {Green}, {Guy}, {Harmer}, {Hernandez}, {Honscheid}, {Huang}, {James}, {Jannuzi}, {Jiang}, {Joyce}, {Karcher}, {Karkar}, {Kehoe}, {Kneib}, {Kueter-Young}, {Lan},
  {Lauer}, {Le Guillou}, {Le Van Suu}, {Lee}, {Lesser}, {Perreault Levasseur}, {Li}, {Mann}, {Marshall}, {Mart{\'\i}nez-V{\'a}zquez}, {Martini}, {du Mas des Bourboux}, {McManus}, {Meier}, {M{\'e}nard}, {Metcalfe}, {Mu{\~n}oz-Guti{\'e}rrez}, {Najita}, {Napier}, {Narayan}, {Newman}, {Nie}, {Nord}, {Norman}, {Olsen}, {Paat}, {Palanque-Delabrouille}, {Peng}, {Poppett}, {Poremba}, {Prakash}, {Rabinowitz}, {Raichoor}, {Rezaie}, {Robertson}, {Roe}, {Ross}, {Ross}, {Rudnick}, {Safonova}, {Saha}, {S{\'a}nchez}, {Savary}, {Schweiker}, {Scott}, {Seo}, {Shan}, {Silva}, {Slepian}, {Soto}, {Sprayberry}, {Staten}, {Stillman}, {Stupak}, {Summers}, {Sien Tie}, {Tirado}, {Vargas-Maga{\~n}a}, {Vivas}, {Wechsler}, {Williams}, {Yang}, {Yang}, {Yapici}, {Zaritsky}, {Zenteno}, {Zhang}, {Zhang}, {Zhou}, \& {Zhou}}]{dey2019}
{Dey}, A., {Schlegel}, D.~J., {Lang}, D., {et~al.} 2019, \aj, 157, 168, \dodoi{10.3847/1538-3881/ab089d}

\bibitem[{{Di Cintio} {et~al.}(2014){Di Cintio}, {Brook}, {Macci{\`o}}, {Stinson}, {Knebe}, {Dutton}, \& {Wadsley}}]{dicintio2014}
{Di Cintio}, A., {Brook}, C.~B., {Macci{\`o}}, A.~V., {et~al.} 2014, \mnras, 437, 415, \dodoi{10.1093/mnras/stt1891}

\bibitem[{{Dickey} {et~al.}(2021){Dickey}, {Starkenburg}, {Geha}, {Hahn}, {Angl{\'e}s-Alc{\'a}zar}, {Choi}, {Dav{\'e}}, {Genel}, {Iyer}, {Maller}, {Mandelker}, {Somerville}, \& {Yung}}]{dickey2021}
{Dickey}, C.~M., {Starkenburg}, T.~K., {Geha}, M., {et~al.} 2021, \apj, 915, 53, \dodoi{10.3847/1538-4357/abc014}

\bibitem[{{Doliva-Dolinsky} {et~al.}(2023){Doliva-Dolinsky}, {Martin}, {Yuan}, {Savino}, {Weisz}, {Ferguson}, {Ibata}, {Kim}, {Lewis}, {McConnachie}, \& {Thomas}}]{dolivadolinsky2023}
{Doliva-Dolinsky}, A., {Martin}, N.~F., {Yuan}, Z., {et~al.} 2023, \apj, 952, 72, \dodoi{10.3847/1538-4357/acdcf6}

\bibitem[{{Driver} {et~al.}(2022){Driver}, {Bellstedt}, {Robotham}, {Baldry}, {Davies}, {Liske}, {Obreschkow}, {Taylor}, {Wright}, {Alpaslan}, {Bamford}, {Bauer}, {Bland-Hawthorn}, {Bilicki}, {Bravo}, {Brough}, {Casura}, {Cluver}, {Colless}, {Conselice}, {Croom}, {de Jong}, {D'Eugenio}, {De Propris}, {Dogruel}, {Drinkwater}, {Dvornik}, {Farrow}, {Frenk}, {Giblin}, {Graham}, {Grootes}, {Gunawardhana}, {Hashemizadeh}, {H{\"a}u{\ss}ler}, {Heymans}, {Hildebrandt}, {Holwerda}, {Hopkins}, {Jarrett}, {Heath Jones}, {Kelvin}, {Koushan}, {Kuijken}, {Lara-L{\'o}pez}, {Lange}, {L{\'o}pez-S{\'a}nchez}, {Loveday}, {Mahajan}, {Meyer}, {Moffett}, {Napolitano}, {Norberg}, {Owers}, {Radovich}, {Raouf}, {Peacock}, {Phillipps}, {Pimbblet}, {Popescu}, {Said}, {Sansom}, {Seibert}, {Sutherland}, {Thorne}, {Tuffs}, {Turner}, {van der Wel}, {van Kampen}, \& {Wilkins}}]{driver2022}
{Driver}, S.~P., {Bellstedt}, S., {Robotham}, A. S.~G., {et~al.} 2022, \mnras, 513, 439, \dodoi{10.1093/mnras/stac472}

\bibitem[{{Efstathiou} {et~al.}(1988){Efstathiou}, {Ellis}, \& {Peterson}}]{efstathiou1988}
{Efstathiou}, G., {Ellis}, R.~S., \& {Peterson}, B.~A. 1988, \mnras, 232, 431, \dodoi{10.1093/mnras/232.2.431}

\bibitem[{{Engler} {et~al.}(2021){Engler}, {Pillepich}, {Joshi}, {Nelson}, {Pasquali}, {Grebel}, {Lisker}, {Zinger}, {Donnari}, {Marinacci}, {Vogelsberger}, \& {Hernquist}}]{engler2021}
{Engler}, C., {Pillepich}, A., {Joshi}, G.~D., {et~al.} 2021, \mnras, 500, 3957, \dodoi{10.1093/mnras/staa3505}

\bibitem[{{Feldmann} {et~al.}(2023){Feldmann}, {Quataert}, {Faucher-Gigu{\`e}re}, {Hopkins}, {{\c{C}}atmabacak}, {Kere{\v{s}}}, {Bassini}, {Bernardini}, {Bullock}, {Cenci}, {Gensior}, {Liang}, {Moreno}, \& {Wetzel}}]{feldmann2023}
{Feldmann}, R., {Quataert}, E., {Faucher-Gigu{\`e}re}, C.-A., {et~al.} 2023, \mnras, 522, 3831, \dodoi{10.1093/mnras/stad1205}

\bibitem[{{Fitts} {et~al.}(2017){Fitts}, {Boylan-Kolchin}, {Elbert}, {Bullock}, {Hopkins}, {O{\~n}orbe}, {Wetzel}, {Wheeler}, {Faucher-Gigu{\`e}re}, {Kere{\v{s}}}, {Skillman}, \& {Weisz}}]{fitts2017}
{Fitts}, A., {Boylan-Kolchin}, M., {Elbert}, O.~D., {et~al.} 2017, \mnras, 471, 3547, \dodoi{10.1093/mnras/stx1757}

\bibitem[{{Foreman-Mackey} {et~al.}(2013){Foreman-Mackey}, {Hogg}, {Lang}, \& {Goodman}}]{foremanmackey2013}
{Foreman-Mackey}, D., {Hogg}, D.~W., {Lang}, D., \& {Goodman}, J. 2013, \pasp, 125, 306, \dodoi{10.1086/670067}

\bibitem[{{Garrison-Kimmel} {et~al.}(2017){Garrison-Kimmel}, {Bullock}, {Boylan-Kolchin}, \& {Bardwell}}]{garrisonkimmel2017}
{Garrison-Kimmel}, S., {Bullock}, J.~S., {Boylan-Kolchin}, M., \& {Bardwell}, E. 2017, \mnras, 464, 3108, \dodoi{10.1093/mnras/stw2564}

\bibitem[{{Geha} {et~al.}(2012){Geha}, {Blanton}, {Yan}, \& {Tinker}}]{geha2012}
{Geha}, M., {Blanton}, M.~R., {Yan}, R., \& {Tinker}, J.~L. 2012, \apj, 757, 85, \dodoi{10.1088/0004-637X/757/1/85}

\bibitem[{{Geha} {et~al.}(2017){Geha}, {Wechsler}, {Mao}, {Tollerud}, {Weiner}, {Bernstein}, {Hoyle}, {Marchi}, {Marshall}, {Mu{\~n}oz}, \& {Lu}}]{sagai}
{Geha}, M., {Wechsler}, R.~H., {Mao}, Y.-Y., {et~al.} 2017, \apj, 847, 4, \dodoi{10.3847/1538-4357/aa8626}

\bibitem[{{Geha} {et~al.}(2024){Geha}, {Mao}, {Wechsler}, {Asali}, {Kado-Fong}, {Kallivayalil}, {Nadler}, {Tollerud}, {Weiner}, {de los Reyes}, {Wang}, \& {Wu}}]{sagaiv}
{Geha}, M., {Mao}, Y.-Y., {Wechsler}, R.~H., {et~al.} 2024, arXiv e-prints, arXiv:2404.14499, \dodoi{10.48550/arXiv.2404.14499}

\bibitem[{{Ginsburg} {et~al.}(2019){Ginsburg}, {Sip{\H{o}}cz}, {Brasseur}, {Cowperthwaite}, {Craig}, {Deil}, {Guillochon}, {Guzman}, {Liedtke}, {Lian Lim}, {Lockhart}, {Mommert}, {Morris}, {Norman}, {Parikh}, {Persson}, {Robitaille}, {Segovia}, {Singer}, {Tollerud}, {de Val-Borro}, {Valtchanov}, {Woillez}, {Astroquery Collaboration}, \& {a subset of astropy Collaboration}}]{astroquery}
{Ginsburg}, A., {Sip{\H{o}}cz}, B.~M., {Brasseur}, C.~E., {et~al.} 2019, \aj, 157, 98, \dodoi{10.3847/1538-3881/aafc33}

\bibitem[{{Gnedin} \& {Zhao}(2002)}]{gnedin2002}
{Gnedin}, O.~Y., \& {Zhao}, H. 2002, \mnras, 333, 299, \dodoi{10.1046/j.1365-8711.2002.05361.x}

\bibitem[{{Governato} {et~al.}(2012){Governato}, {Zolotov}, {Pontzen}, {Christensen}, {Oh}, {Brooks}, {Quinn}, {Shen}, \& {Wadsley}}]{governato2012}
{Governato}, F., {Zolotov}, A., {Pontzen}, A., {et~al.} 2012, \mnras, 422, 1231, \dodoi{10.1111/j.1365-2966.2012.20696.x}

\bibitem[{{Heckman} {et~al.}(2015){Heckman}, {Alexandroff}, {Borthakur}, {Overzier}, \& {Leitherer}}]{heckman2015}
{Heckman}, T.~M., {Alexandroff}, R.~M., {Borthakur}, S., {Overzier}, R., \& {Leitherer}, C. 2015, \apj, 809, 147, \dodoi{10.1088/0004-637X/809/2/147}

\bibitem[{{Hopkins} {et~al.}(2018){Hopkins}, {Wetzel}, {Kere{\v{s}}}, {Faucher-Gigu{\`e}re}, {Quataert}, {Boylan-Kolchin}, {Murray}, {Hayward}, \& {El-Badry}}]{hopkins2018}
{Hopkins}, P.~F., {Wetzel}, A., {Kere{\v{s}}}, D., {et~al.} 2018, \mnras, 477, 1578, \dodoi{10.1093/mnras/sty674}

\bibitem[{Hunter(2007)}]{Hunter:2007}
Hunter, J.~D. 2007, Computing in Science \& Engineering, 9, 90, \dodoi{10.1109/MCSE.2007.55}

\bibitem[{{Jethwa} {et~al.}(2018){Jethwa}, {Erkal}, \& {Belokurov}}]{jethwa2018}
{Jethwa}, P., {Erkal}, D., \& {Belokurov}, V. 2018, \mnras, 473, 2060, \dodoi{10.1093/mnras/stx2330}

\bibitem[{{Jiang} {et~al.}(2021){Jiang}, {Dekel}, {Freundlich}, {van den Bosch}, {Green}, {Hopkins}, {Benson}, \& {Du}}]{jiang2021}
{Jiang}, F., {Dekel}, A., {Freundlich}, J., {et~al.} 2021, \mnras, 502, 621, \dodoi{10.1093/mnras/staa4034}

\bibitem[{{Kado-Fong} {et~al.}(2024{\natexlab{a}}){Kado-Fong}, {Geha}, {Mao}, {de los Reyes}, {Wechsler}, {Asali}, {Kallivayalil}, {Nadler}, {Tollerud}, \& {Weiner}}]{paperone}
{Kado-Fong}, E., {Geha}, M., {Mao}, Y.-Y., {et~al.} 2024{\natexlab{a}}, \apj, 966, 129, \dodoi{10.3847/1538-4357/ad3042}

\bibitem[{{Kado-Fong} {et~al.}(2024{\natexlab{b}}){Kado-Fong}, {Geha}, {Mao}, {de los Reyes}, {Wechsler}, {Weiner}, {Asali}, {Kallivayalil}, {Nadler}, {Tollerud}, \& {Wang}}]{papertwo}
---. 2024{\natexlab{b}}, \apj, 976, 83, \dodoi{10.3847/1538-4357/ad8137}

\bibitem[{{Kado-Fong} {et~al.}(2024{\natexlab{c}}){Kado-Fong}, {Geha}, {Mao}, {de los Reyes}, {Wechsler}, {Weiner}, {Asali}, {Kallivayalil}, {Nadler}, {Tollerud}, \& {Wang}}]{kadofong2024b}
---. 2024{\natexlab{c}}, \apj, 976, 83, \dodoi{10.3847/1538-4357/ad8137}

\bibitem[{{Kado-Fong} {et~al.}(2024{\natexlab{d}}){Kado-Fong}, {Geha}, {Mao}, {de los Reyes}, {Wechsler}, {Asali}, {Kallivayalil}, {Nadler}, {Tollerud}, \& {Weiner}}]{kadofong2024a}
---. 2024{\natexlab{d}}, \apj, 966, 129, \dodoi{10.3847/1538-4357/ad3042}

\bibitem[{{Karachentsev} \& {Kaisina}(2019)}]{karachentsev2019}
{Karachentsev}, I.~D., \& {Kaisina}, E.~I. 2019, Astrophysical Bulletin, 74, 111, \dodoi{10.1134/S1990341319020019}

\bibitem[{{Karachentsev} {et~al.}(2015){Karachentsev}, {Kniazev}, \& {Sharina}}]{karachentsev2015}
{Karachentsev}, I.~D., {Kniazev}, A.~Y., \& {Sharina}, M.~E. 2015, Astronomische Nachrichten, 336, 707, \dodoi{10.1002/asna.201512207}

\bibitem[{{Kim} {et~al.}(2020){Kim}, {Ostriker}, {Somerville}, {Bryan}, {Fielding}, {Forbes}, {Hayward}, {Hernquist}, \& {Pandya}}]{kim2020}
{Kim}, C.-G., {Ostriker}, E.~C., {Somerville}, R.~S., {et~al.} 2020, \apj, 900, 61, \dodoi{10.3847/1538-4357/aba962}

\bibitem[{{Kim} {et~al.}(2024){Kim}, {Read}, {Rey}, {Orkney}, {Nigudkar}, {Pontzen}, {Taylor}, {Agertz}, \& {Das}}]{kim2024}
{Kim}, S.~Y., {Read}, J.~I., {Rey}, M.~P., {et~al.} 2024, arXiv e-prints, arXiv:2408.15214, \dodoi{10.48550/arXiv.2408.15214}

\bibitem[{{Klypin} {et~al.}(2011{\natexlab{a}}){Klypin}, {Trujillo-Gomez}, \& {Primack}}]{klypin2011}
{Klypin}, A.~A., {Trujillo-Gomez}, S., \& {Primack}, J. 2011{\natexlab{a}}, \apj, 740, 102, \dodoi{10.1088/0004-637X/740/2/102}

\bibitem[{{Klypin} {et~al.}(2011{\natexlab{b}}){Klypin}, {Trujillo-Gomez}, \& {Primack}}]{bolshoip}
---. 2011{\natexlab{b}}, \apj, 740, 102, \dodoi{10.1088/0004-637X/740/2/102}

\bibitem[{{Kroupa}(2001)}]{kroupa2001}
{Kroupa}, P. 2001, \mnras, 322, 231, \dodoi{10.1046/j.1365-8711.2001.04022.x}

\bibitem[{{Kulkarni} {et~al.}(2021){Kulkarni}, {Harrison}, {Grefenstette}, {Earnshaw}, {Andreoni}, {Berg}, {Bloom}, {Cenko}, {Chornock}, {Christiansen}, {Coughlin}, {Wuollet Criswell}, {Darvish}, {Das}, {De}, {Dessart}, {Dixon}, {Dorsman}, {El-Badry}, {Evans}, {Ford}, {Fremling}, {Gansicke}, {Gezari}, {Goetberg}, {Green}, {Graham}, {Heida}, {Ho}, {Jaodand}, {Johns-Krull}, {Kasliwal}, {Lazzarini}, {Lu}, {Margutti}, {Martin}, {Masters}, {McKernan}, {Naze}, {Nissanke}, {Parazin}, {Perley}, {Phinney}, {Piro}, {Raaijmakers}, {Rauw}, {Rodriguez}, {Sana}, {Senchyna}, {Singer}, {Spake}, {Stassun}, {Stern}, {Teplitz}, {Weisz}, \& {Yao}}]{kulkarni2021}
{Kulkarni}, S.~R., {Harrison}, F.~A., {Grefenstette}, B.~W., {et~al.} 2021, arXiv e-prints, arXiv:2111.15608, \dodoi{10.48550/arXiv.2111.15608}

\bibitem[{{Kunkel} \& {Demers}(1976)}]{kunkel1976}
{Kunkel}, W.~E., \& {Demers}, S. 1976, in The Galaxy and the Local Group, ed. R.~J. {Dickens}, J.~E. {Perry}, F.~G. {Smith}, \& I.~R. {King}, Vol. 182, 241

\bibitem[{{Li} {et~al.}(2024){Li}, {Greene}, {Carlsten}, \& {Danieli}}]{li2024}
{Li}, J., {Greene}, J.~E., {Carlsten}, S.~G., \& {Danieli}, S. 2024, \apjl, 975, L23, \dodoi{10.3847/2041-8213/ad5b59}

\bibitem[{{Luo} {et~al.}(2024){Luo}, {Leauthaud}, {Greene}, {Huang}, {Kado-Fong}, {Danieli}, {Li}, {Li}, {Blanco}, {Wasleske}, {Wick}, {Mintz}, {Guan}, {Peter}, {Baldassare}, {Brooks}, {Banerjee}, {Bhattacharyya}, {Cai}, {Chen}, {Gunn}, {Johnson}, {Kelvin}, {Li}, {Lin}, {Lupton}, {Mace}, {Medina}, {Read}, {Rosado}, \& {Seifert}}]{luo2024}
{Luo}, Y., {Leauthaud}, A., {Greene}, J., {et~al.} 2024, \mnras, 530, 4988, \dodoi{10.1093/mnras/stae925}

\bibitem[{{Lynden-Bell}(1976)}]{lyndenbell1976}
{Lynden-Bell}, D. 1976, \mnras, 174, 695, \dodoi{10.1093/mnras/174.3.695}

\bibitem[{{Madau} {et~al.}(2014){Madau}, {Shen}, \& {Governato}}]{madau2014}
{Madau}, P., {Shen}, S., \& {Governato}, F. 2014, \apjl, 789, L17, \dodoi{10.1088/2041-8205/789/1/L17}

\bibitem[{{Makarov} {et~al.}(2012){Makarov}, {Makarova}, {Sharina}, {Uklein}, {Tikhonov}, {Guhathakurta}, {Kirby}, \& {Terekhova}}]{makarov2012}
{Makarov}, D., {Makarova}, L., {Sharina}, M., {et~al.} 2012, \mnras, 425, 709, \dodoi{10.1111/j.1365-2966.2012.21581.x}

\bibitem[{{Makarov} {et~al.}(2014){Makarov}, {Prugniel}, {Terekhova}, {Courtois}, \& {Vauglin}}]{hyperleda}
{Makarov}, D., {Prugniel}, P., {Terekhova}, N., {Courtois}, H., \& {Vauglin}, I. 2014, \aap, 570, A13, \dodoi{10.1051/0004-6361/201423496}

\bibitem[{{Manwadkar} \& {Kravtsov}(2022)}]{manwadkar2022}
{Manwadkar}, V., \& {Kravtsov}, A.~V. 2022, \mnras, 516, 3944, \dodoi{10.1093/mnras/stac2452}

\bibitem[{{Mao} {et~al.}(2021){Mao}, {Geha}, {Wechsler}, {Weiner}, {Tollerud}, {Nadler}, \& {Kallivayalil}}]{sagaii}
{Mao}, Y.-Y., {Geha}, M., {Wechsler}, R.~H., {et~al.} 2021, \apj, 907, 85, \dodoi{10.3847/1538-4357/abce58}

\bibitem[{{Mao} {et~al.}(2024){Mao}, {Geha}, {Wechsler}, {Asali}, {Wang}, {Kado-Fong}, {Kallivayalil}, {Nadler}, {Tollerud}, {Weiner}, {de los Reyes}, \& {Wu}}]{sagaiii}
---. 2024, arXiv e-prints, arXiv:2404.14498, \dodoi{10.48550/arXiv.2404.14498}

\bibitem[{{Marasco} {et~al.}(2023){Marasco}, {Belfiore}, {Cresci}, {Lelli}, {Venturi}, {Hunt}, {Concas}, {Marconi}, {Mannucci}, {Mingozzi}, {McLeod}, {Kumari}, {Carniani}, {Vanzi}, \& {Ginolfi}}]{marasco2023}
{Marasco}, A., {Belfiore}, F., {Cresci}, G., {et~al.} 2023, \aap, 670, A92, \dodoi{10.1051/0004-6361/202244895}

\bibitem[{{Martin-Alvarez} {et~al.}(2025){Martin-Alvarez}, {Sijacki}, {Haehnelt}, {Concas}, {Yuan}, {Maiolino}, {Wechsler}, {Rodr{\'\i}guez Montero}, {Farcy}, {Sanati}, {Dubois}, {Rosdahl}, {Lopez-Rodriguez}, \& {Clark}}]{martinalvarez2025}
{Martin-Alvarez}, S., {Sijacki}, D., {Haehnelt}, M.~G., {et~al.} 2025, arXiv e-prints, arXiv:2506.03245, \dodoi{10.48550/arXiv.2506.03245}

\bibitem[{{Mashchenko} {et~al.}(2008{\natexlab{a}}){Mashchenko}, {Wadsley}, \& {Couchman}}]{maschenko2008}
{Mashchenko}, S., {Wadsley}, J., \& {Couchman}, H.~M.~P. 2008{\natexlab{a}}, Science, 319, 174, \dodoi{10.1126/science.1148666}

\bibitem[{{Mashchenko} {et~al.}(2008{\natexlab{b}}){Mashchenko}, {Wadsley}, \& {Couchman}}]{mashchenko2008}
---. 2008{\natexlab{b}}, Science, 319, 174, \dodoi{10.1126/science.1148666}

\bibitem[{{McAlpine} {et~al.}(2016){McAlpine}, {Helly}, {Schaller}, {Trayford}, {Qu}, {Furlong}, {Bower}, {Crain}, {Schaye}, {Theuns}, {Dalla Vecchia}, {Frenk}, {McCarthy}, {Jenkins}, {Rosas-Guevara}, {White}, {Baes}, {Camps}, \& {Lemson}}]{mcalpine2016}
{McAlpine}, S., {Helly}, J.~C., {Schaller}, M., {et~al.} 2016, Astronomy and Computing, 15, 72, \dodoi{10.1016/j.ascom.2016.02.004}

\bibitem[{{McGaugh}(2012)}]{mcgaugh2012}
{McGaugh}, S.~S. 2012, \aj, 143, 40, \dodoi{10.1088/0004-6256/143/2/40}

\bibitem[{{McQuinn} {et~al.}(2019){McQuinn}, {van Zee}, \& {Skillman}}]{mcquinn2019}
{McQuinn}, K. B.~W., {van Zee}, L., \& {Skillman}, E.~D. 2019, \apj, 886, 74, \dodoi{10.3847/1538-4357/ab4c37}

\bibitem[{{Mishra} {et~al.}(2024){Mishra}, {Johnson}, {Rudie}, {Chen}, {Schaye}, {Qu}, {Zahedy}, {Boettcher}, {Cantalupo}, {Chen}, {Faucher-Gigu{\'e}re}, {Greene}, {Li}, {Liu}, {Lopez}, \& {Petitjean}}]{mishra2024}
{Mishra}, N., {Johnson}, S.~D., {Rudie}, G.~C., {et~al.} 2024, \apj, 976, 149, \dodoi{10.3847/1538-4357/ad7b0a}

\bibitem[{{Monzon} {et~al.}(2024){Monzon}, {van den Bosch}, \& {Mitra}}]{monzon2024}
{Monzon}, J.~S., {van den Bosch}, F.~C., \& {Mitra}, K. 2024, \apj, 976, 197, \dodoi{10.3847/1538-4357/ad834e}

\bibitem[{{Morgan} {et~al.}(2025){Morgan}, {Sazonova}, {Roberts}, {Balogh}, {Roediger}, {Ferrarese}, {C{\^o}t{\'e}}, {Boselli}, {Fossati}, {Cuillandre}, \& {Gwyn}}]{morgan2025}
{Morgan}, C.~R., {Sazonova}, E., {Roberts}, I.~D., {et~al.} 2025, arXiv e-prints, arXiv:2505.13605, \dodoi{10.48550/arXiv.2505.13605}

\bibitem[{{Moster} {et~al.}(2018){Moster}, {Naab}, \& {White}}]{moster2018}
{Moster}, B.~P., {Naab}, T., \& {White}, S. D.~M. 2018, \mnras, 477, 1822, \dodoi{10.1093/mnras/sty655}

\bibitem[{{Mostow} {et~al.}(2024){Mostow}, {Torrey}, {Rose}, {Garcia}, {Ahvazi}, {Lisanti}, \& {Kallivayalil}}]{mostow2024}
{Mostow}, O., {Torrey}, P., {Rose}, J.~C., {et~al.} 2024, arXiv e-prints, arXiv:2412.09566, \dodoi{10.48550/arXiv.2412.09566}

\bibitem[{{Muni} {et~al.}(2025){Muni}, {Pontzen}, {Read}, {Agertz}, {Rey}, {Taylor}, {Kim}, \& {Gray}}]{muni2025}
{Muni}, C., {Pontzen}, A., {Read}, J.~I., {et~al.} 2025, \mnras, 536, 314, \dodoi{10.1093/mnras/stae2748}

\bibitem[{{Munshi} {et~al.}(2021){Munshi}, {Brooks}, {Applebaum}, {Christensen}, {Quinn}, \& {Sligh}}]{munshi2021}
{Munshi}, F., {Brooks}, A.~M., {Applebaum}, E., {et~al.} 2021, \apj, 923, 35, \dodoi{10.3847/1538-4357/ac0db6}

\bibitem[{{Muzzin} {et~al.}(2013){Muzzin}, {Marchesini}, {Stefanon}, {Franx}, {McCracken}, {Milvang-Jensen}, {Dunlop}, {Fynbo}, {Brammer}, {Labb{\'e}}, \& {van Dokkum}}]{muzzin2013}
{Muzzin}, A., {Marchesini}, D., {Stefanon}, M., {et~al.} 2013, \apj, 777, 18, \dodoi{10.1088/0004-637X/777/1/18}

\bibitem[{{Naab} \& {Ostriker}(2017)}]{naab2017}
{Naab}, T., \& {Ostriker}, J.~P. 2017, \araa, 55, 59, \dodoi{10.1146/annurev-astro-081913-040019}

\bibitem[{{Nadler} {et~al.}(2020){Nadler}, {Wechsler}, {Bechtol}, {Mao}, {Green}, {Drlica-Wagner}, {McNanna}, {Mau}, {Pace}, {Simon}, {Kravtsov}, {Dodelson}, {Li}, {Riley}, {Wang}, {Abbott}, {Aguena}, {Allam}, {Annis}, {Avila}, {Bernstein}, {Bertin}, {Brooks}, {Burke}, {Rosell}, {Kind}, {Carretero}, {Costanzi}, {da Costa}, {De Vicente}, {Desai}, {Evrard}, {Flaugher}, {Fosalba}, {Frieman}, {Garc{\'\i}a-Bellido}, {Gaztanaga}, {Gerdes}, {Gruen}, {Gschwend}, {Gutierrez}, {Hartley}, {Hinton}, {Honscheid}, {Krause}, {Kuehn}, {Kuropatkin}, {Lahav}, {Maia}, {Marshall}, {Menanteau}, {Miquel}, {Palmese}, {Paz-Chinch{\'o}n}, {Plazas}, {Romer}, {Sanchez}, {Santiago}, {Scarpine}, {Serrano}, {Smith}, {Soares-Santos}, {Suchyta}, {Tarle}, {Thomas}, {Varga}, {Walker}, \& {DES Collaboration}}]{nadler2020}
{Nadler}, E.~O., {Wechsler}, R.~H., {Bechtol}, K., {et~al.} 2020, \apj, 893, 48, \dodoi{10.3847/1538-4357/ab846a}

\bibitem[{{Navarro} {et~al.}(1996){Navarro}, {Eke}, \& {Frenk}}]{navarro1996}
{Navarro}, J.~F., {Eke}, V.~R., \& {Frenk}, C.~S. 1996, \mnras, 283, L72, \dodoi{10.1093/mnras/283.3.L72}

\bibitem[{{Navarro} {et~al.}(1997){Navarro}, {Frenk}, \& {White}}]{nfw}
{Navarro}, J.~F., {Frenk}, C.~S., \& {White}, S. D.~M. 1997, \apj, 490, 493, \dodoi{10.1086/304888}

\bibitem[{{O{\~n}orbe} {et~al.}(2015){O{\~n}orbe}, {Boylan-Kolchin}, {Bullock}, {Hopkins}, {Kere{\v{s}}}, {Faucher-Gigu{\`e}re}, {Quataert}, \& {Murray}}]{onorbe2015}
{O{\~n}orbe}, J., {Boylan-Kolchin}, M., {Bullock}, J.~S., {et~al.} 2015, \mnras, 454, 2092, \dodoi{10.1093/mnras/stv2072}

\bibitem[{{Obreschkow} {et~al.}(2018){Obreschkow}, {Murray}, {Robotham}, \& {Westmeier}}]{obreschkow2018}
{Obreschkow}, D., {Murray}, S.~G., {Robotham}, A.~S.~G., \& {Westmeier}, T. 2018, \mnras, 474, 5500, \dodoi{10.1093/mnras/stx3155}

\bibitem[{{Oh} {et~al.}(2015){Oh}, {Hunter}, {Brinks}, {Elmegreen}, {Schruba}, {Walter}, {Rupen}, {Young}, {Simpson}, {Johnson}, {Herrmann}, {Ficut-Vicas}, {Cigan}, {Heesen}, {Ashley}, \& {Zhang}}]{oh2015}
{Oh}, S.-H., {Hunter}, D.~A., {Brinks}, E., {et~al.} 2015, \aj, 149, 180, \dodoi{10.1088/0004-6256/149/6/180}

\bibitem[{{Peng} {et~al.}(2010){Peng}, {Lilly}, {Kova{\v{c}}}, {Bolzonella}, {Pozzetti}, {Renzini}, {Zamorani}, {Ilbert}, {Knobel}, {Iovino}, {Maier}, {Cucciati}, {Tasca}, {Carollo}, {Silverman}, {Kampczyk}, {de Ravel}, {Sanders}, {Scoville}, {Contini}, {Mainieri}, {Scodeggio}, {Kneib}, {Le F{\`e}vre}, {Bardelli}, {Bongiorno}, {Caputi}, {Coppa}, {de la Torre}, {Franzetti}, {Garilli}, {Lamareille}, {Le Borgne}, {Le Brun}, {Mignoli}, {Perez Montero}, {Pello}, {Ricciardelli}, {Tanaka}, {Tresse}, {Vergani}, {Welikala}, {Zucca}, {Oesch}, {Abbas}, {Barnes}, {Bordoloi}, {Bottini}, {Cappi}, {Cassata}, {Cimatti}, {Fumana}, {Hasinger}, {Koekemoer}, {Leauthaud}, {Maccagni}, {Marinoni}, {McCracken}, {Memeo}, {Meneux}, {Nair}, {Porciani}, {Presotto}, \& {Scaramella}}]{peng2010}
{Peng}, Y.-j., {Lilly}, S.~J., {Kova{\v{c}}}, K., {et~al.} 2010, \apj, 721, 193, \dodoi{10.1088/0004-637X/721/1/193}

\bibitem[{P\'erez \& Granger(2007)}]{PER-GRA:2007}
P\'erez, F., \& Granger, B.~E. 2007, Computing in Science and Engineering, 9, 21, \dodoi{10.1109/MCSE.2007.53}

\bibitem[{{Pillepich} {et~al.}(2019){Pillepich}, {Nelson}, {Springel}, {Pakmor}, {Torrey}, {Weinberger}, {Vogelsberger}, {Marinacci}, {Genel}, {van der Wel}, \& {Hernquist}}]{pillepich2019}
{Pillepich}, A., {Nelson}, D., {Springel}, V., {et~al.} 2019, \mnras, 490, 3196, \dodoi{10.1093/mnras/stz2338}

\bibitem[{{Polzin} {et~al.}(2021){Polzin}, {van Dokkum}, {Danieli}, {Greco}, \& {Romanowsky}}]{polzin2021}
{Polzin}, A., {van Dokkum}, P., {Danieli}, S., {Greco}, J.~P., \& {Romanowsky}, A.~J. 2021, \apjl, 914, L23, \dodoi{10.3847/2041-8213/ac024f}

\bibitem[{{Pontzen} {et~al.}(2021){Pontzen}, {Rey}, {Cadiou}, {Agertz}, {Teyssier}, {Read}, \& {Orkney}}]{pontzen2021}
{Pontzen}, A., {Rey}, M.~P., {Cadiou}, C., {et~al.} 2021, \mnras, 501, 1755, \dodoi{10.1093/mnras/staa3645}

\bibitem[{{Posti} {et~al.}(2019){Posti}, {Fraternali}, \& {Marasco}}]{posti2019}
{Posti}, L., {Fraternali}, F., \& {Marasco}, A. 2019, \aap, 626, A56, \dodoi{10.1051/0004-6361/201935553}

\bibitem[{{Price-Whelan} {et~al.}(2018){Price-Whelan}, {Sip{\H{o}}cz}, {G{\"u}nther}, {Lim}, {Crawford}, {Conseil}, {Shupe}, {Craig}, {Dencheva}, {Ginsburg}, {VanderPlas}, {Bradley}, {P{\'e}rez-Su{\'a}rez}, {de Val-Borro}, {Paper Contributors}, {Aldcroft}, {Cruz}, {Robitaille}, {Tollerud}, {Coordination Committee}, {Ardelean}, {Babej}, {Bach}, {Bachetti}, {Bakanov}, {Bamford}, {Barentsen}, {Barmby}, {Baumbach}, {Berry}, {Biscani}, {Boquien}, {Bostroem}, {Bouma}, {Brammer}, {Bray}, {Breytenbach}, {Buddelmeijer}, {Burke}, {Calderone}, {Cano Rodr{\'\i}guez}, {Cara}, {Cardoso}, {Cheedella}, {Copin}, {Corrales}, {Crichton}, {D{\textquoteright}Avella}, {Deil}, {Depagne}, {Dietrich}, {Donath}, {Droettboom}, {Earl}, {Erben}, {Fabbro}, {Ferreira}, {Finethy}, {Fox}, {Garrison}, {Gibbons}, {Goldstein}, {Gommers}, {Greco}, {Greenfield}, {Groener}, {Grollier}, {Hagen}, {Hirst}, {Homeier}, {Horton}, {Hosseinzadeh}, {Hu}, {Hunkeler}, {Ivezi{\'c}}, {Jain}, {Jenness}, {Kanarek}, {Kendrew}, {Kern}, {Kerzendorf}, {Khvalko},
  {King}, {Kirkby}, {Kulkarni}, {Kumar}, {Lee}, {Lenz}, {Littlefair}, {Ma}, {Macleod}, {Mastropietro}, {McCully}, {Montagnac}, {Morris}, {Mueller}, {Mumford}, {Muna}, {Murphy}, {Nelson}, {Nguyen}, {Ninan}, {N{\"o}the}, {Ogaz}, {Oh}, {Parejko}, {Parley}, {Pascual}, {Patil}, {Patil}, {Plunkett}, {Prochaska}, {Rastogi}, {Reddy Janga}, {Sabater}, {Sakurikar}, {Seifert}, {Sherbert}, {Sherwood-Taylor}, {Shih}, {Sick}, {Silbiger}, {Singanamalla}, {Singer}, {Sladen}, {Sooley}, {Sornarajah}, {Streicher}, {Teuben}, {Thomas}, {Tremblay}, {Turner}, {Terr{\'o}n}, {van Kerkwijk}, {de la Vega}, {Watkins}, {Weaver}, {Whitmore}, {Woillez}, {Zabalza}, \& {Contributors}}]{astropy:2018}
{Price-Whelan}, A.~M., {Sip{\H{o}}cz}, B.~M., {G{\"u}nther}, H.~M., {et~al.} 2018, \aj, 156, 123, \dodoi{10.3847/1538-3881/aabc4f}

\bibitem[{{Read} {et~al.}(2016){Read}, {Agertz}, \& {Collins}}]{read2016}
{Read}, J.~I., {Agertz}, O., \& {Collins}, M.~L.~M. 2016, \mnras, 459, 2573, \dodoi{10.1093/mnras/stw713}

\bibitem[{{Read} {et~al.}(2017){Read}, {Iorio}, {Agertz}, \& {Fraternali}}]{read2017}
{Read}, J.~I., {Iorio}, G., {Agertz}, O., \& {Fraternali}, F. 2017, \mnras, 467, 2019, \dodoi{10.1093/mnras/stx147}

\bibitem[{{Rey} {et~al.}(2025){Rey}, {Taylor}, {Gray}, {Kim}, {Andersson}, {Pontzen}, {Agertz}, {Read}, {Cadiou}, {Yates}, {Orkney}, {Scholte}, {Saintonge}, {Breneman}, {McQuinn}, {Muni}, \& {Das}}]{rey2025}
{Rey}, M.~P., {Taylor}, E., {Gray}, E.~I., {et~al.} 2025, arXiv e-prints, arXiv:2503.03813, \dodoi{10.48550/arXiv.2503.03813}

\bibitem[{{Rhee} {et~al.}(2024){Rhee}, {Yi}, {Ko}, {Contini}, {Jang}, {Jeon}, {Han}, {Pichon}, {Dubois}, {Kraljic}, \& {Peirani}}]{rhee2024}
{Rhee}, J., {Yi}, S.~K., {Ko}, J., {et~al.} 2024, \apj, 971, 111, \dodoi{10.3847/1538-4357/ad5a83}

\bibitem[{{Rodr{\'\i}guez-Cardoso} {et~al.}(2025){Rodr{\'\i}guez-Cardoso}, {Roca-F{\`a}brega}, {Jung}, {Nguyễn}, {Kim}, {Primack}, {Agertz}, {Barrow}, {Gallego}, {Nagamine}, {Powell}, {Revaz}, {Vel{\'a}zquez}, {Genina}, {Kim}, {Lupi}, {Abel}, {Cen}, {Ceverino}, {Dekel}, {Oh}, {Quinn}, \& {The Agora Collaboration}}]{rodriguezcardoso2025}
{Rodr{\'\i}guez-Cardoso}, R., {Roca-F{\`a}brega}, S., {Jung}, M., {et~al.} 2025, \aap, 698, A303, \dodoi{10.1051/0004-6361/202453639}

\bibitem[{{Romeo} {et~al.}(2020){Romeo}, {Agertz}, \& {Renaud}}]{romeo2020}
{Romeo}, A.~B., {Agertz}, O., \& {Renaud}, F. 2020, \mnras, 499, 5656, \dodoi{10.1093/mnras/staa3245}

\bibitem[{{Sales} {et~al.}(2022){Sales}, {Wetzel}, \& {Fattahi}}]{sales2022}
{Sales}, L.~V., {Wetzel}, A., \& {Fattahi}, A. 2022, Nature Astronomy, 6, 897, \dodoi{10.1038/s41550-022-01689-w}

\bibitem[{{Sandage} {et~al.}(1979){Sandage}, {Tammann}, \& {Yahil}}]{sandage1979}
{Sandage}, A., {Tammann}, G.~A., \& {Yahil}, A. 1979, \apj, 232, 352, \dodoi{10.1086/157295}

\bibitem[{{Sawala} {et~al.}(2015){Sawala}, {Frenk}, {Fattahi}, {Navarro}, {Bower}, {Crain}, {Dalla Vecchia}, {Furlong}, {Jenkins}, {McCarthy}, {Qu}, {Schaller}, {Schaye}, \& {Theuns}}]{sawala2015}
{Sawala}, T., {Frenk}, C.~S., {Fattahi}, A., {et~al.} 2015, \mnras, 448, 2941, \dodoi{10.1093/mnras/stu2753}

\bibitem[{{Sbaffoni} {et~al.}(2025){Sbaffoni}, {Liske}, {Driver}, {Robotham}, \& {Taylor}}]{sbaffoni2025}
{Sbaffoni}, A., {Liske}, J., {Driver}, S.~P., {Robotham}, A.~S.~G., \& {Taylor}, E.~N. 2025, \aap, 696, A89, \dodoi{10.1051/0004-6361/202453570}

\bibitem[{{Schaller} {et~al.}(2015){Schaller}, {Dalla Vecchia}, {Schaye}, {Bower}, {Theuns}, {Crain}, {Furlong}, \& {McCarthy}}]{schaller2015}
{Schaller}, M., {Dalla Vecchia}, C., {Schaye}, J., {et~al.} 2015, \mnras, 454, 2277, \dodoi{10.1093/mnras/stv2169}

\bibitem[{{Schaye} {et~al.}(2015){Schaye}, {Crain}, {Bower}, {Furlong}, {Schaller}, {Theuns}, {Dalla Vecchia}, {Frenk}, {McCarthy}, {Helly}, {Jenkins}, {Rosas-Guevara}, {White}, {Baes}, {Booth}, {Camps}, {Navarro}, {Qu}, {Rahmati}, {Sawala}, {Thomas}, \& {Trayford}}]{schaye2015}
{Schaye}, J., {Crain}, R.~A., {Bower}, R.~G., {et~al.} 2015, \mnras, 446, 521, \dodoi{10.1093/mnras/stu2058}

\bibitem[{{Smith} {et~al.}(2018){Smith}, {Sijacki}, \& {Shen}}]{smith2018}
{Smith}, M.~C., {Sijacki}, D., \& {Shen}, S. 2018, \mnras, 478, 302, \dodoi{10.1093/mnras/sty994}

\bibitem[{{Stark} {et~al.}(2009){Stark}, {McGaugh}, \& {Swaters}}]{stark2009}
{Stark}, D.~V., {McGaugh}, S.~S., \& {Swaters}, R.~A. 2009, \aj, 138, 392, \dodoi{10.1088/0004-6256/138/2/392}

\bibitem[{{The pandas development Team}(2024)}]{pandas2022}
{The pandas development Team}. 2024, {pandas-dev/pandas: Pandas}, v2.2.1,  Zenodo, \dodoi{10.5281/zenodo.3509134}

\bibitem[{{Thompson} \& {Heckman}(2024)}]{thompson2024}
{Thompson}, T.~A., \& {Heckman}, T.~M. 2024, \araa, 62, 529, \dodoi{10.1146/annurev-astro-041224-011924}

\bibitem[{{Thornton} {et~al.}(2024){Thornton}, {Amon}, {Wechsler}, {Adhikari}, {Mao}, {Myles}, {Geha}, {Kallivayalil}, {Tollerud}, \& {Weiner}}]{thornton2024}
{Thornton}, J., {Amon}, A., {Wechsler}, R.~H., {et~al.} 2024, \mnras, 535, 1, \dodoi{10.1093/mnras/stae2040}

\bibitem[{{Tollerud} {et~al.}(2014){Tollerud}, {Boylan-Kolchin}, \& {Bullock}}]{tollerud2014}
{Tollerud}, E.~J., {Boylan-Kolchin}, M., \& {Bullock}, J.~S. 2014, \mnras, 440, 3511, \dodoi{10.1093/mnras/stu474}

\bibitem[{{van den Bosch} {et~al.}(2001){van den Bosch}, {Burkert}, \& {Swaters}}]{vandenbosch2001}
{van den Bosch}, F.~C., {Burkert}, A., \& {Swaters}, R.~A. 2001, \mnras, 326, 1205, \dodoi{10.1046/j.1365-8711.2001.04656.x}

\bibitem[{Van Der~Walt {et~al.}(2011)Van Der~Walt, Colbert, \& Varoquaux}]{van2011numpy}
Van Der~Walt, S., Colbert, S.~C., \& Varoquaux, G. 2011, Computing in Science \& Engineering, 13, 22

\bibitem[{{Virtanen} {et~al.}(2020){Virtanen}, {Gommers}, {Oliphant}, {Haberland}, {Reddy}, {Cournapeau}, {Burovski}, {Peterson}, {Weckesser}, {Bright}, {van der Walt}, {Brett}, {Wilson}, {Millman}, {Mayorov}, {Nelson}, {Jones}, {Kern}, {Larson}, {Carey}, {Polat}, {Feng}, {Moore}, {VanderPlas}, {Laxalde}, {Perktold}, {Cimrman}, {Henriksen}, {Quintero}, {Harris}, {Archibald}, {Ribeiro}, {Pedregosa}, {van Mulbregt}, \& {SciPy 1. 0 Contributors}}]{scipy2020}
{Virtanen}, P., {Gommers}, R., {Oliphant}, T.~E., {et~al.} 2020, Nature Methods, 17, 261, \dodoi{10.1038/s41592-019-0686-2}

\bibitem[{{Wang} {et~al.}(2015){Wang}, {Dutton}, {Stinson}, {Macci{\`o}}, {Penzo}, {Kang}, {Keller}, \& {Wadsley}}]{wang2015}
{Wang}, L., {Dutton}, A.~A., {Stinson}, G.~S., {et~al.} 2015, \mnras, 454, 83, \dodoi{10.1093/mnras/stv1937}

\bibitem[{{Wang} {et~al.}(2024{\natexlab{a}}){Wang}, {Zhu}, {Li}, {Hong}, \& {Feng}}]{wang2024}
{Wang}, T.-R., {Zhu}, W., {Li}, X.-F., {Hong}, W.-S., \& {Feng}, L.-L. 2024{\natexlab{a}}, arXiv e-prints, arXiv:2412.09452, \dodoi{10.48550/arXiv.2412.09452}

\bibitem[{{Wang} {et~al.}(2024{\natexlab{b}}){Wang}, {Nadler}, {Mao}, {Wechsler}, {Abel}, {Behroozi}, {Geha}, {Asali}, {de los Reyes}, {Kado-Fong}, {Kallivayalil}, {Tollerud}, {Weiner}, \& {Wu}}]{sagav}
{Wang}, Y., {Nadler}, E.~O., {Mao}, Y.-Y., {et~al.} 2024{\natexlab{b}}, arXiv e-prints, arXiv:2404.14500, \dodoi{10.48550/arXiv.2404.14500}

\bibitem[{{Watts} {et~al.}(2024){Watts}, {Cortese}, {Catinella}, {Fraser-McKelvie}, {Emsellem}, {Coccato}, {van de Sande}, {Brown}, {Ascasibar}, {Battisti}, {Boselli}, {Davis}, {Groves}, \& {Thater}}]{watts2024}
{Watts}, A.~B., {Cortese}, L., {Catinella}, B., {et~al.} 2024, \mnras, 530, 1968, \dodoi{10.1093/mnras/stae898}

\bibitem[{{Weigel} {et~al.}(2016){Weigel}, {Schawinski}, \& {Bruderer}}]{weigel2016}
{Weigel}, A.~K., {Schawinski}, K., \& {Bruderer}, C. 2016, \mnras, 459, 2150, \dodoi{10.1093/mnras/stw756}

\bibitem[{{Willmer}(1997)}]{willmer1997}
{Willmer}, C.~N.~A. 1997, \aj, 114, 898, \dodoi{10.1086/118522}

\bibitem[{{Wright} {et~al.}(2017){Wright}, {Robotham}, {Driver}, {Alpaslan}, {Andrews}, {Baldry}, {Bland-Hawthorn}, {Brough}, {Brown}, {Colless}, {da Cunha}, {Davies}, {Graham}, {Holwerda}, {Hopkins}, {Kafle}, {Kelvin}, {Loveday}, {Maddox}, {Meyer}, {Moffett}, {Norberg}, {Phillipps}, {Rowlands}, {Taylor}, {Wang}, \& {Wilkins}}]{wright2017}
{Wright}, A.~H., {Robotham}, A.~S.~G., {Driver}, S.~P., {et~al.} 2017, \mnras, 470, 283, \dodoi{10.1093/mnras/stx1149}

\bibitem[{{Wu} {et~al.}(2022){Wu}, {Peek}, {Tollerud}, {Mao}, {Nadler}, {Geha}, {Wechsler}, {Kallivayalil}, \& {Weiner}}]{wu2022}
{Wu}, J.~F., {Peek}, J.~E.~G., {Tollerud}, E.~J., {et~al.} 2022, \apj, 927, 121, \dodoi{10.3847/1538-4357/ac4eea}

\bibitem[{{Xu} {et~al.}(2025){Xu}, {Jing}, {Cole}, {Frenk}, {Bose}, {Elbers}, {Wang}, {Wang}, {Moore}, {Aguilar}, {Ahlen}, {Bianchi}, {Brooks}, {Claybaugh}, {de la Macorra}, {Dey}, {Forero-Romero}, {Gazta{\~n}aga}, {Gontcho}, {Gutierrez}, {Honscheid}, {Ishak}, {Kisner}, {Koposov}, {Landriau}, {Le Guillou}, {Miquel}, {Moustakas}, {Poppett}, {Prada}, {P{\'e}rez-R{\`a}fols}, {Rossi}, {Sanchez}, {Sprayberry}, {Tarl{\'e}}, {Weaver}, \& {Zou}}]{xu2025}
{Xu}, K., {Jing}, Y.~P., {Cole}, S., {et~al.} 2025, \mnras, 540, 1635, \dodoi{10.1093/mnras/staf782}

\bibitem[{{Zaritsky} {et~al.}(2023){Zaritsky}, {Donnerstein}, {Dey}, {Karunakaran}, {Kadowaki}, {Khim}, {Spekkens}, \& {Zhang}}]{zaritsky2023}
{Zaritsky}, D., {Donnerstein}, R., {Dey}, A., {et~al.} 2023, \apjs, 267, 27, \dodoi{10.3847/1538-4365/acdd71}

\bibitem[{{Zheng} {et~al.}(2024){Zheng}, {Tchernyshyov}, {Olsen}, {Choi}, {Bustard}, {Roman-Duval}, {Zhu}, {Di Teodoro}, {Werk}, {Putman}, {McLeod}, {Faerman}, {Simons}, \& {Peek}}]{zheng2024}
{Zheng}, Y., {Tchernyshyov}, K., {Olsen}, K., {et~al.} 2024, \apj, 974, 22, \dodoi{10.3847/1538-4357/ad64d2}

\end{thebibliography}
\bibliographystyle{aasjournal}

\appendix
\makeatletter
\@twocolumnfalse
\@firstcolumnfalse
\if@twocolumn\@twocolumnfalse\fi
\makeatother

\section{The schematic methodology}\label{s:appendix:schematic}

\begin{figure*}
    \centering
    \includegraphics[width=\linewidth]{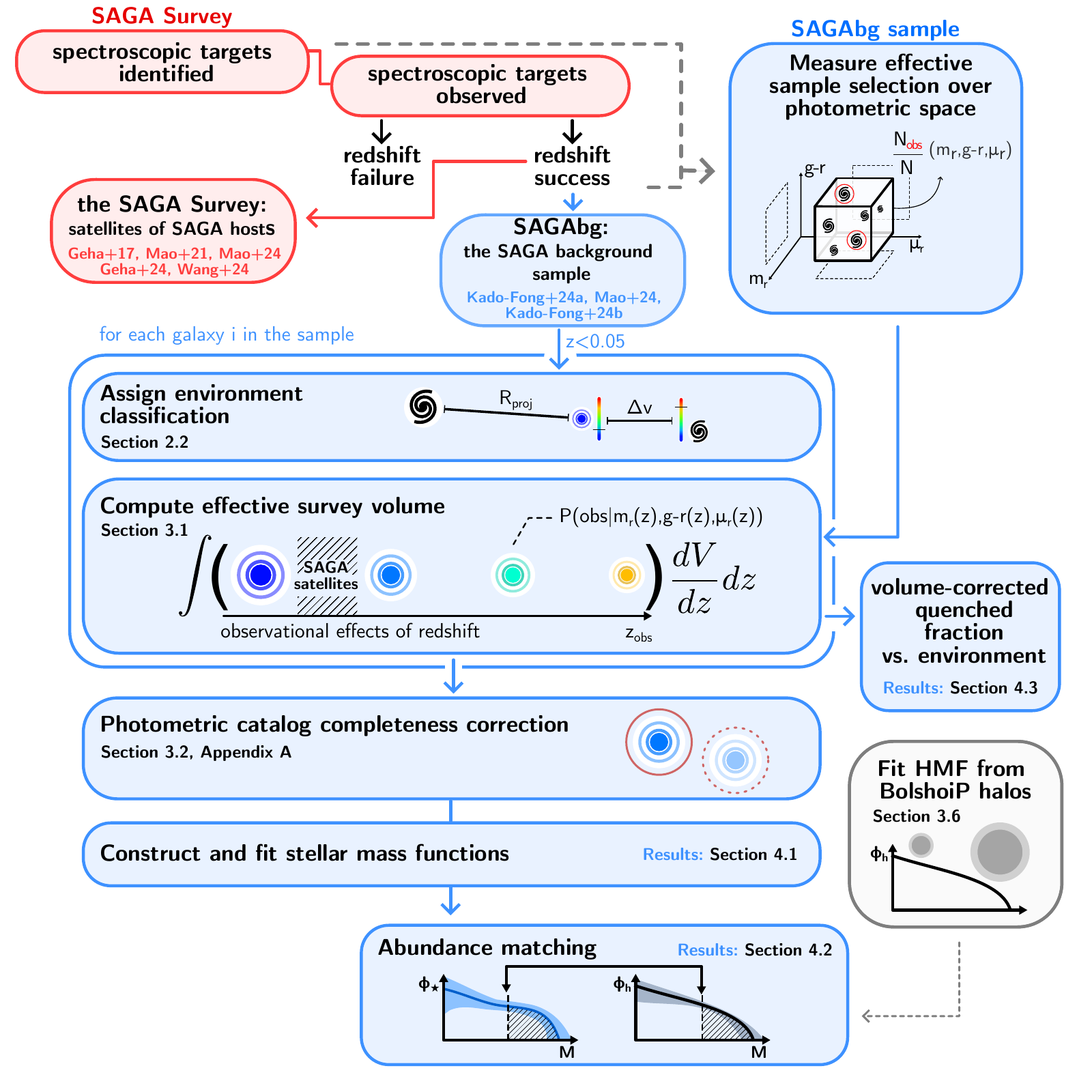}
    \caption{
        A schematic overview of the methodology employed to measure the \sagalocal{} SMF, SHMR, and quenched fraction.
    }
    \label{f:smf_selection}
\end{figure*}

To guide the reader in both the flow of the paper and the connection between the SAGA Survey and the \sagalocal{} sample, we provide a schematic description of our approach to deriving each of these quantities in \autoref{f:smf_selection}. From top to bottom, the figure walks through the process from sample selection and collection, selection function correction, and data product generation. The red panels show processing related primarily to the SAGA Survey itself, while blue panels show processing and analysis related specifically to the SAGAbg sample. The grey panel denotes the use of the \bolshoip{} dark-matter only cosmological simulation to measure halo mass functions for abundance matching. Each panel notes both the relevant section and, where applicable, a graphical representation of the process step.

\section{Photometric Incompleteness Corrections}\label{s:appendix:sbdetection}

As discussed in \autoref{s:smf:photcomplete}, the absolute completeness of the \sagalocal{} must be evaluated over both spectroscopic and photometric completeness. The DESI Legacy Imaging survey, from which the SAGA photometric catalogs are constructed, has a 5$\sigma$ point source depth of $m_r=23.9$ with an average 1.2'' seeing in the $r$-band \citep{dey2019}. 

For the sake of a rough estimate of whether photometric incompleteness should be considered, let us adopt the limiting surface brightness of compact objects to be approximately $\mu_{r,\lim}=24.9$ mag arcsec$^{-2}$. The $r-$band luminosity at which a galaxy reaches the surface brightness limit, then, can be written as:
\begin{equation}
    L_{r,\lim} = 3.4\times10^{12} (1+z)^4 R_{\rm eff}^2 g_{AB}^\nu 10^{-0.4 \mu_{r,\lim}}
\end{equation}
For a galaxy of size $R_{\rm eff} = 1$ kpc, this corresponds to a limiting $r-$band luminosity of $L_{r,\lim} = 2.8\times10^7\ L_\odot$ given a solar absolute magnitude of $M_{r,\odot}=4.61$.
The \logmstar[$<9$] galaxies in the \sagalocal{} have a median mass-to-light ratio in the $r-$band of ${\rm M}_\star/L_r = 0.9 {\rm M}_\odot/L_\odot$, implying that a photometric correction term may be necessary for the low-mass end of our sample.

To actually compute this term, we must model the allowable number density galaxies in the \sagalocal{} volume that are both missing from the photometric targeting catalog, \textit{and} at an apparent magnitude brighter than our $m_r<20.75$ limit. We choose a forward modeling approach to this problem, wherein we sample 
the observed stellar mass-to-light relation of the \sagalocal{} sample and the mass-size relation of Asali et al., 2025 (submitted) 10,000 times over a grid of stellar masses at $6<$\logmstar{}$<10$ and redshifts $0.0015<z<0.05$. We find the mass-to-light relation of the \sagalocal{} galaxies to be well-described by a normally distributed linear relation wherein \logmstar[]$-M_r = 3.21 (\pm0.01)\log_{10}(\rm M_\star/M_\odot) \ -2.2(\pm0.1)$ with $\sigma_{M/L}=0.24\pm0.01$.

For a given instantiation, the mass, redshift, size and stellar mass-to-light ratio of the model galaxy immediately yields the apparent magnitude and $r$-band surface brightness. We discard galaxies fainter than $m_r=20.75$ from the sample, and define the probability that a random galaxy at a given mass and redshift to be detected in the photometric targeting catalog as the fraction of model galaxies brighter than the surface brightness detection limit. Finally, we marginalize over redshift assuming a uniformly distributed galaxy density within the sphere to arrive at the probability that a galaxy of a given stellar mass will be detected in the photometric targeting catalog. 

\begin{figure*}
    \centering
    \includegraphics[width=\linewidth]{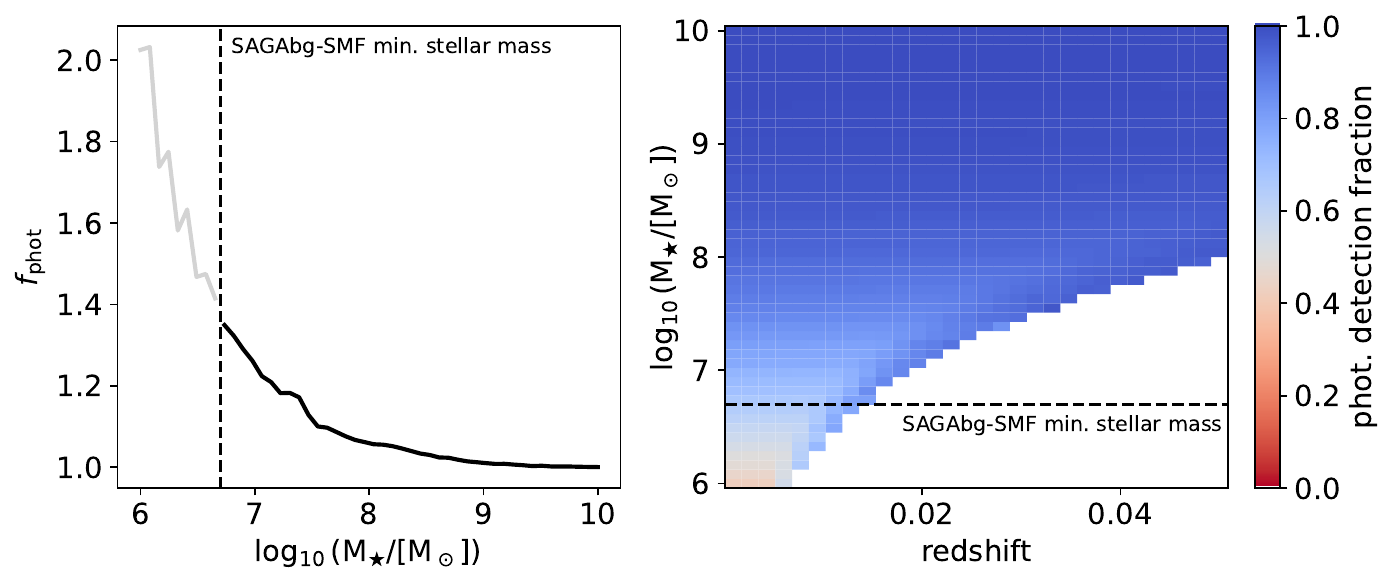}
    \caption{
        We model the distribution of the \sagalocal{} galaxies in the absolute magnitude and size space to arrive at the photometric completeness correction,
        $f_{\rm phot}$, as a function of stellar mass (left), and as a function of stellar mass and redshfit (right). The dashed black line in both panels shows the minimum stellar mass considered in this work. 
    }
    \label{f:appendix:fphot}
\end{figure*}

The results of this simple exercise are shown in \autoref{f:appendix:fphot}. At a given mass, the completeness correction small compared to the counting uncertainty incurred in that bin. We thus conclude that photometric completeness is a secondary consideration for this sample as compared to spectroscopic volume completeness.

\section{The mass-dependence of the \sagalocal{} effective volume}\label{s:appendix:veff}
The $V_{\rm eff}$ volume correction is fundamental to accurately estimating differential number densities from galaxy samples because it quantifies the volume fraction over which a given galaxy is observable in the survey, compensating for our intrinsic property-dependent ability to observe galaxies at a given distance.

\begin{figure*}[ht!]
    \centering
    \includegraphics[width=.9\linewidth]{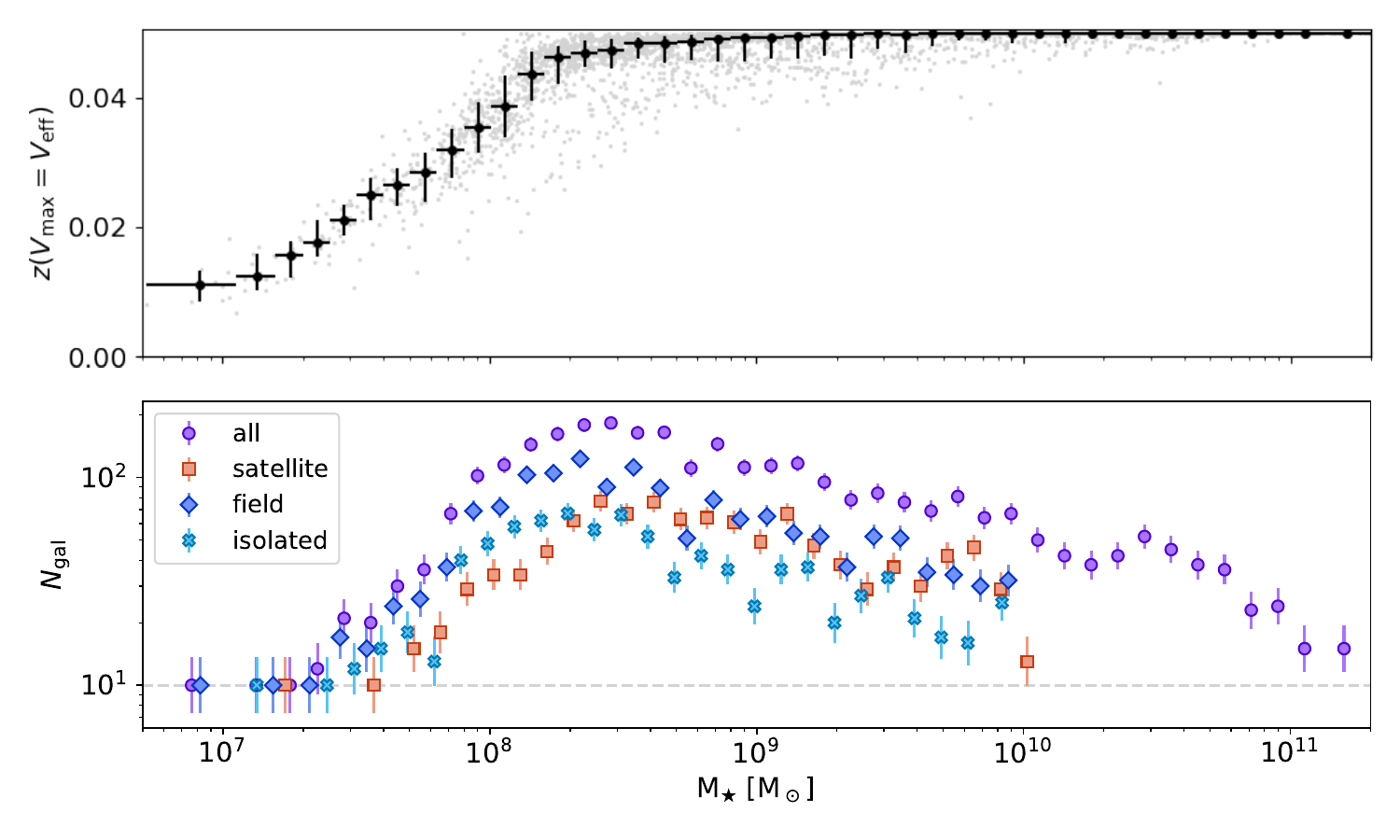}
    \caption{
        \textit{Top:} the maximum detectable redshift, $z_{\rm max}$, that our $V_{\rm eff}$ measurements would correspond to in a magnitude-limited sample. Black errorbars show the binned median (errorbars show the 16\textsuperscript{th} to 84\textsuperscript{th} percentile range) as a function of stellar mass.
        \textit{Bottom:} the raw galaxy counts for each environmental subsample of the \sagalocal{} galaxies as a function of stellar mass. The dashed grey line shows $N_{\rm gal}=10$, the minimum allowed bin count. Errorbars show the 68\% CI derived from Poissonian counting statistics assuming a Jeffreys prior.
    }
    \label{f:appendix:smf_counts}
\end{figure*}

To give the reader a sense of the comoving volume probed by the \sagalocal{} sample as a function of stellar mass, in the top panel of \autoref{f:appendix:smf_counts} we convert our $V_{\rm eff}$ values to the redshift that they would correspond to in a magnitude-limited sample -- that is, the redshift $z_{\rm max}$ such that $V_{\rm eff} = \int_0^{z_{\rm max}} \frac{dV}{dz} dz$. This shows that galaxies above $\rm M_\star \gtrsim 10^8 M_\odot$ are generally complete across the full redshift range of the \sagalocal{} sample. This can also be seen in the lower panel of 
\autoref{f:appendix:smf_counts}, which shows that the raw counts in the sample are consistent with a turnover at $\rm M_\star \sim 10^8 M_\odot$ for each of the sub-samples considered in this work.

Below this stellar mass, our volume-completeness drops substantially -- the corresponding maximum detectable redshift $z_{\rm max}$ drops to around $z=0.01$ at the lowest stellar masses considered. However, we note that it is important to remember that the $V_{\rm eff}$ method is not a way to account for galaxies that the sample has missed (as is the case for the much-smaller photometric completeness correction described in \autoref{s:smf:photcomplete} and \autoref{s:appendix:sbdetection}), but a correction to the volume over which we compute the number density of the galaxies that the sample \textit{does} contain.

\section{\rrr{The modified STY method}}

\rrr{To test our implementation of the $1/V_{\rm eff}$ method, we also implement a modification of the STY parametric fitting method \citep{sandage1979, efstathiou1988, willmer1997}  as described below in \autoref{s:model:fit}.}
\rrr{We adopt the $1/V_{\rm eff}$ method as our fiducial fitting approach: as detailed below, the modified STY approach requires stronger assumptions about how the photometric properties of galaxies in the sample change as a function of mass, whereas the $1/V_{\rm eff}$ method estimates this completeness as a function of distance, which is more directly predictable from observables. However, we implement both methods to show that our inferred SMF fitting parameters are consistent between the two methods.
}

\rrr{
In the original STY method it is assumed that there exists a sample-wide, redshift-dependent mass completeness limit, such that the probability of any object $i$ in a given galaxy sample being drawn from a parametric SMF $\Phi({\rm M_\star}|\vec X)$ described by parameters $\vec X$ can be written as
}

\begin{equation}\label{e:sty}
    \mathrrr{ {\rm Pr}[{\rm M_{\star,i}, z}|\vec X] \propto \frac{\phi({\rm M_{\star,i}}|\vec X)}{\int_{\rm M_{\star, \rm min}(z)}^{\rm M_{\star, \rm max} (z)} \phi({\rm M_\star}|\vec X) d{\rm M_\star}}}.
\end{equation}
\rrr{where $\rm M_{\star, min}(z)$ and $\rm M_{\star, max}(z)$ are the minimum and maximum stellar masses for which the sample is complete as a function of redshift $z$. The non-monotonicity of the mass completeness of the SAGAbg selection function at fixed redshift means that we cannot exactly define these minimum and maximum stellar masses. As such, we adopt a more general form of \autoref{e:sty} such that: 
}

\begin{equation}
    \mathrrr{ {\rm Pr}[{\rm M_{\star,i}, z},\vec X] \propto \frac{\phi({\rm M_{\star,i}},\vec X)}{\int_{0}^{\infty} {\rm Pr}[{\rm obs}|\vec \theta_i(\rm M_\star)]\phi({\rm M_{\star}},\vec X)d{\rm M_\star}}},
\end{equation}
\rrr{
where ${\rm Pr}[{\rm obs}|\vec \theta_i(\rm M_\star)]$ takes the same for as \autoref{e:veff}, and is the probability that a galaxy will observed with photometric properties $\vec \theta_i(\rm M_\star)$ as a function of stellar mass $\rm M_\star$. Here, we calculate the expected photometric properties of the $i$\textsuperscript{th} galaxy as a function of its stellar mass, as opposed to as a function of redshift as in \autoref{e:veff}. 
}

\rrr{
In practice, we hold the color fixed and allow the apparent magnitude and surface brightness to vary as a function of stellar mass, assuming a fixed mass-to-light ratio. We adopt the scaling of the Asali et al. 2025 (submitted) mass-size relation (where effective radius scales with stellar mass approximately as $\rm R_{eff} \propto M^{0.3}$) to estimate a rough scaling between stellar mass and r-band surface brightness of $\mu_r \propto -\log_{10}(\rm M_\star/M_\odot)$. 
We can then write the likelihood of all galaxies in the sample being drawn from $\phi({\rm M_\star},\vec X)$ as:
}

\begin{equation}
    \mathrrr{} \mathcal{L}(\text{SAGAbg-SMF}|\vec X) \propto \prod_{i}^{N_{\rm gal}} {\rm Pr}[{\rm M}_{\star,i}|\vec X].
\end{equation}

\rrr{
Due to the construction of the likelihood for the STY method, the fitting procedure cannot be used directly to constrain the absolute scale of the SMF. Although there are methods to compute this normalization after fitting, we forgo this step because the $1/V_{\rm eff}$ method includes this normalization. We adopt the same priors for the STY method as assumed for the $1/V_{\rm eff}$ method.
}

\rrr{In \autoref{t:smf} we report fitting parameters and uncertainties from the fiducial approach described in \autoref{s:model:fit}. We also report the inferred parameters obtained via the modified STY approach in \autoref{t:smfsty}, and here note that all inferred parameters are within the 68\% credible interval of our fiducial inference.}

\section{The field dwarf SHMR}\label{s:appendix:field_shmr}
In addition to the abundance matching presented in the main text (\autoref{s:model:am}), we infer the SHMR that best describes the field dwarfs of the \sagalocal{} sample.
To estimate the halo mass function as a function of the environment, we apply a modified version of the environmental classification scheme presented in \autoref{s:data:environment} to the $z=0$ snapshot of the \textsf{BolshoiP} \citep{bolshoip} dark-matter only simulations. 
We define potential host halos to be those with $\log_{10}(\rm M_{vir}/M_\odot)>12.14$, which corresponds to the lowest estimated halo mass of the HyperLEDA hosts used for the observational sample. For a randomly drawn sample of 100,000 halos in the simulation, we compute the projected radius and line-of-sight velocity difference to each potential host halo as:

\begin{equation*}
    \begin{split}
        R_{\rm proj}(\vec r_{i},\vec r_{j}) &= \sqrt{(x_i-x_j)^2+(y_i-y_j)^2}\\
        \Delta v_{i,j} &= H_0(z_i-z_j) + (v_{i,z}-v_{j,z})
    \end{split}
\end{equation*}

Where the $i$ and $j$ subscripts indicate the test object and potential host halo, respectively, $\vec r=(x,y,z)$ are the simulated positions of each halo, and $v_{z}$ is the peculiar velocity of the halo in the $z$-direction. To be consistent with our observed halo classification, we also use the halo mass-based estimate of $V_{\rm 200}$ given in \autoref{e:v200} for the simulated host halos, as we find that it is 16\% (median fractional offset) systematically higher than the true simulation maximum circular velocity reported in the simulation. We then use the simulated estimates of $R_{\rm proj}$ and $\Delta v$ to determine environment in the same way as for the observed galaxies in the sample.

We find that the fraction of satellite and field galaxies in the simulation closely matches the observed sample ($f_{\rm sample}^{\rm sim}=[0.47,0.53]$ for simulated satellite and field halos, respectively, compared to $f_{\rm sample}^{\rm obs}=[0.45,0.55]$ in the observed sample). However, the simulated halos predict a substantially lower fraction of isolated dwarfs ($f^{\rm sim}_{\rm sample}=0.16$, compared to $f^{\rm obs}_{\rm sample}=0.32$). This discrepancy is because the isolated criterion uses a constant velocity difference cutoff to determine environment (unlike the field classification, where the minimum velocity difference required to trigger the satellite condition scales with the potential host's mass). As a result, the isolated classifier is more sensitive to the lower halo mass limit chosen for potential hosts.

\begin{figure*}[ht!]
    \centering
    \includegraphics[width=\linewidth]{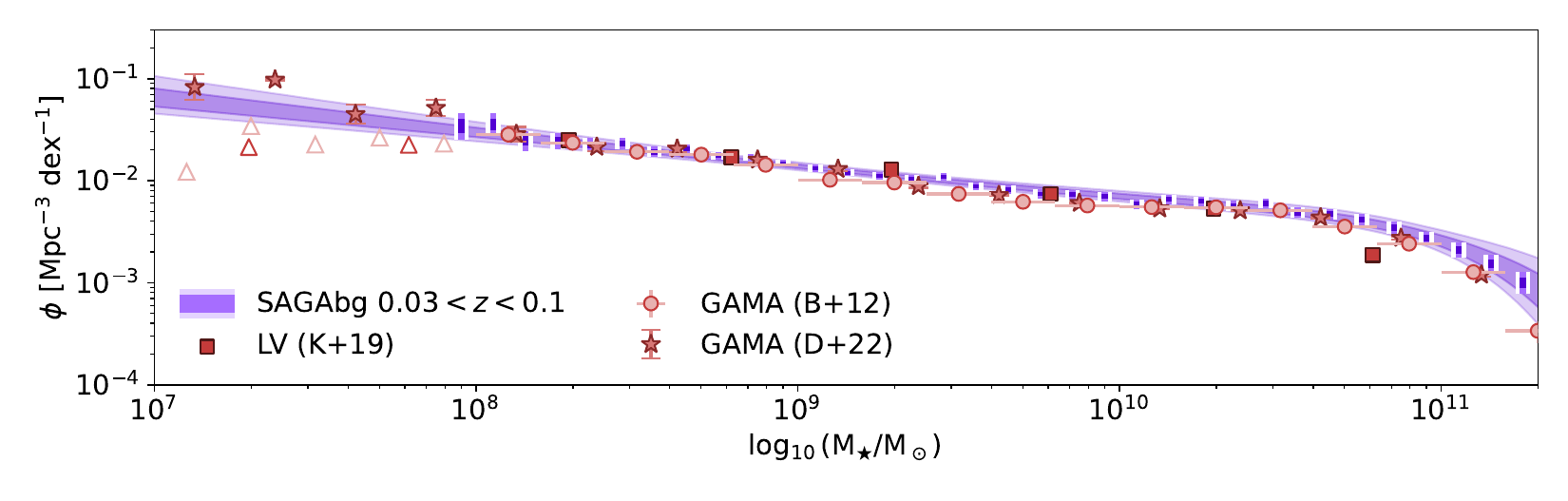}
    \caption{
        The stellar mass function of the SAGAbg sample, as in the top panel of \autoref{f:smf}, but in the redshift range $0.03<z<0.1$.
    }
    \label{f:appendix:smf_0p03_to_0p1}
\end{figure*}

\begin{figure*}[ht!]
    \centering
    \includegraphics[width=\linewidth]{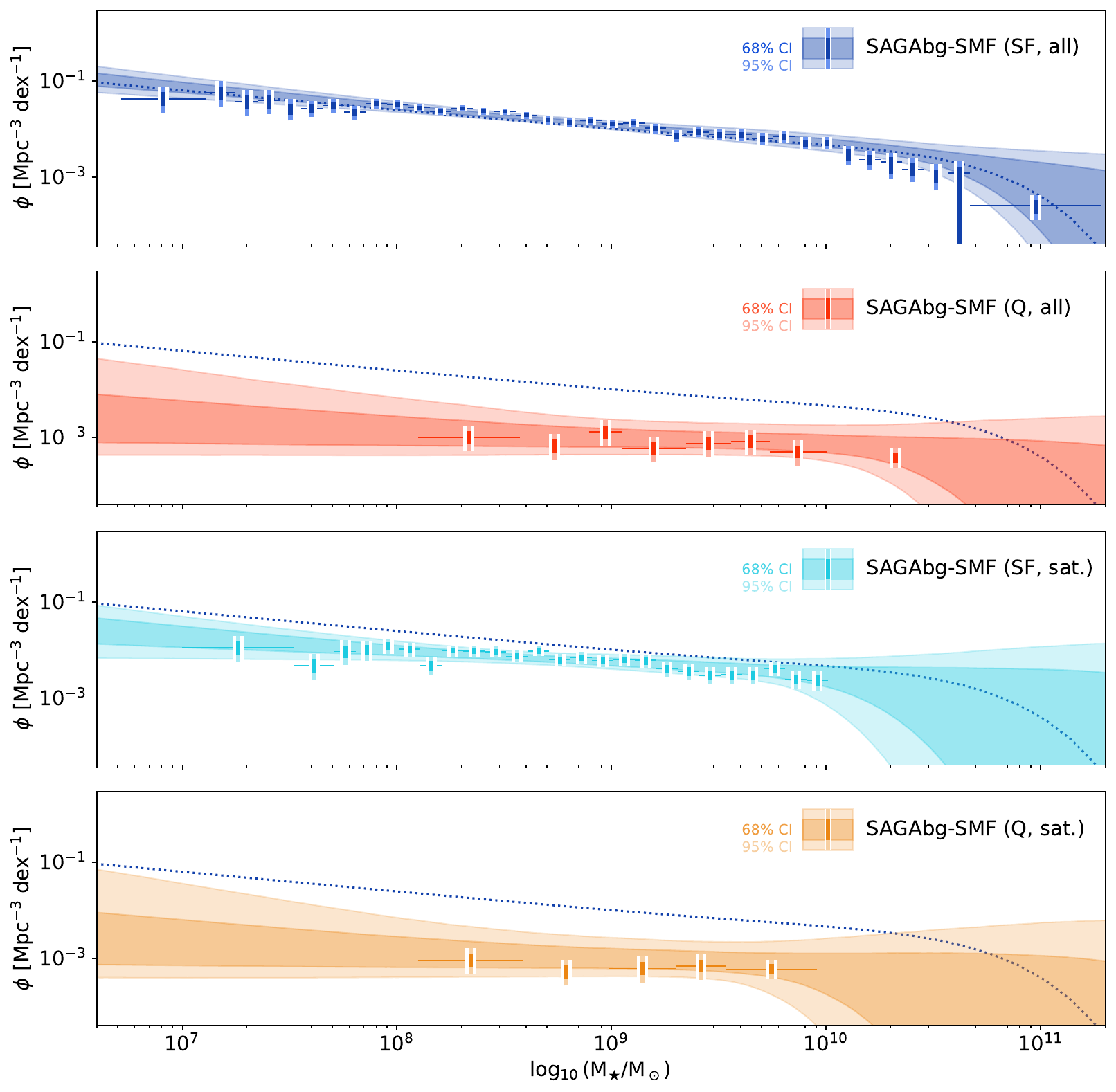}
    \caption{
        The stellar mass function of the SAGAbg sample as a function of star formation activity. From top to bottom, the panels show all star-forming galaxies, all quenched galaxies, star-forming satellites, and quenched satellites. The dotted curve in each panel shows the best-fit double Schechter function for the star-forming \sagalocal{} subsample.
    }
    \label{f:appendix:smf_sf}
\end{figure*}

The slope of the low-mass field galaxy HMF and isolated galaxy HMF are nearly identical in our test ($\beta(\text{field})=-0.96$, $\beta(\text{isolated})=-1.02$); the slope of the satellite HMF (for our definition of ``satellite'') is slightly shallower, at $\beta(\text{satellite})=-0.86$. As expected, the slope of the environmentally-averaged HMF is between these values, at $\beta(\text{all})=-0.91$. Our form for the HMF is not a good approximation for the HMF functional form of \cite{klypin2011} for very massive halos ($\rm M_h\gtrsim 10^{13} M_{\odot}$), where the true HMF declines more steeply than the low-mass power-law slope. To verify that the divergence in our approximation to the HMF does not affect our abundance matching results, we also perform abundance matching for our environmentally-averaged sample against the analytic form of the HMF given by \cite{klypin2011}, and verify that our inferred SHMR does not significantly change. 

We find that the SHMR implied via this exercise for the field dwarfs alone is statistically consistent with the abundance matching-based SHMR of the full \sagalocal{} sample, supporting the idea that the presence of satellite galaxies in the \sagalocal{} sample is not expected to fully explain the consistency between the SHMR we estimate for the \sagalocal{} sample and previous measurements of the SHMR for Local Group and Local Volume satellite systems.

\section{Additional SMFs of the SAGAbg sample}\label{s:appendix:highzsmf}
In this final appendix section, we present machine-readable tabulations of our measured SMFs and consider additional subsamples of the SAGAbg sample over which SMFs can be constructed. In particular, we compute the SMF of the SAGAbg sample at higher redshift ($0.03<z<0.1$), and as a function of star formation activity. In all cases we use the same methodology as outlined in \autoref{s:smf}, and classify star formation activity in the same way as described in \autoref{s:model:fq}.

First, we present the machine-readable SMFs measured from the \sagalocal{} sample as a function of environment in \autoref{t:smfmrt}. Only the first few rows of the table are shown here; the reader can find a full version of the table associated with the online version of this article. 

The redshift range of the \sagalocal{} sample, $z<0.05$, corresponds to a maximum comoving distance of $d\approx 210$ Mpc; the exclusion of the SAGA host volume at the low-redshift end of this range further reduces the effective comoving volume probed by the sample. To check whether considerations such as cosmic variance \citep{obreschkow2018} substantially affect our measured stellar mass functions, we additionally measure the stellar mass function of SAGAbg galaxies at $0.03<z<0.1$ using the same methodology presented in \autoref{s:smf}. 

Because we do not have a complete census of \logmstar[$\gtrsim10.5$] hosts in this redshift range, we only measure the environmentally-averaged SMF at $0.03<z<0.1$. As shown in \autoref{f:appendix:smf_0p03_to_0p1}, our measurement of the SMF out to $z=0.1$ is consistent both with our \sagalocal{} SMF and with previous literature measurements down to our limiting stellar mass in this redshift range (\logmstar[$\sim8$]). We thus conclude that volume effects such as cosmic variance do not play a substantial role in setting the shape or normalization of the \sagalocal{} stellar mass functions.

We now consider the \sagalocal{} SMF as a function of star formation activity for both the environmentally-averaged sample and the satellite subsample. Because we have very few quenched field dwarfs in the sample, the field and isolated dwarf SMFs would be unaffected by a cut in SFR; we thuse do not recalculate the field or isolated dwarf SMF as a function of star formation activity.

The resulting SMFs are shown in \autoref{f:appendix:smf_sf}; their accompanying double Schechter fits are detailed in \autoref{t:smfsf}. We find that the star-forming SMF of both the environmentally-averaged and satellite samples are consistent in their low-mass slope with that of the full, field, and isolated \sagalocal{} SMF. The quenched subsamples, meanwhile, suggest a slightly shallower SMF -- though it should be noted that the dynamic range over which the quenched SMFs are fit is around an order of magnitude smaller than that of the star-forming galaxies due to a lack of low-mass quenched systems in the sample.  

\setlength{\tabcolsep}{3pt}
\begin{deluxetable*}{cccccccc}
\tablecaption{Galaxy Stellar Mass Functions by Environment}
\tablewidth{0pt}
\tablehead{
\colhead{$\log M_{\star,\mathrm{min}}$} & 
\colhead{$\log M_{\star,\mathrm{max}}$} & 
\colhead{SMF$_{5}$} & 
\colhead{SMF$_{16}$} & 
\colhead{SMF$_{50}$} & 
\colhead{SMF$_{84}$} & 
\colhead{SMF$_{95}$} & 
\colhead{Environment} \\
\colhead{(dex)} & 
\colhead{(dex)} & 
\colhead{(Mpc$^{-3}$ dex$^{-1}$)} & 
\colhead{(Mpc$^{-3}$ dex$^{-1}$)} & 
\colhead{(Mpc$^{-3}$ dex$^{-1}$)} & 
\colhead{(Mpc$^{-3}$ dex$^{-1}$)} & 
\colhead{(Mpc$^{-3}$ dex$^{-1}$)} & 
\colhead{} 
}
\startdata
6.71 & 7.05 & 2.655e-02 & 3.754e-02 & 5.340e-02 & 7.301e-02 & 9.474e-02 & all \\
7.05 & 7.20 & 4.753e-02 & 6.748e-02 & 9.378e-02 & 1.285e-01 & 1.665e-01 & all \\
7.20 & 7.30 & 2.775e-02 & 3.950e-02 & 5.413e-02 & 7.401e-02 & 9.604e-02 & all \\
\vdots & \vdots & \vdots & \vdots & \vdots & \vdots & \vdots & \vdots \\
\enddata
\tablecomments{SMF values represent $dn/d\log_{10}(M_\star)$ in units of Mpc$^{-3}$ dex$^{-1}$. Subscripts indicate percentiles of the uncertainty distribution. Environment classifications include all galaxies, satellites, field (central) galaxies, and isolated galaxies. Only a portion of this table is shown; the full version of this table is available in machine-readable format with the online version of this article.}
\end{deluxetable*}\label{t:smfmrt}

\setlength{\tabcolsep}{3pt}
\begin{deluxetable*}{ccccccc}
\tablecaption{\rrr{Modified STY method inference of double Schechter parameters}}
\tablewidth{0pt}
\tablehead{
    \colhead{} &
    \colhead{\rrr{Environment}}&
    \colhead{\rrr{$\alpha_1$}} & 
    \colhead{\rrr{$\alpha_2$}} &     
    \colhead{\rrr{$\log_{10}\left(\frac{\rm M_{ch}}{\rm M_\odot}\right)$}} &
    \colhead{} & \colhead{} \\
}
\startdata
\midrule
\phantom{text} & \rrr{All} & \rrr{$-1.46^{+0.03}_{-0.07}$} & \rrr{$-0.3^{+0.2}_{-0.3}$} & \rrr{$10.8^{+0.3}_{-0.1}$} & \phantom{text} & \phantom{text} \\
\phantom{text} & \rrr{Satellite} & \rrr{$-1.49^{+0.03}_{-0.03}$} & \rrr{$-0.93^{+0.43}_{-0.06}$} & \rrr{$10.52^{+0.29}_{-0.02}$} & \phantom{text} & \phantom{text} \\
\phantom{text} & \rrr{Field} & \rrr{$-1.51^{+0.04}_{-0.04}$} & \rrr{$-0.39^{+0.24}_{-0.29}$} & \rrr{$10.74^{+0.21}_{-0.16}$} & \phantom{text} & \phantom{text} \\
\phantom{text} & \rrr{Isolated} & \rrr{$-1.55^{+0.11}_{-0.05}$} & \rrr{$-0.58^{+0.27}_{-0.26}$} & \rrr{$10.8^{+0.2}_{-0.2}$} & \phantom{text} & \phantom{text} \\
\enddata
\tablecomments{\rrr{Best-fit parameters for the stellar mass functions presented in \autoref{f:appendix:smf_sf} as inferred via the modified STY method presented in \autoref{s:model:fit}.}}
\end{deluxetable*}\label{t:smfsty}

\begin{deluxetable*}{ccccccc}
\tablecaption{\sagalocal{} Stellar mass function parameters as a function of star formation activity}
\tablewidth{0pt}
\tablehead{
    \colhead{Classification}&
    \colhead{$N_{\rm gal}$}&
    \colhead{$10^{3}\phi_{0,1}$} & 
    \colhead{$\alpha_1$} & 
    \colhead{$10^{3}\phi_{0,2}$} & 
    \colhead{$\alpha_2$} &     
    \colhead{$\log_{10}\left(\frac{\rm M_{ch}}{\rm M_\odot}\right)$} \\
    \colhead{} &   
    \colhead{} &
    \colhead{[Mpc$^{-3}$]} &
    \colhead{} &
    \colhead{[Mpc$^{-3}$]} &
    \colhead{} &    
    \colhead{} 
}
\startdata
\midrule
Star-forming, all & 2322 & $0.98^{+0.65}_{-0.52}$ & $-1.41^{+0.06}_{-0.07}$ & $1.91^{+1.33}_{-0.92}$ & $-0.56^{+0.39}_{-0.33}$ & $10.51^{+0.85}_{-0.24}$ \\
Quenched, all & 90 & $0.15^{+0.09}_{-0.04}$ & $-1.17^{+0.11}_{-0.18}$ & $0.49^{+0.72}_{-0.26}$ & $-0.37^{+0.26}_{-0.35}$ & $10.58^{+0.89}_{-0.58}$ \\
Star-forming, satellites & 921 & $0.50^{+0.56}_{-0.28}$ & $-1.31^{+0.12}_{-0.13}$ & $2.44^{+2.89}_{-1.42}$ & $-0.49^{+0.35}_{-0.37}$ & $10.78^{+0.86}_{-0.99}$ \\
Quenched, satellites & 71 & $0.16^{+0.11}_{-0.05}$ & $-1.16^{+0.12}_{-0.22}$ & $0.61^{+1.14}_{-0.35}$ & $-0.37^{+0.26}_{-0.35}$ & $10.38^{+1.04}_{-0.80}$ \\
\enddata
\tablecomments{Best-fit parameters for the stellar mass functions presented in \autoref{f:appendix:smf_sf}, where the marginalized posterior of each parameter is summarized by the 50\textsuperscript{th} percentile in each table entry, with the super- and subscript values showing the 68\% credible intervals.}
\end{deluxetable*}\label{t:smfsf}

\end{document}